\documentclass[a4paper,12pt]{article}              

\usepackage{graphics}
\usepackage{fancyhdr}
\usepackage{amssymb,amsfonts,amsmath,amsthm}
\usepackage{natbib}
\usepackage{bstnotations}
\usepackage[margin=30mm]{geometry}

\usepackage{colortbl}
\usepackage{tikz}
\usepackage{subfigure}

\tikzstyle{block} = [rectangle, draw, fill=blue!10, 
    text width=4cm, text centered, minimum height=0.5cm]
\tikzstyle{blockpck} = [rectangle, draw, fill=blue!10, 
    text width=6cm, text centered, minimum height=0.5cm]
\tikzstyle{blockpck2} = [rectangle, draw, fill=blue!10, 
    text width=8cm, text centered, minimum height=0.5cm]
\tikzstyle{blockdots} = [rectangle, draw, fill=blue!10, 
    text width=1cm, text centered, minimum height=0.5cm]
\tikzstyle{blockyellow} = [rectangle, draw, fill=yellow!1, 
    text width=4cm, text centered, minimum height=0.5cm]
\tikzstyle{blockred} = [rectangle, draw, fill=red!10, 
    text width=4cm, text centered, minimum height=0.5cm]
\tikzstyle{Block} = [rectangle, draw, fill=blue!20, 
    text width=15cm, text centered, minimum height=4em]
\tikzstyle{blockmin} = [rectangle, draw, fill=blue!10, 
    text width=3cm, text centered, minimum height=0.5cm]   
\tikzstyle{blockminmin} = [rectangle, draw, fill=blue!10, 
    text width=1cm, text centered, minimum height=0.5cm]  
\tikzstyle{line} = [draw, -latex']
\tikzstyle{node} = [circle, minimum width=3pt,fill, inner sep=0pt]
\usetikzlibrary{shapes,arrows}

	\newcommand{\cA}{{\mathcal A}}
	\newcommand{\cN}{{\mathcal N}}
	\newcommand{\cU}{{\mathcal U}}
	\newcommand{\vX}{\ve{X}}
	
	\newcommand{\vt}{\ve{\theta}}
	\newcommand{\vC}{\ve{\mathcal{X}}}
	\newcommand{\vc}{\ve{\mathcal{\chi}}}
	\newcommand{\cY}{\mathcal{Y}}
\DeclareMathOperator*{\argmin}{arg\,min}
\newcommand*{\argminl}{\argmin\limits}

\usepackage{authblk}

\begin{document}
\title{Polynomial-Chaos-based Kriging} 

\author[1]{Roland Sch\"obi} \author[1]{Bruno Sudret} \author[2]{Joe
  Wiart} 

\affil[1]{Chair of Risk, Safety and Uncertainty Quantification,
  Department of Civil Engineering, ETH Zurich, Stefano-Franscini-Platz
  5, 8093 Zurich, Switzerland} 

\affil[2]{WHIST Lab, Institut Mines Telecom, 46 rue Barrault
75634 Paris Cedex 13, France}

\date{}
\maketitle

\abstract{Computer simulation has become the standard tool in many
  engineering fields for designing and optimizing systems, as well as
  for assessing their reliability. Optimization and uncertainty
  quantification problems typically require a large number of runs of
  the computational model at hand, which may not be feasible with
  high-fidelity models directly. Thus surrogate models (a.k.a
  meta-models) have been increasingly investigated in the last decade.
  Polynomial Chaos Expansions (PCE) and Kriging are two popular
  non-intrusive meta-modelling techniques. PCE surrogates the
  computational model with a series of orthonormal polynomials in the
  input variables where polynomials are chosen in coherency with the
  probability distributions of those input variables. A least-square
  minimization technique may be used to determine the coefficients of
  the PCE. On the other hand, Kriging assumes that the computer model
  behaves as a realization of a Gaussian random process whose parameters
  are estimated from the available computer runs, \emph{i.e.} input
  vectors and response values. These two techniques have been developed
  more or less in parallel so far with little interaction between the
  researchers in the two fields. In this paper, \emph{PC-Kriging} is
  derived as a new non-intrusive meta-modeling approach combining PCE
  and Kriging. A sparse set of orthonormal polynomials (PCE)
  approximates the global behavior of the computational model whereas
  Kriging manages the local variability of the model output. An adaptive
  algorithm similar to the least angle regression algorithm determines
  the optimal sparse set of polynomials.  PC-Kriging is validated on
  various benchmark analytical functions which are easy to sample for
  reference results. From the numerical investigations it is concluded
  that PC-Kriging performs better than or at least as good as the two
  distinct meta-modeling techniques. A larger gain in accuracy is
  obtained when the experimental design has a limited size, which is an
  asset when dealing with demanding computational models.  

  }

\maketitle

\clearpage

\section{Introduction}
Modern engineering makes a large use of computer simulation in order to
design systems of ever increasing complexity and assess their
performance. As an example, let us consider a structural engineer
planning a new structure. An essential part of his work is to predict
the behavior of the not-yet-built structure based on some assumptions
and available information about \eg acceptable dimensions,
loads to be applied to the structure and material properties. These
ingredients are basically the input parameters of computational models
that predict the performance of the system under various conditions. The
models help the engineer analyze/understand the behavior of the
structure and eventually optimize the design in order to comply with
safety and serviceability constraints.

Similar conditions can also be found in many other scientific and
engineering disciplines. The common point is the simulation of the
behavior of a physical process or system by dedicated algorithms which
provide a numerical solution to the governing equations. These
simulation algorithms aim at reproducing the physical process with the
highest possible fidelity. As an example, finite element models have
become a standard tool in modern civil and mechanical engineering. Due
to high fidelity, such models typically exploit the available computer
power, meaning that a single run of the model may take hours to days of
computing, even when using a large number of CPUs.

An additional layer of complexity comes from the fact that most input
parameters of such computational models are not perfectly known in
practice. Some parameters (\eg material properties, applied loads, etc.)
may exhibit natural variability so that the exact value to be used for
simulating the behavior of a particular system is not known in advance
(this is referred to as {\em aleatory uncertainty}). Some others may
have a unique value which is however not directly measurable and prone
to lack of knowledge ({\em epistemic uncertainty}). In a probabilistic
setup these parameters are modelled by random variables with prescribed
joint probability density function, or more generally, by random fields.
The goal of {\em uncertainty propagation} is to assess the effect of the
input uncertainty onto the model output, and consequently onto the
performance of the system under consideration
\citep{Derocquigny2008,SudretHDR}.

Propagating uncertainties usually requires a large number of repeated
calls to the model for different values of the input parameters, for
instance through a Monte Carlo simulation procedure. Such an approach
usually require thousands to millions of runs which is not affordable
even with modern high performance computing architectures. To circumvent
this problem, {\em surrogate models} may be used, which replace the {\em
  original} computational model by an easy-to-evaluate function
\citep{storlie2009,Hastie:2001,forrester2008}. These surrogate models,
also known as \emph{response surfaces} or \emph{meta-models}, are capable
of quickly predicting responses to new input realizations. This allows
for conducting analyses which require a large number of model
evaluations, such as structural reliability and optimization, in a
reasonable time.

Among the various options for constructing meta-models, this paper
focuses on {\em non-intrusive approaches}, meaning that the
computational model is considered as a ``black-box model'': once an
input vector (\ie a realization of the random input in the process of
uncertainty propagation) is selected, the model is run and provides an
output vector of {\em quantities of interest}. No additional knowledge
on the inner structure of the computer code is assumed. Popular types of
meta-models that may be built from a limited set of runs of the original
model (called the {\em experimental design} of computations) include
Polynomial Chaos Expansions (PCE) \cite{Ghanembook2003}, Gaussian
process modeling (also called Kriging)
\cite{Sacks1989,Rasmussen2006,Santner2003,Stein:1999}, and support
vector machines \cite{Gunn1998,Smola2006,
  Vazquez:Walter:2003,Clarke:Griebsch:2003}, which have been extensively
investigated in the last decade. Polynomial chaos expansions and Kriging
are specifically of interest in this paper.

Polynomial chaos expansions (PCE), also know as spectral expansions,
approximate the computational model by a series of multivariate
polynomials which are orthogonal with respect to the distributions of
the input random variables.  Traditionally, spectral expansions have
been used to solve partial differential equations in an \emph{intrusive}
manner \cite{Ghanembook2003}. In this setup truncated expansions are
inserted into the governing equations and the expansion coefficients are
obtained using a Galerkin scheme.  This pioneering approach called
\emph{spectral stochastic finite element method} (SSFEM), was later
developed by \cite{Xiu2002,
  Sudret2004,Wan2005,Wan:Karniadakis:2006,SudretREEF2006}, among others.
These intrusive methods require specific, problem-dependent algorithmic
developments though.  Because it is not always possible and/or feasible
to treat a computational model intrusively, especially when legacy codes
are at hand in an industrial context, non-intrusive polynomial chaos
expansions were developed. So-called {\em projections methods} were
developed by \cite{Ghoicel2002,Lemaitre02,keese2005,Xiu:Hesthaven:2005},
see a review in \cite{Xiu2009a}. {\em Least-square minimization}
techniques have been introduced by \cite{Choi2004a,Berveiller2006a,
  Sudret2008b}.  Further developments which combine spectral expansions
and compressive sensing ideas have lead to so-called {\em sparse
  polynomial chaos expansions} \cite{BlatmanCras2008,BlatmanPEM2010,
  BlatmanRESS2010, Doostan2011, BlatmanJCP2011,
  Doostan2013,Jakeman2014a}.  This is the approach followed in this
paper. Further recent applications of PCE to structural reliability
analysis and design optimization can be found in
\cite{eldred2009,eldred2008,sarangi2014}.

The second meta-modeling technique of interest in this paper is Kriging,
which originates from interpolating geographical data in mining
\cite{Krige1951} and is today also known as Gaussian process modeling
\cite{Rasmussen2006,Santner2003}. The Kriging meta-model is interpreted
as the realization of a Gaussian process. Practical applications can be
found in many fields, such as structural reliability analysis
\cite{Kaymaz2005,Bect2012,Echard2011,Bichon2008,Dubourg2013,Dubourg2014a}
and design optimization \cite{Jones1998,Dubourg2011a,DubourgThesis}.
The implementation of the Kriging meta-modeling technique can be found
in \emph{e.g.} the Matlab toolbox \emph{DACE} \cite{Lophaven2002} and
the more recent R toolbox \emph{DiceKriging}
\cite{roustant2012,roustant2013}.

So far the two distinct meta-modeling approaches have been applied in
various fields rather independently. To our knowledge, there has not
been any attempt to combine PCE and Kriging in a systematic way yet. To
bridge the gap between the two communities, this paper aims at combining
the two distinct approaches into a new and more powerful meta-modeling
technique called \emph{Polynomial-Chaos-Kriging} (PC-Kriging). As seen
in the sequel the combination of the characteristics and the advantages
of both approaches leads to a more accurate and flexible algorithm which
will be detailed in this paper.

The paper is organized as follows. PCE and Kriging are first summarized
in Section~\ref{sec:pce} and Section~\ref{sec:k} respectively.
Section~\ref{sec:pck} introduces the new meta-modeling approach
PC-Kriging as the combination of the two distinct approaches. The
accuracy of the new and traditional meta-modeling approaches is compared
in Section~\ref{sec:ana} on a set of benchmark analytical functions.

\section{Polynomial Chaos Expansions} \label{sec:pce}
\subsection{Problem definition}
Consider the probability space $(\Omega,\mathcal{F},\mathcal{P})$, where
$\Omega$ denotes the event space equipped with $\sigma$-algebra
$\mathcal{F}$ and the probability measure $\mathcal{P}$. Random
variables are denoted by capital letters $X(\omega): \Omega\mapsto
\mathcal{D}_X\subset \Rr$ and their realizations denoted by the
corresponding lower case letters, \emph{e.g.} $x$. Random vectors
(\emph{e.g.} $\ve{X} = \{ X_1,\ldots,X_M \}\tr$) and their realizations
(\emph{e.g.} $\ve{x} = \{ x_1,\ldots,x_M \}\tr$) are denoted by bold
faced capital and lower case letters, respectively.

In this context, consider a system whose behavior is represented by a
computational model $\cm$ which maps the $M$-dimensional input parameter
space to the $1$-dimensional output space, \emph{i.e.} $\cm: \ve{x} \in
\cd_X \subset \Rr^M \mapsto y\in \Rr$ where $\ve{x}=\{x_1,\ldots,x_M\}\tr$.
As the input vector $\ve{x}$ is assumed to be affected by uncertainty, a
probabilistic framework is introduced. Due to uncertainties in the input
vector, it is represented by a random vector $\ve{X}$ with given joint
probability density function (PDF) $f_{\ve{X}}$. For the sake of
simplicity the components are assumed independent throughout the paper,
so that the joint PDF may be written as the product of the marginal PDFs
denoted by $f_{X_i}$, $i=1,\ldots,M$. Note that the case of dependent
input variables can easily be addressed by using an isoprobabilistic
transform first, such as the Nataf or Rosenblatt transform
\cite{BlatmanPEM2010}. The output of the model is a random variable $Y$
obtained by propagating the input uncertainty in $\ve{X}$ through the
computational model $\cm$:
\begin{equation}
 Y = \cm(\vX).
\end{equation} 
In this paper we consider that the computational model is
a \emph{deterministic} mapping from the input to the output space,
\emph{i.e.} repeated evaluations with the same input value $\ve{x}_0\in
\mathcal{D}_X$ lead to the same output value $y_0=\cm(\ve{x}_0)$.

Provided that the output random variable $Y$ is a second-order variable
(\emph{i.e.} $\Esp{Y^2}< +\infty$), it can be cast as the following
polynomial chaos expansion (PCE) \cite{Ghanembook2003,Soize2004}:
\begin{equation} \label{eq:ypce} Y \equiv \cm(\vX) = \sum_{\ua \in
\Nn^M} \ve{a}_{\ua} \,\psi_{\ua}(\vX),
\end{equation} where $\{\ve{a}_{\ua},\, \ua \in \Nn^M\}$ are
coefficients of the multivariate orthonormal polynomials
$\psi_{\ua}(\vX)$ in coherency with the distribution of the input random
vector $\vX$, $\ua=\{\alpha_1,\ldots,\alpha_M\}$ is the multi-index and
$M$ is the number of input variables (dimensions). Since the components
of $\vX$ are independent, the joint probability density function
$f_{\ve{X}}$ is the product of the margins $f_{X_i}$. Then a
\emph{functional inner product} for each marginal PDF $f_{X_i}$ is
defined by
\begin{equation} 
\langle \phi_1,\phi_2 \rangle_i = \int_{\mathcal{D}_i}
\phi_1(x) \,\phi_2(x) \,f_{X_i}(x)dx,
\end{equation} 
for any two functions $\{\phi_i,\phi_2\}$ such that the integral exists.
For each variable $i=1,\ldots,M$ an orthonormal polynomial basis can be
constructed which satisfies \cite{Xiu2002}:
\begin{equation} 
\langle P_j^{(i)},P_k^{(i)} \rangle =
\int_{\mathcal{D}_i} P_j^{(i)}(x) \,P_k^{(i)}(x) \,f_{X_i}(x)dx =
\delta_{jk},
\end{equation}
where $P_j^{(i)},\,P_k^{(i)}$ are two candidate univariate polynomials
in the $i$-th variable, $\cd_i$ is the support of the random variable
$X_i$ and $\delta_{jk}$ is the Kronecker delta which is equal to 1 for
$j=k$ and equal to $0$ otherwise. \cite{Xiu2002} summarize various
orthonormal bases for some classical PDFs, some of which are summarized
in Tab.~\ref{tab:orthobasis}.

\begin{table} [ht]
  \caption{\label{tab:orthobasis} Classical orthogonal/orthonormal polynomials (as presented in \cite{SudretBookPhoon2014})}
\centering
\vspace*{0.1in} \small
\begin{tabular}{l l l l }
  \hline
  Distribution & PDF & Orthogonal polynomials & Orthonormal basis \\
  \hline
  Uniform & $\mathbf{1}_{]-1,1[}(x)/2$ & Legendre $P_k(x)$ & $P_k(x)/\sqrt{\frac{1}{2k+1}}$ \\
  Gaussian & $\frac{1}{\sqrt{2 \pi}}e^{-x^2/2}$ & Hermite $H_{e_k}(x)$ & $H_{e_k}(x)/\sqrt{k!}$ \\
  Gamma & $x^a e^{-x} \mathbf{1}_{\mathbb{R}^+}(x)$ & Laguerre $L_k^a(x)$ & $L_k^a(x)/\sqrt{\frac{\Gamma(k+a+1)}{k!}}$ \\
  Beta & $\mathbf{1}_{]-1,1[}(x)\frac{(1-x)^a(1+x)^b}{B(a)B(b)}$ & Jacobi $J_k^{a,b}(x)$ & $J_k^{a,b}(x)/ \mathcal{J}_{a,b,k}$ \\
   & & &{$\mathcal{J}_{a,b,k}^2= \frac{2^{a+b+1}}{2k+a+b+1} \frac{\Gamma(k+a+1)\Gamma(k+b+1)}{\Gamma(k+a+b+1)\Gamma(k+1)}$}\\
  \hline
  \\
\end{tabular}
\end{table}

The multivariate polynomials in Eq.~(\ref{eq:ypce}) are then composed of
univariate polynomials by tensor product, \ie by multiplying the various
polynomials in each input variable:
\begin{equation}
\psi_{\ua}(\vX) = \prod_{i=1}^M \psi_{\alpha_i}^{(i)}(X_i),
\end{equation}
where $\psi_{\alpha_i}^{(i)}$ is the polynomial of degree $\alpha_i$ in the $i$-th variable.

The main idea of polynomial chaos expansion (PCE) is then to surrogate
the computational model by an infinite series of polynomials as shown in
Eq.~(\ref{eq:ypce}). In practice, it is not feasible to handle infinite
series, thus the need of a truncation scheme. Such a truncation scheme
corresponds to a set of multi-indexes $\ua\in \cA \subset \Nn^M$ such
that the system response is accurately approximated with respect to some
error measure \cite{BlatmanPEM2010,BlatmanJCP2011}:
\begin{equation}\label{eq:pce}
  Y \approx Y^{\text{(PCE)}} \eqdef\cm^{\text{(PCE)}}(\vX) = \sum_{\ua
    \in \cA} \ve{a}_{\ua} \,\psi_{\ua}(\vX). 
\end{equation}
There are several ways to select \emph{a priori} a truncation set $\cA$.
A simple and commonly applied scheme consists in upper-bounding the
total degree of polynomials to a maximal value $p$. The total degree of
polynomials is defined by
\begin{equation} \label{eq:total}
|\ua| = \sum_{i=1}^M \alpha_i.
\end{equation}
In this case the set of multi-indices is denoted by $\cA^{M,p} = \{ \ua
\in \Nn^M : |\ua| \leq p \}$ where $p$ is the maximal total polynomial
degree. The cardinality of the set $\cA$ reads:
\begin{equation}
\left|\cA^{M,p}\right| = \frac{(M+p)!}{M!\,p!}.
\end{equation}
This cardinality grows polynomially with both $M$ and $p$. Such a
truncation scheme thus leads to non tractable problems if the response
is highly nonlinear in its input parameters (need for a large $p$)
and/or if the size of the input vector $\vX$ is large (say, $M>10$).
This problem is referred to as the \emph{curse of dimensionality}.

\begin{figure} [!ht]
  \centering
    \includegraphics[width=0.7\linewidth, angle=0]{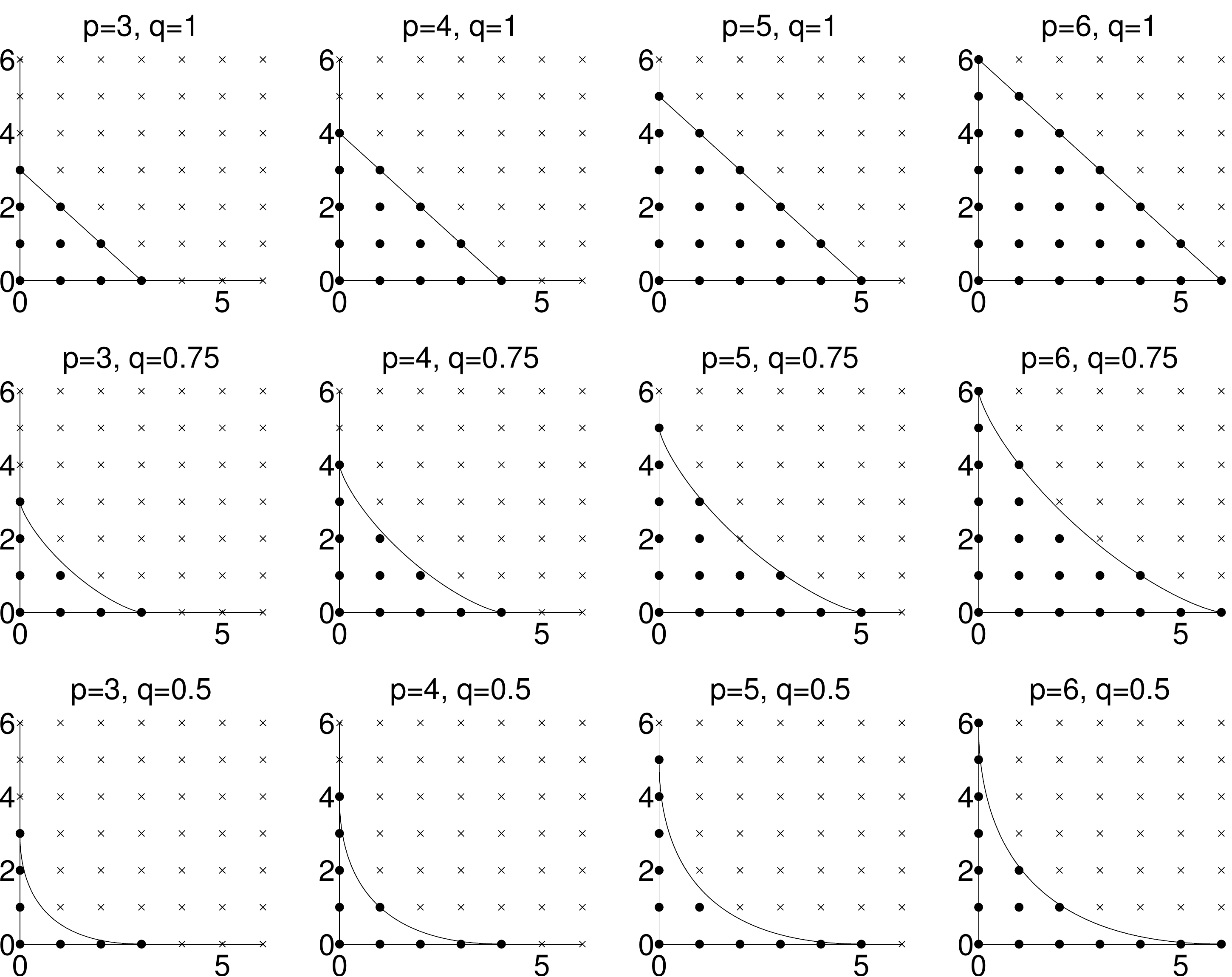}
  \caption{\label{fig:index} Representation of a hyperbolic index set $\ua\in\cA_q^{M,p}$ for various $p$ and $q$ ($M=2$)}
\end{figure}

\cite{BlatmanPEM2010,BlatmanThesis} proposed a more restrictive
truncation scheme called \emph{hyperbolic truncation set}. The authors
observed that many systems tend to have only low-degree interaction
polynomials and thus it is not necessary to compute all interaction
terms of higher polynomial degree. The hyperbolic index set is based on
the following $q$-norm:
\begin{equation}
\cA_q^{M,p} \equiv \{ \ua \in \Nn^M : \|\ua \|_q \leq p \},
\end{equation}
where
\begin{equation}
\| \ua \|_q \equiv \left( \sum_{i=1}^M \alpha_i^q\right)^{\frac{1}{q}},
\end{equation}
$0<q\leq1$ is a tuning parameter and $p$ is the maximal total degree of
the polynomials. A decreasing $q$ leads to a smaller number of
interactive polynomials, \emph{i.e.} a smaller set of polynomials. When
$q \rightarrow 0$, only univariate polynomials are left in the set of
polynomials which is called an additive model
\cite{SudretBookPhoon2014}. For the sake of illustration, the retained
polynomial indices $\ua\in \cA_q^{M,p}$ of a 2-dimensional input space
($M=2$) and varying $p$ and $q$ are illustrated in
Figure~\ref{fig:index}. The indices denoted by $\bullet$ are part of
$\cA_q^{M,p}$ and the solid black line represents $\| \ua \|_q=p$. Note
that for $q=1$, the hyperbolic index sets are equivalent to the total
degree index set (see Eq.~(\ref{eq:total})).

\subsection{Computation of the coefficients}
After defining the set of candidate polynomials, the next step is to determine the expansion coefficients $\ve{a}_{\ua}$ of each multivariate polynomial $\psi_{\ua}(\ve{x})$. In this paper we consider only \emph{non-intrusive} methods which are based on repeatedly evaluating the model $\cm$ over a set of input realizations $\vC=\{  \vc^{(1)},\ldots,\vc^{(N)} \}$, the so-called \emph{experimental design}. Different non-intrusive methods have been proposed in the last decade to calibrate PC meta-models, namely projection \cite{Ghoicel2002,Lemaitre02,keese2005}, stochastic collocation \cite{Xiu:Hesthaven:2005,Xiu2009a} and least-square minimization methods \cite{Chkifa2013,Migliorati2014,Berveiller2006a,BlatmanPEM2010,BlatmanJCP2011}. In this paper we adopt the least-square minimization method. The expansion coefficients $\ve{a}=\{ \ve{a}_{\ua},\, \ua \in \cA \subset \Nn^M\}$ are calculated by minimizing the expectation of the least-squares residual:
\begin{equation}\label{eq:argmin}
\ve{a} = \argminl_{\ve{a} \in \Rr^{|\cA|}} \Esp{ \left( Y-\sum_{\ua \in \cA} \ve{a}_{\ua}\, \psi_{\ua}(\vX) \right)^2 }.
\end{equation}
In practice the expectation in Eq.~(\ref{eq:argmin}) is evaluated by an
empirical sample-based estimator. Denoting by $\cY=\{
\cm(\vc^{(1)}),\ldots,\cm(\vc^{(N)}) \} \equiv \{
\cY^{(1)},\ldots,\cY^{(N)} \}$ the set of outputs of the exact model $\cm$ for
each point in the experimental design $\vC$, the discretized
least-squares error minimization problem derived from
Eq.~(\ref{eq:argmin}) reads
\begin{equation}
\hat{\ve{a}} = \argminl_{\ve{a} \in \Rr^{|\cA|}} \frac{1}{N} \sum_{i=1}^N \left( \cY^{(i)}-\sum_{\ua \in \cA} \ve{a}_{\ua} \,\psi_{\ua}(\vc^{(i)}) \right)^2.
\end{equation}
The optimal expansion coefficients $\hat{\ve{a}}$ may be computed by solving the linear system
\begin{equation} \label{eq:apce}
\hat{\ve{a}} = (\mathbf{F}\tr \mathbf{F})^{-1}\mathbf{F}\tr \cY,
\end{equation}
where $\mathbf{F}$ is the information matrix of size $N \times |\cA|$ whose generic term reads:
\begin{equation} \label{eq:f}
F_{ij} = \psi_j(\vc^{(i)}),\ \ \ i=1,\ldots,N,\ \ j=1,\ldots,|\cA|.
\end{equation}

Typically for smooth functions, a small number of polynomials is able to represent accurately the output of the computational model. Thus a further reduction of the set of predictors in high-dimensional space is possible. Various types of generalized regression algorithms have been proposed in the literature, namely the \emph{least absolute shrinkage operator} (LASSO) \cite{tibshirani1996},  the \emph{Least Angle Regression} (LAR) \cite{Efron2004}, \emph{low-rank approximations} \cite{Doostan2013,Peng2014} and compressive sensing \cite{Sargsyan2014}. 
In the applications of this paper, the LAR algorithm is used in combination with hyperbolic index sets.

Given a PCE meta-model, \emph{i.e.} a set of polynomials $\psi_{\ua}, \,
\ua\in\cA$ and the corresponding parameters $\ve{a}_{\ua}$, the response
of a new sample $\ve{x}\in \cd_X$ may be eventually predicted by
Eq.~(\ref{eq:pce}):
\begin{equation}
y^{(\text{PCE})} = \cm^{(\text{PCE})}(\ve{x}).
\end{equation}

\subsection{Error estimation} \label{sec:errorestimation}
As seen in Eq.~(\ref{eq:pce}), polynomial chaos expansions (PCE) are approximations of the exact computational model and thus the prediction at new input samples leads to some residual error. That is why error measures are developed to quantify the deviation between the exact output $\cY=\{ \cY^{(i)},i=1,\ldots,N \}$ and the meta-model output $\cY^{\textsf{(PCE)}} = \cm^{\textsf{(PCE)}}(\vC)$. The \emph{generalization error} (also called $L^2$-error) is the expectation of the squared output residuals \cite{Vapnik1995}, \emph{i.e.}
\begin{equation}\label{eq:errGen}
Err_{gen} = \Esp{ \left( Y-Y^{\textsf{(PCE)}} \right)^2},
\end{equation}
where $Y^{\text{(PCE)}}$ corresponds to the truncated series (see
Eq.~(\ref{eq:pce})) and the expectation is defined with respect to the
PDF of the input variables $\ve{X}$. If the computational model $\cm$ is
inexpensive to evaluate, the generalization error can be estimated
accurately using an auxiliary validation set $\mathbb{X} = \{
\ve{x}^{(1)},\ldots,\ve{x}^{(n)} \}$, which is sampled from the input
distribution $f_{\vX}$. The estimate of the generalization error then
reads:
\begin{equation}\label{eq:errGengen}
\widehat{Err}_{gen} = \frac{1}{n} \sum_{i=1}^n \left( \cm(\ve{x}^{(i)})-\cm^{\text{(PCE)}}(\ve{x}^{(i)}) \right)^2.
\end{equation}
However, this rarely happens in real applications since the very purpose
of building a meta-model is to avoid evaluating $\cm$ on a large sample
set $\mathbb{X}$. Note that in Section~\ref{sec:ana} though, the various
meta-modeling techniques are compared on analytical benchmark functions,
making it possible to use such an estimate of the generalization error.

When the use of a large validation set is not affordable, the {\em
  empirical error} based on the available experimental design $\vC$ may
be defined:
\begin{equation} \label{eq:err}
Err_{emp} \equiv \frac{1}{N} \sum_{i=1}^N  \left( \cY^{(i)}-\cm^{\textsf{(PCE)}}(\vc^{(i)}) \right)^2.
\end{equation}
Normalizing the empirical error by the variance of the output values leads to the \emph{relative empirical error} which is defined as
\begin{equation} \label{eq:erremp}
\epsilon_{emp} \equiv \frac{\sum_{i=1}^N  \left( \cY^{(i)}-\cm^{\textsf{(PCE)}}(\vc^{(i)}) \right)^2}{\sum_{i=1}^N  \left( \cY^{(i)}- \mu_{\cY} \right)^2},
\end{equation}
where $\mu_{\cY}$ is the mean value of the output values $\cY$. The empirical error which is based on the experimental design generally underestimates the generalization error. In particular if the number of polynomials $|\cA|$ is close to the number of samples $N$ in the experimental design, the empirical error tends to zero whereas the true (generalization) error does not. This phenomenon is called \emph{overfitting} (in the extreme case where $N=|\cA|$ the predictors may interpolate the experimental design points and thus the empirical error vanishes). Hence the \emph{leave-one-out} (LOO) error has been proposed as an estimate of the generalization error \cite{Stone1974,Geisser:1975}. The general formulation of the leave-one-out error is
\begin{equation} \label{eq:pceloo}
Err_{LOO}^{\text{(PCE)}} \equiv \frac{1}{N} \sum_{i=1}^N \left( \cY^{(i)}-\cm^{\textsf{(PCE)}}_{(-i)}(\vc^{(i)}) \right)^2,
\end{equation}
where $\cm^{\textsf{(PCE)}}_{(-i)}(\cdot)$ is a PCE model built from the sample set $\vC^{(-i)} = \vC \backslash \vc^{(i)} \equiv \{ \vc^{(j)},j=1,\ldots,i-1,i+1,\ldots,N \}$ and $\cY= \{ \cY^{(i)},i=1,\ldots,N \}$ are the response values of the exact computational model. The LOO error is a special case of the \emph{leave-$k$-out cross-validation} error \cite{Allen1971} which discards $k$ samples from the initial experimental design to build up a model and predict the error at the $k$ discarded samples.

In theory the computational cost of the LOO error is proportional to the
number of samples $N$ since it would require the determination of $N$
PCE meta-models corresponding to each experimental design $\vC^{(-i)}$.
In the special case of linearly parameterized regression, which is the
case for PCE, it is possible to calculate the LOO error analytically
\emph{without} building $N$ separate models. The LOO error reads (see
\emph{e.g.} \cite{Saporta2006,BlatmanThesis} for the proof)
\begin{equation} \label{eq:errpce}
Err_{LOO}^{\text{(PCE)}} = \frac{1}{N} \sum_{i=1}^N \left( \frac{\cY^{(i)}-\cm^{\textsf{(PCE)}}(\vc^{(i)})}{1-h_i} \right)^2,
\end{equation}
where $h_i$ is the $i^{\textsf{th}}$ diagonal term of the matrix $
  \mathbf{F}\left( \mathbf{F}\tr \mathbf{F} \right)^{-1}\mathbf{F}\tr
$ and the information matrix $\mathbf{F}$ is defined in
Eq.~(\ref{eq:f}). Note that the PCE used in Eq.~(\ref{eq:errpce}) is
built only once from the full experimental design $\vC$.

\section{Kriging} \label{sec:k}
\subsection{Problem definition} \label{sec:k:problem}
The second meta-modeling technique in this paper is Kriging, also known as \emph{Gaussian process modeling}, which assumes that the response of a computational model is a realization of a Gaussian random process \cite{Santner2003}, \emph{i.e.} 
\begin{equation} \label{eq:kriging}
\cm(\ve{x}) \approx \cm^{\textsf{(K)}}(\ve{x}) = \ub\tr \ve{f}(\ve{x}) + \sigma^2 \, Z(\ve{x}),
\end{equation}
where $\ub\tr \ve{f}(\ve{x}) = \sum_{j=1}^P \beta_jf_j(\ve{x})$ is the mean value of the Gaussian process, also called trend, with coefficients $\ub$, $\sigma^2$ is the Gaussian process variance and $Z(\ve{x})$ is a zero-mean, unit-variance stationary Gaussian process. The zero-mean Gaussian process $Z(\ve{x})$ is fully determined by the auto-correlation function between two input sample points $R(\ve{x},\ve{x}') = R(|\ve{x}-\ve{x}'|;\vt)$ due to stationarity, where $\vt$ are hyper-parameters to be computed.

Various correlation functions can be found in the literature \cite{Rasmussen2006,Santner2003}, some of which are the \emph{linear}, \emph{exponential}, \emph{Gaussian} (also called \emph{squared exponential}) and \emph{Mat\'ern} autocorrelation function. In this paper the Mat\'ern autocorrelation function is mainly used as it is a generalization of the exponential and the Gaussian autocorrelation functions. The general Mat\'ern kernel of degree $\nu$ is defined as \cite{Matern1986}
\begin{equation}\label{eq:matern}
  R(|\ve{x}-\ve{x}'|;\ve{l}, \nu) = \prod_{i=1}^M \frac{1}{2^{\nu-1}
    \Gamma(\nu)} \left( \sqrt{2\nu} \frac{|x_i-x_i'|}{l_i}
  \right)^{\nu} \, \kappa_{\nu} \left(\sqrt{2\nu}
    \frac{|x_i-x_i'|}{l_i} \right), 
\end{equation}
where $\ve{x}$ and $\ve{x}'$ are two sample points in the input space $\cd_X$, $\ve{l}=\{ l_i>0,\, i=1,\ldots,M \}$ are the scale parameters (also called \emph{correlation lengths}), $\nu \geq 1/2$ is the shape parameter, $\Gamma(\cdot)$ is the Euler Gamma function and $\kappa_{\nu}(\cdot)$ is the modified Bessel function of the second kind (also known as Bessel function of the third kind). In many publications the shape parameter is set to either $\nu=3/2$ or $\nu=5/2$ which simplifies Eq.~(\ref{eq:matern}) to \cite{roustant2012}:
\begin{equation}
    R(|\ve{x}-\ve{x}'|;\ve{l},\nu=3/2) = \prod_{i=1}^M \left( 1+ \frac{\sqrt{3}\,|x_i-x_i'|}{l_i}\right) \exp \left( - \frac{\sqrt{3}\, |x_i-x_i'|}{l_i}\right),
\end{equation}
\begin{equation}\label{eq:mat52}
    R(|\ve{x}-\ve{x}'|;\ve{l},\nu=5/2) = \prod_{i=1}^M \left( 1+ \frac{\sqrt{5}\,|x_i-x_i'|}{l_i} + \frac{5(x_i-x_i')^2}{3 \,l_i^2}\right) \exp \left( - \frac{\sqrt{5}\, |x_i-x_i'|}{l_i}\right).
\end{equation}


Apart from the correlation part in Eq.~(\ref{eq:kriging}) there is also a trend part $\ub\tr\ve{f}(\ve{x})$. Three different flavors of Kriging are defined in the literature \cite{Rasmussen2006,Santner2003,Stein:1999}, namely simple, ordinary and universal Kriging according to the choice of the trend. \emph{Simple Kriging} assumes that the trend has a known constant value, \emph{i.e.} $\ub\tr\ve{f}(\ve{x}) = \beta_0$. In \emph{ordinary Kriging} the trend has a constant but unknown value, \emph{i.e.} $P=1$, $f_1(\ve{x}) = 1$ and $\beta_1$ is unknown. The most general and flexible formulation is \emph{universal Kriging} which assumes that the trend is composed of a sum of $P$ pre-selected functions $f_k(\ve{x})$, \emph{i.e.}
\begin{equation} \label{eq:trend}
 \ub\tr\ve{f}(\ve{x}) = \sum_{k=1}^P \beta_k f_k(\ve{x}),
\end{equation}
where $\beta_k$ is the trend coefficient of each function. Note that
simple and ordinary Kriging are special cases of universal Kriging. As
discussed later in this paper, one approach to set up a trend is to use
a sparse set of polynomials, which defines a new variant of universal
Kriging.

\subsection{Calibration of the Kriging model}
Given a value for the auto-correlation hyper-parameters $\hat{\vt}$, the calibration of the Kriging model parameters $\{\ub(\hat{\vt}),\sigma^2_y(\hat{\vt})\}$ may be computed using an \emph{empirical best linear unbiased estimator} (BLUE). The optimization yields an analytical expression as a function of $\hat{\vt}$:
\begin{equation} \label{eq:kbeta}
\ub(\hat{\vt}) = \left( \mathbf{F}\tr \mathbf{R}^{-1} \mathbf{F} \right)^{-1} \mathbf{F}\, \mathbf{R}^{-1} \mathcal{Y},
\end{equation}
\begin{equation} \label{eq:ksigma}
\sigma_y^2(\hat{\vt}) = \frac{1}{N} \left( \mathcal{Y}-\mathbf{F}\, \ub \right)\tr \mathbf{R}^{-1} \left( \mathcal{Y}-\mathbf{F}\, \ub \right),
\end{equation}
where $\cY = \{\cY^{(i)}, \,i=1,\ldots,N\}$ are model responses of the exact computational model on the experimental design $\vC=\{\vc^{(i)}, \, i=1,\ldots,N\}$, $\mathbf{R}_{ij} = R(|\vc^{(i)}-\vc^{(j)}|;\hat{\vt})$ is the correlation matrix and $\mathbf{F}_{ij}=f_j(\vc^{(i)})$ is the information matrix.

In recent developments, the optimal correlation parameters $\hat{\vt}$ may be determined by either a maximum-likelihood-estimate (denoted by \emph{ML}) \cite{Marrel2008,DubourgThesis} or by leave-one-out cross-validation (\emph{CV}) \cite{Bachoc2012}. The optimal parameters are determined through a minimization which reads:
\begin{equation}\label{eq:ml}
\hat{\ve{\theta}}_{ML} = \argminl_{\vt} \left[ \frac{1}{N} \left( \cY-\mathbf{F} \, \ub \right)\tr \mathbf{R}^{-1} \left( \cY-\mathbf{F} \, \ub \right) \left(\det \mathbf{R} \right)^{1/N} \right],
\end{equation}
\begin{equation} \label{eq:cv}
\hat{\ve{\theta}}_{CV} = \argminl_{\vt} \left[\cY\tr \mathbf{R}(\ve{\theta})^{-1} {\rm diag}\left(\mathbf{R}(\ve{\theta})^{-1}\right)^{-2} \, \mathbf{R}(\ve{\theta})^{-1}  \,\cY\right].
\end{equation}
The comparison of both approaches shows that ML is preferable to CV in
well-specified cases, \emph{i.e.} when the meta-model autocorrelation
function family is identical to the autocorrelation function of the
computational model. For practical problems, \emph{i.e.} assuming a
black-box model, the autocorrelation function family is not known with
certainty. In this case CV shall lead to more robust results than ML, as
discussed in \cite{Bachoc2012}.

{ Determining the optimal correlation parameters in  Eq.~(\ref{eq:ml}) and (\ref{eq:cv}) is a complex multi-dimensional minimization problem. Optimization algorithms can be cast into two distinct categories: local and global optimization algorithms. Local methods are usually gradient based algorithms such as the quasi-Newton Broyden-Fletcher-Goldfarb-Shanno (BFGS) algorithm \cite{goldfarb1970,Fletcher1970,shanno1970} and its modifications \cite{Byrd1999}. Global methods are algorithms such as genetic algorithms \cite{Goldberg1989} and differential evolution algorithms \cite{Storn1997,Deng2013}.  The best optimization algorithm is problem dependent and in many cases not known a-priori.}

{ The optimal correlation parameters are then used for predicting
  the model response at new samples of the input space.} By assumption,
the prediction of a Kriging model of a new point $\ve{x}$ is a Gaussian
random variable with mean $\mu_{\hat{y}}(\ve{x})$ and variance
$\sigma_{\hat{y}}^2(\ve{x})$:
\begin{equation}\label{eq:mpred}
\mu_{\hat{y}}(\ve{x}) = \ve{f}(\ve{x})\tr \ub + \ve{r}(\ve{x})\tr \mathbf{R}^{-1} \left( \mathcal{Y}-\mathbf{F} \ub \right),
\end{equation}
\begin{equation}\label{eq:spred}
\sigma_{\hat{y}}^2(\ve{x}) = \sigma_y^2 \left( 1- \langle \ve{f}(\ve{x})\tr \ve{r}(\ve{x})\tr \rangle
\left[
\begin{array}{cc}
  \ve{0} & \mathbf{F}\tr \\
  \mathbf{F} & \mathbf{R}
\end{array}
\right]^{-1}
\left[
\begin{array}{c}
  \ve{f}(\ve{x}) \\
  \ve{r}(\ve{x})
\end{array}
\right]
\right),
\end{equation}
where $r_i(\ve{x})= R(|\ve{x}-\vc^{(i)}|;\vt)$ is the correlation between the new sample $\ve{x}$ and the sample $\vc^{(i)}$ of the experimental design.  The prediction mean is used as the surrogate to the original model $\cm$, whereas the variance gives a \emph{local} error indicator about the precision. It is important to note that the Kriging model interpolates the data, \emph{i.e.}
$$\mu_{\hat{y}}(\vc^{(i)}) = \cm(\vc^{(i)}), \ \ \  \sigma_{\hat{y}}^2(\vc^{(i)})= 0, \ \ \ \forall \vc^{(i)}\in \vC.$$

Apart from this procedure to calibrate the meta-model and predict model
responses, the Kriging meta-modeling technique has been developed
further in recent years. The latest developments in Kriging are
contributed in the aspects of optimal estimation of the hyper-parameters
\cite{Bachoc2012,Bachoc2013,Bachoc2014}, the use of adaptive kernels
\cite{Duvenaud2012,ginsbourger2013} and the use of additive
auto-correlation kernels \cite{Durrande2012,
  Durrande2013,ginsbourger2012}

\subsection{Error estimation} \label{sec:k:error} A local error measure
for any sample $\ve{x}$ is given by the prediction variance
$\sigma^2_{\widehat{y}}(\ve{x})$ in Eq.~(\ref{eq:spred}). This
information is useful to detect regions where the prediction accuracy is
low. Adding new samples to the experimental design $\vC$ in the regions
with high prediction variance may lead to an overall increase in the
accuracy of the meta-model in that region. This characteristics is
exploited when devising adaptive experimental designs in structural
reliability analysis, see \cite{Echard2013,Bichon2008,
  Bichon2011,Dubourg2011a}.

 A simple global error measure of the accuracy of the meta-model (such
 as Eq.~(\ref{eq:erremp}) for PC expansions) is not available for
 Kriging due to its interpolating properties, which make the empirical
 error vanish (considering no nugget effect in its auto-correlation
 function). Thus one approach to a global error measure is the
 \emph{leave-one-out} (LOO) error
\begin{equation} \label{eq:LOOkriging}
Err_{LOO}^{\text{(K)}} = \frac{1}{N} \sum_{i=1}^N \left(\cY^{(i)}- \mu_{\hat{y},(-i)}(\vc^{(i)}) \right)^2,
\end{equation}
where $\mu_{\hat{y},(-i)}(\vc^{(i)})$ is the prediction mean $\mu_{\hat{y}}$ of sample $\vc^{(i)}$ by a Kriging meta-model based on the experimental design $\vC^{(-i)}=\vC \backslash \vc^{(i)}$ and $\cY= \{\cY^{(i)}, i=1,\ldots,N \}$ is the exact model response. \cite{Dubrule1983} derived an analytical solution for the LOO error for universal Kriging without computing the $N$ meta-models explicitly in the same spirit as Eq.~(\ref{eq:errpce}) for PC expansions. The prediction mean and variance are given by
\begin{equation} \label{eq:look1}
\mu_{\hat{y},(-i)} = -\sum_{j=1, j \neq i}^N \frac{\mathbf{B}_{ij}}{\mathbf{B}_{ii}} \cY^{(j)} = -\sum_{j=1}^N \frac{\mathbf{B}_{ij}}{\mathbf{B}_{ii}} \cY^{(j)} + \cY^{(i)},
\end{equation}
\begin{equation} \label{eq:look2}
\sigma_{\hat{y},(-i)}^2 = \frac{1}{\mathbf{B}_{ii}},
\end{equation}
where $\mathbf{B}$ is a square matrix of size $(N+P)$ with $N$ and $P$
denoting the number of samples in the experimental design and the number
of polynomials in the trend part, respectively:
\begin{equation} \label{eq:look3}
\mathbf{B} = \left[
\begin{array}{cc}
  \sigma^2\mathbf{R} & \mathbf{F} \\
  \mathbf{F}\tr & \ve{0}
\end{array} \right]^{-1},
\end{equation}
where $\sigma^2$ is the Kriging variance for the full experimental design $\vC$ estimated by Eq.~(\ref{eq:ksigma}). A generalized version of this algorithm called \emph{v-fold cross-correlation error} can be found in \cite{Dubrule1983}.

\subsection{PCE as a particular case of universal Kriging}
\label{sec:k:conn}
PC expansions can be interpreted as Kriging models where the samples of
the experimental design are uncorrelated, \ie where the autocorrelation
function consists of the Dirac function:
\begin{equation}
R(|\ve{x}-\ve{x}'|)=\delta(\ve{x}-\ve{x}').
\end{equation}
The correlation matrix is then the identity matrix of size $N$, \ie
$\mathbf{R} = \mathbf{I}_N$. This reduces Eq.~(\ref{eq:kbeta}) to
Eq.~(\ref{eq:apce}) and Eq.~(\ref{eq:mpred}) to Eq.~(\ref{eq:ypce}).

Further it can be shown that the leave-one-out error in
Eq.~(\ref{eq:LOOkriging})-(\ref{eq:look2}) reduces to
Eq.~(\ref{eq:errpce}) by the following derivations. Consider the
symmetric partitioned matrix $\mathbf{C}$ and the corresponding inverse
$\mathbf{D}$, \ie{} $\mathbf{D} = \mathbf{C}^{-1}$ which are defined as:
$$\mathbf{C} = \left[ \begin{array}{cc}
\mathbf{C}_{11} & \mathbf{C}_{12} \\ 
\mathbf{C}_{12}\tr & \mathbf{C}_{22}
\end{array}  \right], \qquad \qquad \mathbf{D} = \left[ \begin{array}{cc}
  \mathbf{D}_{11} & \mathbf{D}_{12} \\ 
  \mathbf{D}_{12}\tr & \mathbf{D}_{22}
\end{array}  \right],$$ 
where $\mathbf{C}_{11},\, \mathbf{D}_{11}$ (resp. $\mathbf{C}_{22},\,
\mathbf{D}_{22}$) are square matrices with
dimension $N$  (resp. $P$). Using block matrix inversion one can derive: 
\begin{equation}
\mathbf{D}_{11} = \mathbf{C}_{11}^{-1}+\mathbf{C}_{11}^{-1}\mathbf{C}_{12}\left( \mathbf{C}_{22}-\mathbf{C}_{12}\tr \mathbf{C}_{11}^{-1}\mathbf{C}_{12} \right)^{-1} \mathbf{C}_{12}\tr \mathbf{C}_{11}^{-1}.
\end{equation}
In the context of the leave-one-out error, taking $ \mathbf{C} \equiv
\mathbf{B}$ in Eq.~(\ref{eq:look3}), $\mathbf{C}_{11} = \sigma^2
\mathbf{I}_{N}$, $\mathbf{C}_{12} = \mathbf{F}$, $\mathbf{C}_{22} =
\mathbf{0}_{P}$:
\begin{equation}
  \begin{split}
    \mathbf{D}_{11} &= \frac{1}{\sigma^2}\mathbf{I}_{N}+
    \frac{1}{\sigma^2}\mathbf{I}_{N}\mathbf{F}\left(
      \mathbf{0}_{P}-\mathbf{F}\tr\frac{1}{\sigma^2}\mathbf{I}_{N}
      \mathbf{F} \right)^{-1}
    \mathbf{F}\tr\frac{1}{\sigma^2}\mathbf{I}_{N}\\
    &= \frac{1}{\sigma^2}\mathbf{I}+\frac{1}{\sigma^2}\mathbf{F}\left(
      -\mathbf{F}\tr\mathbf{F} \right)^{-1}\mathbf{F}\tr =
    \frac{1}{\sigma^2}\left( \mathbf{I}-\mathbf{F} \left(
        \mathbf{F}\tr\mathbf{F} \right)^{-1}\mathbf{F}\tr \right).
  \end{split}
\end{equation}
Then, the leave-one-out error in Eq.~(\ref{eq:LOOkriging}) combined with
Eq.~(\ref{eq:look1}) and the above inverse formulation of the
$\mathbf{B}$ matrix reads:
\begin{equation}
  \begin{split}
    Err_{LOO}^{\text{K}} &= \frac{1}{N}\sum_{i=1}^{N} \left( \cY^{(i)}+
      \sum_{j=1}^N \frac{B_{ij}}{B_{ii}}\cY^{(j)} -\cY^{(i)} \right)^2
= \frac{1}{N} \sum_{i=1}^N \left( \frac{1}{B_{ii}} \sum_{j=1}^N
      B_{ij}\cY^{(j)} \right)^2\\
    &= \frac{1}{N} \sum_{i=1}^N \left( \frac{1}{B_{ii}\sigma^2}
      \sum_{j=1}^N \left( \mathbf{I}-\mathbf{F}\left( \mathbf{F}\tr
          \mathbf{F} \right)^{-1}\mathbf{F}\tr \right)_{ij} \cY^{(j)}
    \right)^2\\
    &= \frac{1}{N} \sum_{i=1}^N \left(
      \frac{1}{1-\left(\text{diag}\left[ \mathbf{F}\left( \mathbf{F}\tr
              \mathbf{F} \right)^{-1}\mathbf{F}\tr \right]\right)_i}
      \cdot \left[ \cY^{(i)}-\sum_{j=1}^N \left[\mathbf{F}\left(
            \mathbf{F}\tr \mathbf{F}
          \right)^{-1}\mathbf{F}\tr\right]_{ij} \cY^{(j)} \right]\right)^2 \\
    &= \frac{1}{N} \sum_{i=1}^{N} \left( \frac{ \cY^{(i)}- \left[\left(
            \mathbf{F}\tr \mathbf{F} \right)^{-1}\mathbf{F}\tr
          \cY\right]\tr f(\vc^{(i)}) }{1-\left(\text{diag}\left[
            \mathbf{F}\left( \mathbf{F}\tr \mathbf{F}
            \right)^{-1}\mathbf{F}\tr \right]\right)_i} \right)^2,
  \end{split}
\end{equation}
which is equivalent to the formulation of the leave-one-out error in
Eq.~(\ref{eq:errpce}) for the case of PCE and $f(\vc^{(i)})\equiv
\psi(\vc^{(i)})$. Thus the leave-one-out error in PCE can be seen as a
special case of the leave-one-out error in the Kriging framework.  

\section{PC-Kriging} \label{sec:pck}
\subsection{Principle}
The characteristic of Kriging is to interpolate local variations of the output of the computational model as a function of the neighboring experimental design points. In contrast, polynomial chaos expansions (PCE) are used for approximating the \emph{global} behavior of $\cm$ using a set of orthogonal polynomials. By combining the two techniques we aim at capturing the global behavior of the computational model with the set of orthogonal polynomials in the trend of a universal Kriging model and the local variability with the Gaussian process. The new approach called \emph{Polynomial-Chaos-Kriging} (PC-Kriging) combines these two distinct meta-modeling techniques and their characteristics.

Using now the standard notation for truncated polynomial chaos expansions (see Eq.~(\ref{eq:pce})), we cast the PC-Kriging meta-model as follows \cite{SchoebiRouen2014}:
\begin{equation}\label{eq:pck}
\cm(\ve{x})\approx \cm^{\textsf{(PCK)}}(\ve{x}) = 
\sum_{\ua\in \cA} \ve{a}_{\ua}\psi_{\ua}(\ve{x})+ \sigma^2 Z(\ve{x}),
\end{equation}
where $\sum_{\ua\in \cA} \ve{a}_{\ua}\psi_{\ua}(\ve{x})$ is a weighted sum of orthonormal polynomials describing the mean value of the Gaussian process and $\cA$ is the index set of the polynomials. $Z(\ve{x})$ is a zero-mean, unit-variance stationary Gaussian process defined by an autocorrelation function $R(|\ve{x}-\ve{x}'|;\vt)$ and is parametrized by a set of hyper-parameters $\vt$.  

Building a PC-Kriging meta-model consists of two parts: the
determination of the optimal set of polynomials contained in the
regression part (\emph{i.e.} the truncation set $\cA$) and the
calibration of the correlation hyper-parameters $\vt$ as well as the
Kriging parameters $\{\sigma^2,\ve{a_{\ua}}\}$. The set of polynomials
is determined using the Least-Angle-Regression (LAR) algorithm as in
\cite{BlatmanJCP2011} together with hyperbolic index sets to obtain
sparse sets of polynomials. After the set of polynomials is fixed, the
trend and correlation parameters are evaluated using the universal
Kriging equations (Eq.~(\ref{eq:kbeta})-(\ref{eq:cv})).

\subsection{Algorithm}
\begin{figure}[!ht]
\centering
\begin{tikzpicture}[node distance = 1cm, auto]

    \node [blockyellow] (Input) {Input Distributions\\ $f_{\ve{X}}$};
    \node [blockyellow, right of=Input, node distance=5cm] (DOE) {Experimental design\\ $\{ \vC,\,\cY \}$};
    \node [blockyellow, right of=DOE, node distance=5cm] (corr) {Autocorrelation function\\ $R(|\ve{x}-\ve{x}'|;\vt)$};
    \node [node, below of=DOE, node distance =1cm](node1){};
    \node [Block, below of=DOE, node distance=4.5cm] (OPCK) {Optimal PC-Kriging\\
    $  $\\
    \begin{tikzpicture}[node distance = 1cm, auto]
    \node [block] (I1){
        \begin{tikzpicture} [node distance = 1cm, auto]
        \node [blockmin] (lar1) {LAR (iteration 1)};
        \node [blockmin, below of=lar1] (i1) {$y = \cm^{\text{(K)}(Q=1)}(\ve{x})$};
        \node [blockminmin, below of=i1] (LOO1) {LOO$_1$};
        \path [line] (i1) -- (LOO1);
        \path [line] (lar1) -- (i1);
        \end{tikzpicture}
    };
    \node [blockred, right of=I1, node distance=4.5cm] (I2){
        \begin{tikzpicture} [node distance = 1cm, auto]
        \node [blockmin] (lar2) {LAR (iteration 2)};
        \node [blockmin, below of=lar2] (i2) {$y = \cm^{\text{(K)}(Q=2)}(\ve{x})$};
        \node [blockminmin, below of=i2] (LOO2) {LOO$_2$};
        \path [line] (i2) -- (LOO2);
        \path [line] (lar2) -- (i2);
        \end{tikzpicture}
    };
    \node [blockdots, right of=I2, node distance=3cm] (I3){
	$\cdots$
    };
    \node [block, right of=I3, node distance=3cm] (I4){
        \begin{tikzpicture} [node distance = 1cm, auto]
        \node [blockmin] (lar4) {LAR (iteration $P$)};
        \node [blockmin, below of=lar4] (i4) {$y = \cm^{\text{(K)}(Q=P)}(\ve{x})$};
        \node [blockminmin, below of=i4] (LOO4) {LOO$_P$};
        \path [line] (i4) -- (LOO4);
        \path [line] (lar4) -- (i4);
        \end{tikzpicture}
    };    
    \node [blockpck, below of= I2, node distance=3cm] (pckriging) {PC-Kriging model\\ $\cm^{\text{(OPCK)}}=\argmin_{\cm^{\text{(K)}(Q=P)}} LOO_P$};
    \path [line] (I2) -- (pckriging);
    \path [line] (I1) |- (pckriging);
    \path [line] (I4) |- (pckriging);
    \end{tikzpicture}   
    };
    \node [block, below of=OPCK, node distance=4cm] (prediction) {Prediction\\ $\{\mu_{\widehat{y}}(\ve{x}),\, \sigma_{\widehat{y}}^2(\ve{x})\}$};
	
	\path [line] (Input) |- (node1);
	\path [line] (OPCK) -- (prediction);
	\path [line] (DOE) -- (node1);
	\path [line] (corr) |- (node1);
	\path [line] (node1) -- (OPCK);

\end{tikzpicture}
\caption{\label{fig:flowOPC} Flowchart for Optimal PC-Kriging (OPC-Kriging)}
\end{figure}
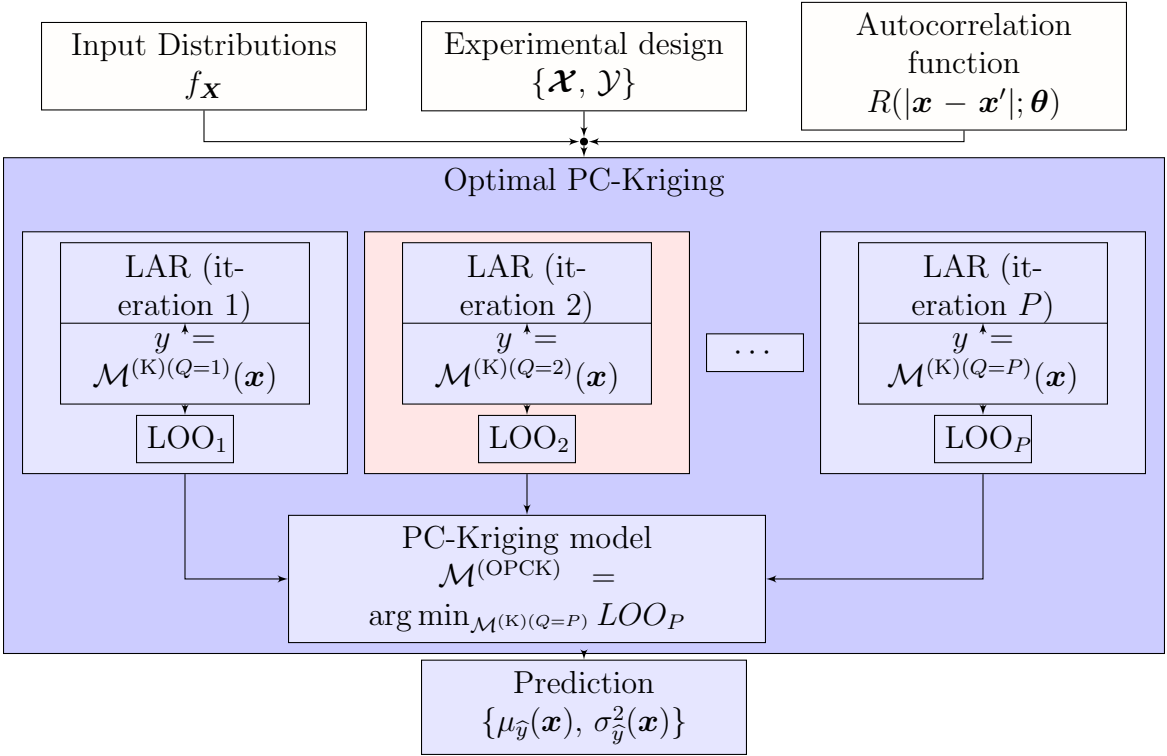

The two distinct frameworks for PCE and Kriging can be combined in
various ways. In this paper two approaches will be explained in detail,
\emph{i.e.} the Sequential PC-Kriging (SPC-Kriging) and the Optimal
PC-Kriging (OPC-Kriging). Both approaches are based on the same input
information, namely the experimental design $\vC$, the corresponding
response values $\cY$ obtained from the computational model $\cm(\vC)$,
the description of the stochastic input variables (joint PDF
$f_{\ve{X}}$) and the parametric expression of the an auto-correlation
function $R(|\ve{x}-\ve{x}'|;\vt)$. The two approaches are defined as
follows:
\begin{itemize}
\item \emph{Sequential PC-Kriging} (SPC-Kriging): in this approach, the
  set of polynomials and the Kriging meta-model are determined
  \emph{sequentially}. The assumption behind this procedure is that the
  optimal set of polynomials found by the LAR algorithm in the context of
  pure PCE can be used directly as an optimal trend for the universal
  Kriging.  In a first step the optimal set of polynomials is determined
  using the PCE framework: $\cA$ is found by applying the LAR procedure
  as in \cite{BlatmanJCP2011}.  The set of multivariate orthonormal
  polynomials $\cA$ is then embedded into a universal Kriging model as
  the trend. The universal Kriging meta-model is calibrated using
  Eq.~(\ref{eq:kbeta})-(\ref{eq:cv}).

  At the end of the algorithm the accuracy of the meta-model can be
  measured by the leave-one-out error given in Eq.~(\ref{eq:LOOkriging})
  or, when using a validation set, by
  the sample-based generalization error in Eq.~(\ref{eq:errGengen}).\\
  The SPC-Kriging algorithm is illustrated in Figure~\ref{fig:flowSPC}
  in which the white boxes represent the required input information and
  the blue boxes represent the computational tasks. Given a calibrated
  SPC-Kriging model, the response of new input realizations (\emph{i.e.}
  the prediction) is computed by Eq.~(\ref{eq:mpred}) and
  Eq.~(\ref{eq:spred}).
  
  \begin{figure}[!ht]
  \centering
  \begin{tikzpicture}[node distance = 1cm, auto]
  
      \node [blockyellow] (Input) {Input Distributions\\ $f_{\ve{X}}$};
      \node [blockyellow, right of=Input, node distance=5cm] (DOE) {Experimental design \\ $\{\vC,\,\cY\}$};
      \node [block, below of=Input] (LARS) {LAR};
      \node [blockyellow, right of=DOE, node distance=5cm] (corr) {Autocorrelation function\\ $R(|\ve{x}-\ve{x}'|;\vt)$};
      \node [node, below of=DOE] (node1) {};
      \node [Block, below of=DOE, node distance=2.5cm] (SPCK) {Sequential-PC-Kriging\\
      $  $\\
      \begin{tikzpicture}[node distance = 1cm, auto]
      \node [blockpck2, node distance=3cm] (I4){$\cm^{\textsf{(SPCK)}}(\ve{x}) = 
      \sum_{\ua\in \cA} \ve{a}_{\ua}\psi_{\ua}(\ve{x})+ \sigma^2 Z(\ve{x},\omega)$
      };    
      \end{tikzpicture}   
      };
      \node [block, below of=SPCK, node distance=2cm] (prediction) {Prediction \\ $\{\mu_{\widehat{y}}(\ve{x}),\, \sigma_{\widehat{y}}^2(\ve{x})\}$};

  	\path [line] (Input) -- (LARS);
  	\path [line] (LARS) -- (node1);
  	\path [line] (SPCK) -- (prediction);
  	\path [line] (DOE) -- (node1);
  	\path [line] (corr) |- (node1);
  	\path [line] (DOE) -- (LARS);
  	\path [line] (node1) -- (SPCK);
  
  \end{tikzpicture}
  \caption{\label{fig:flowSPC} Flowchart for Sequential-PC-Kriging (SPC-Kriging)}
  \end{figure}
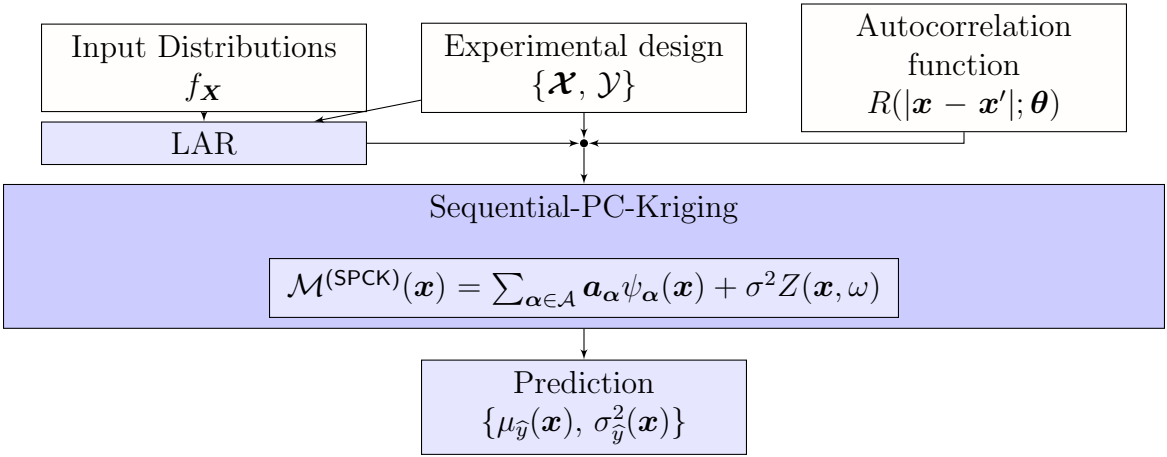  
  
\item \emph{Optimal PC-Kriging} (OPC-Kriging): in this approach, the
  PC-Kriging meta-model is obtained iteratively. The set of orthonormal
  multivariate polynomials is determined by LAR algorithm in the same
  way as in SPC-Kriging. Yet the LAR algorithm results in a list of
  ranked polynomials which are chosen depending on their correlation to
  the current residual at each iteration in decreasing order.
  OPC-Kriging consists of an iterative algorithm where each polynomial
  is added one-by-one to the trend part. In each iteration, the
  coefficients of the trend and the parameters of the auto-correlation
  function are calibrated. In the end, a number $|\cA|$ of different
  PC-Kriging models are available. The $|\cA|$ meta-models are then
  compared in terms of the LOO error (Eq.~(\ref{eq:LOOkriging})). The
  optimal PC-Kriging meta-model is then chosen as the one with minimal
  leave-one-out error.

  Figure~\ref{fig:flowOPC} illustrates the OPC-Kriging algorithm in a
  flowchart. The notation $\cm^{(\text{K})(Q)}(\ve{x})$ means a
  universal Kriging meta-model where the trend is modeled by the $Q$
  first polynomials selected in $\cA$ according to the ranking obtained
  by LAR.  Note that the red box represents the universal Kriging model
  which eventually minimizes the LOO error ($Q=P^*$) and which is thus
  finally chosen.  The final box marks the prediction of new model
  responses which is computed by Eq.~(\ref{eq:mpred}) and
  Eq.~(\ref{eq:spred}).
\end{itemize}

As a summary PC-Kriging can be viewed as a universal Kriging meta-model
with a non-standard trend part. Thus the error estimates from Kriging in
Section~\ref{sec:k:error} are valid here with no modification. In
particular, the LOO error in Eq.~(\ref{eq:LOOkriging}) is computed to
compare different PC-Kriging models and to compute the optimal
meta-model in OPC-Kriging. The performance of both approaches and the
comparison to the traditional PCE and Kriging approaches is now
illustrated for a variety of benchmark analytical functions.

\clearpage

\section{Analytical benchmark functions} \label{sec:ana}
\subsection{Setup}
Easy-to-evaluate functions are tested to verify and validate the new
PC-Kriging approach by comparing the meta-model to the exact model
response. In this paper, six analytical functions of different
dimensionality are illustrated, namely four with uniformly distributed
input variables (\emph{i.e.} Ishigami, Sobol', Rosenbrock and Morris
functions) and two with Gaussian input variables (\emph{i.e.} Rastrigin
and O'Hagan function). References to the original use of the benchmark
functions in the literature are given below.

The first four analytical functions use uniformly distributed random
variables in the input space. The \emph{Ishigami} function is a smooth
function with three independent input parameters commonly used for
benchmarking methods in global sensitivity analysis.
\begin{equation}
f_1(\ve{x}) = \sin x_1 + 7\, \sin^2 x_2 + 0.1\,x_3^4 \, \sin x_1,
\end{equation}
where $X_i\sim\cU(-\pi,\pi), \ i=1,2,3$.  The \emph{Sobol}' function is
also well-known sensitivity analysis because the Sobol' indices are easy
to derive analytically \cite{Sobol1993}:
\begin{equation}
f_2(\ve{x}) = \prod_{i=i}^8 \frac{|4\,x_i-2|+c_i}{1+c_i},
\end{equation}
where $X_i\sim \cU(0,1), \ i=1,\ldots,8$ and
$c=(1,2,5,10,20,50,100,500)\tr$ as in \cite{Sudret2008b}. Due to the
absolute value operator in the enumerator the function behaves
non-smoothly at the point $x_i=0.5$. The \emph{Rosenbrock} function is a
polynomial function with a 2-dimensional input space
\cite{Rosenbrock1960}:
\begin{equation}
f_3(\ve{x}) = 100 \,\left(x_2-x_1^2\right)^2+\left(1-x_1\right)^2,
\end{equation} 
where $X_i\sim\cU(-2,2), \ i=1,2$. The last function considered is the \emph{Morris} function which is defined by \cite{Morris1991}
\begin{equation}\label{eq:f4}
f_4(\ve{x})= \sum_{i=1}^{20} \beta_i\, w_i + \sum_{i<j}^{20} \beta_{ij} \,w_i w_j + \sum_{i<j<l}^{20} \beta_{ijl}\, w_i w_j w_l + 5\, w_1 w_2 w_3 w_4,
\end{equation}
where $X_i\sim\cU(0,1), \ i=1,\ldots,20$ and $w_i = 2\,(x_i-1/2)$ for all $i$ except for $i=3,5,7$ where $w_i =\displaystyle{ 2\,(\frac{1.1\,x_i}{x_i+0.1}-1/2)}$. The coefficients are defined as:
        $\beta_i = 20,\ i=1,\ldots, 10$;
        $\beta_{ij} = -15,\ i,j = 1, \ldots, 6$;
        $\beta_{ijl} = -10,\ i,j,l=1,\ldots, 5$.
        The remaining coefficients are set equal to $\beta_i = (-1)^i$ and $\beta_{ij} = (-1)^{i+j}$ as in \cite{BlatmanThesis}.
        
Two other benchmark functions of independent Gaussian variables are also studied. The \emph{Rastrigin} function has a two-dimensional input space and is defined by \cite{Rastrigin1974}
\begin{equation}
f_5(\ve{x}) = 10-\sum_{i=1}^2 (x_i^2-5 \,\cos(2\pi \,x_i)),
\end{equation}
where $X_i\sim\cN(0,1), \ i=1,2$. The last function is the \emph{O'Hagan} function which is defined by \cite{Oakley2004}
\begin{equation}
f_6(\ve{x}) = \ve{a_1}\tr \ve{x} + \ve{a_2}\tr \sin(\ve{x}) + \ve{a_3}\tr \cos(\ve{x}) + \ve{x}\tr \ve{Q}\ve{x},
\end{equation}
where $X_i\sim\cN(0,1), \ i=1,\ldots,15$. The vectors $\{ \ve{a}_1,\ve{a}_2.\ve{a}_3 \}$ and matrix $\ve{Q}$ are defined in \cite{Oakley2004}. 

Note that the functions $f_1-f_4$ have uniform input random variables.
Accordingly the PC trend in PC-Kriging is built up from multivariate
Legendre polynomials. In contrast $f_5,f_6$ have Gaussian input random
variables. Thus the PC trend is modeled by multivariate Hermite
polynomials (see Tab.~\ref{tab:orthobasis}).

\subsection{Analysis}
At the beginning of each algorithm the experimental design is generated
with the Latin-hypercube sampling technique \cite{McKay1979}. Then the
meta-modeling is processed applying the four previously discussed
meta-modeling techniques, \emph{i.e.} ordinary Kriging, PCE, SPC-Kriging
and OPC-Kriging. { Note that for the Kriging meta-models, the maximum likelihood formulation (Eq.~(\ref{eq:ml})) in combination with the gradient based BFGS optimization algorithm is used in order to compute the optimal correlation parameters.} 

Their performance is compared by means of the
\emph{relative generalization error} which is defined as the ratio
between the generalization error (Eq.~(\ref{eq:errGen})) and the output
variance:
\begin{equation}
\epsilon_{gen} = \frac{\Esp{ \left( Y-Y^{\textsf{(PCE)}} \right)^2}}{\text{Var}\left[Y\right]}.
\end{equation}
The error is estimated here using a large validation set $\mathbb{X}=\{
\ve{x}^{(1)},\ldots,\ve{x}^{(n)} \}$ of size $n=10^5$, which results in
\begin{equation}
 \label{eq:errgen}
\widehat{\epsilon}_{gen} \approx \frac{\sum_{i=1}^n  \left(
    \cm(\ve{x}^{(i)})-\widehat{\cm}(\ve{x}^{(i)})
  \right)^2}{\sum_{i=1}^n  \left( \cm(\ve{x}^{(i)})- \mu_{y} \right)^2},
\end{equation}
where $\mu_y$ is the mean value of the set of exact model responses over
the validation set $\mathbb{Y} = \{y^{(1)},\ldots,y^{(n)}\} \equiv \{
\cm(\ve{x}^{(1)}),\ldots,\cm(\ve{x}^{(n)}) \}$. For all Kriging-based
approaches, the meta-model $\widehat{\cm}(\ve{x})$ is the prediction
mean $\mu_{\hat{y}}(\ve{x})$. Note that the samples in $\mathbb{X}$
follow the distribution of the input variables $\ve{X}$ in order to
obtain a reliable error estimate.

For each experimental setup, the analysis is replicated to account for
the statistical uncertainties in the experimental design. 50~independent
runs of the full analysis are carried out and the results are
represented using boxplots. In a box plot the central mark represents
the median value of the 50~runs, the edges are the 25th and 75th
percentile denoted by $q_{25}$ and $q_{75}$. The whiskers describe the
boundary to the outliers. Outliers are defined as the values smaller
than $q_{25}-1.5\,(q_{75}-q_{25})$ or larger than
$q_{75}+1.5\,(q_{75}-q_{25})$.

\subsection{Results}
\subsubsection{Visualization of PC-Kriging's behavior}
\label{sec:results:viz} 
\begin{figure} [!ht]
  \centering
  \subfigure[Rastrigin function \label{fig:behaviour:a}]{
    \includegraphics[width=0.23\linewidth, angle=0]{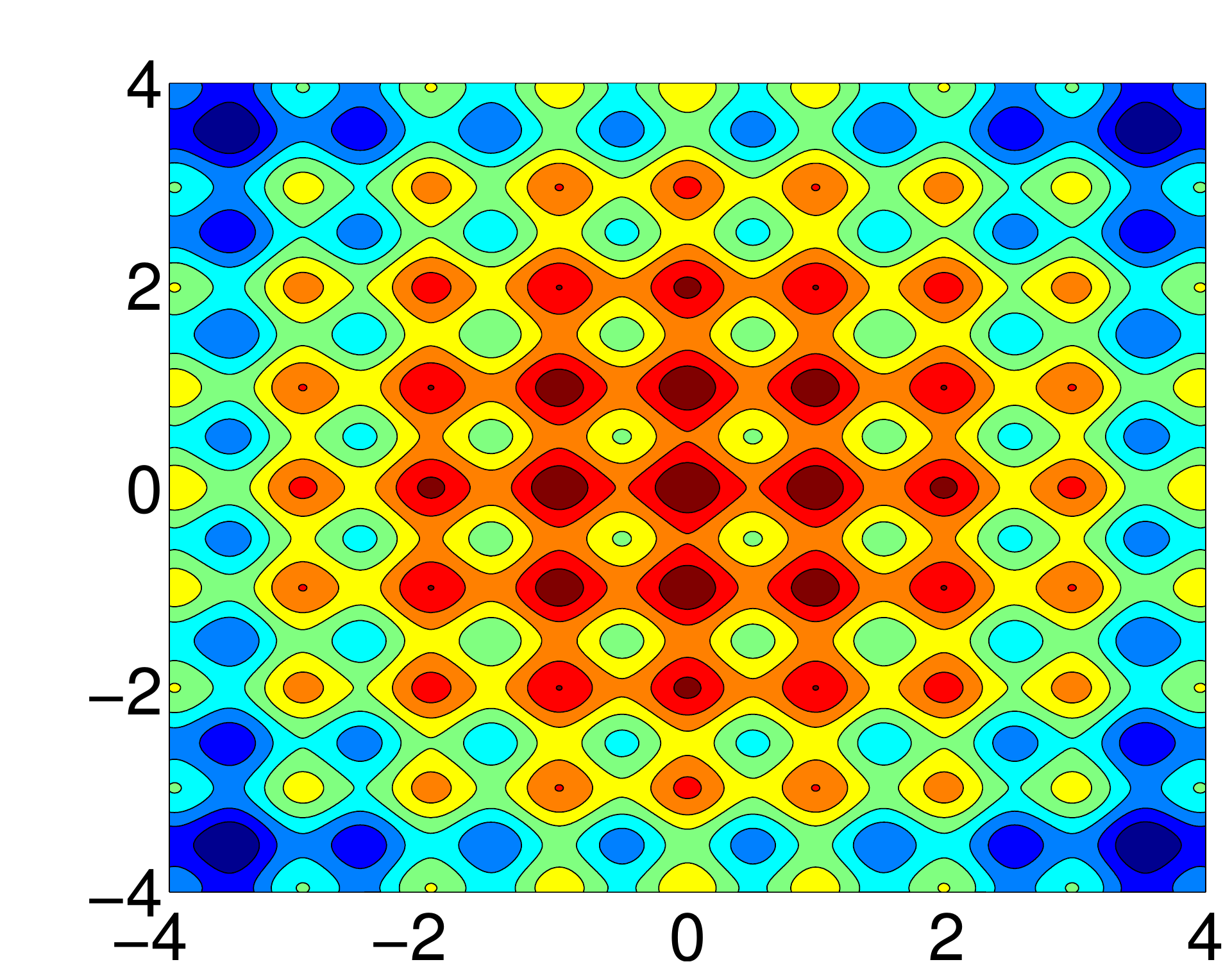}
  }
  \subfigure[Ordinary Kriging \label{fig:behaviour:b}]{
    \includegraphics[width=0.23\linewidth, angle=0]{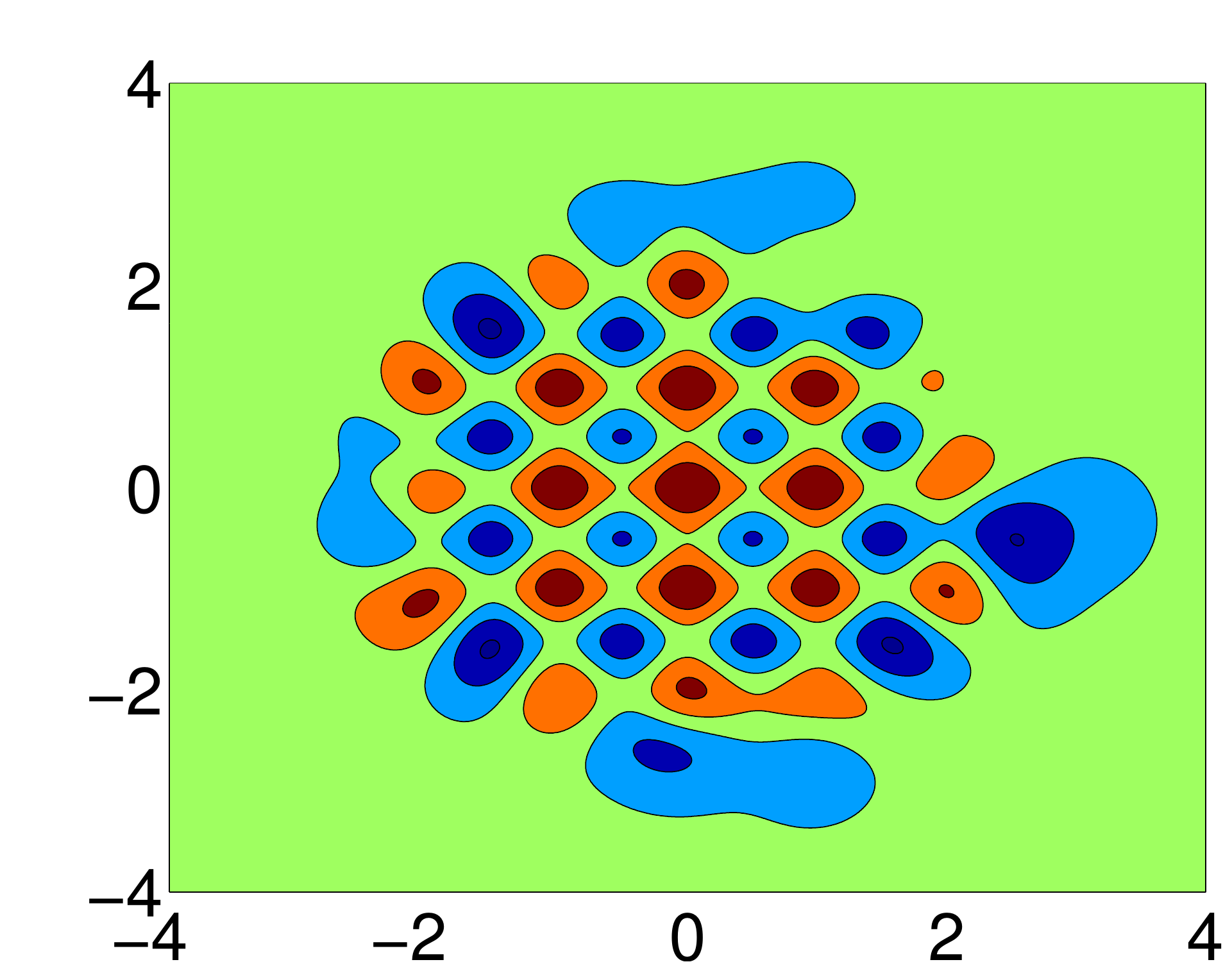}
  } 
  \subfigure[PCE \label{fig:behaviour:c}] {
    \includegraphics[width=0.23\linewidth, angle=0]{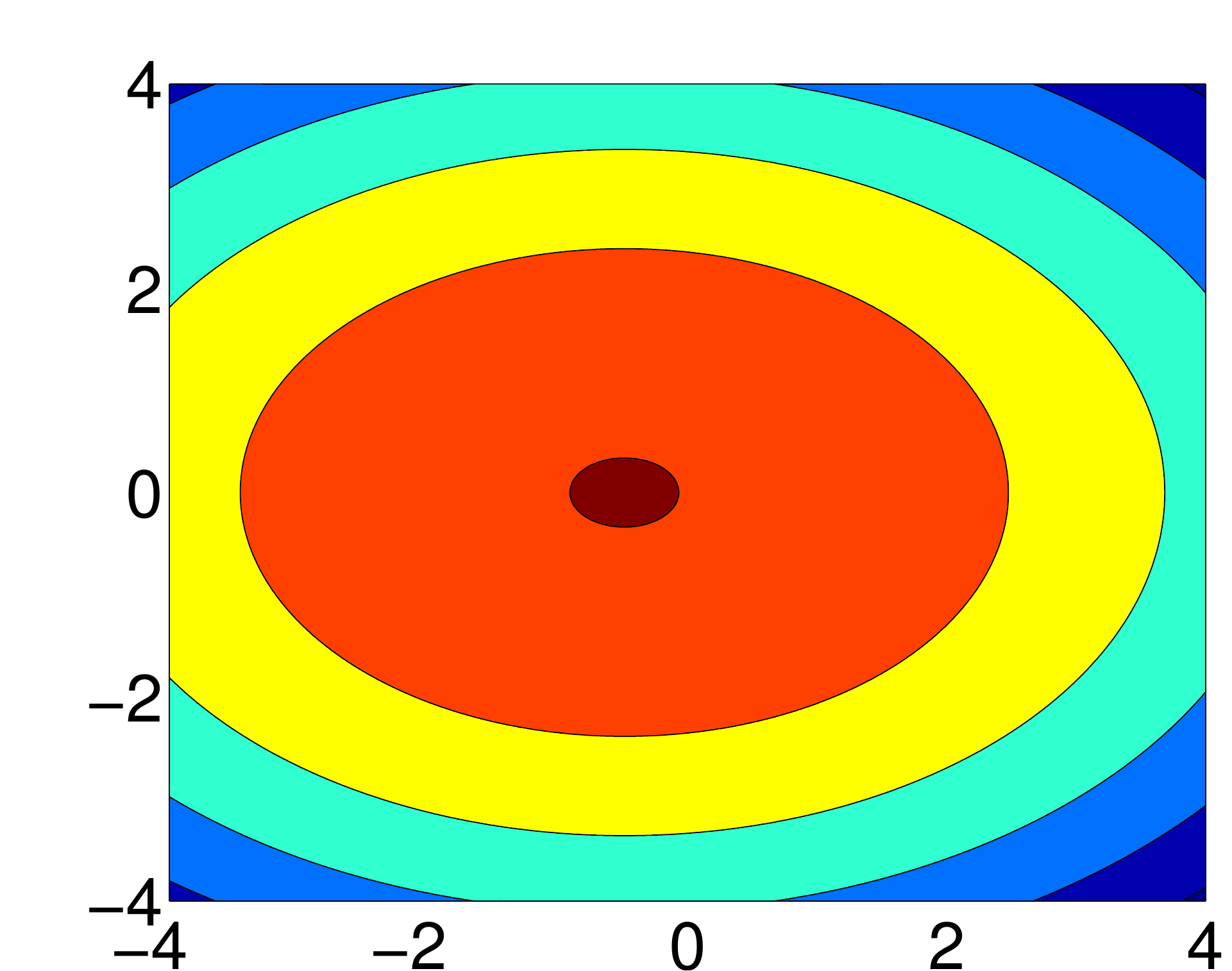}
  }
  \subfigure[PC-Kriging \label{fig:behaviour:d}]{
    \includegraphics[width=0.23\linewidth, angle=0]{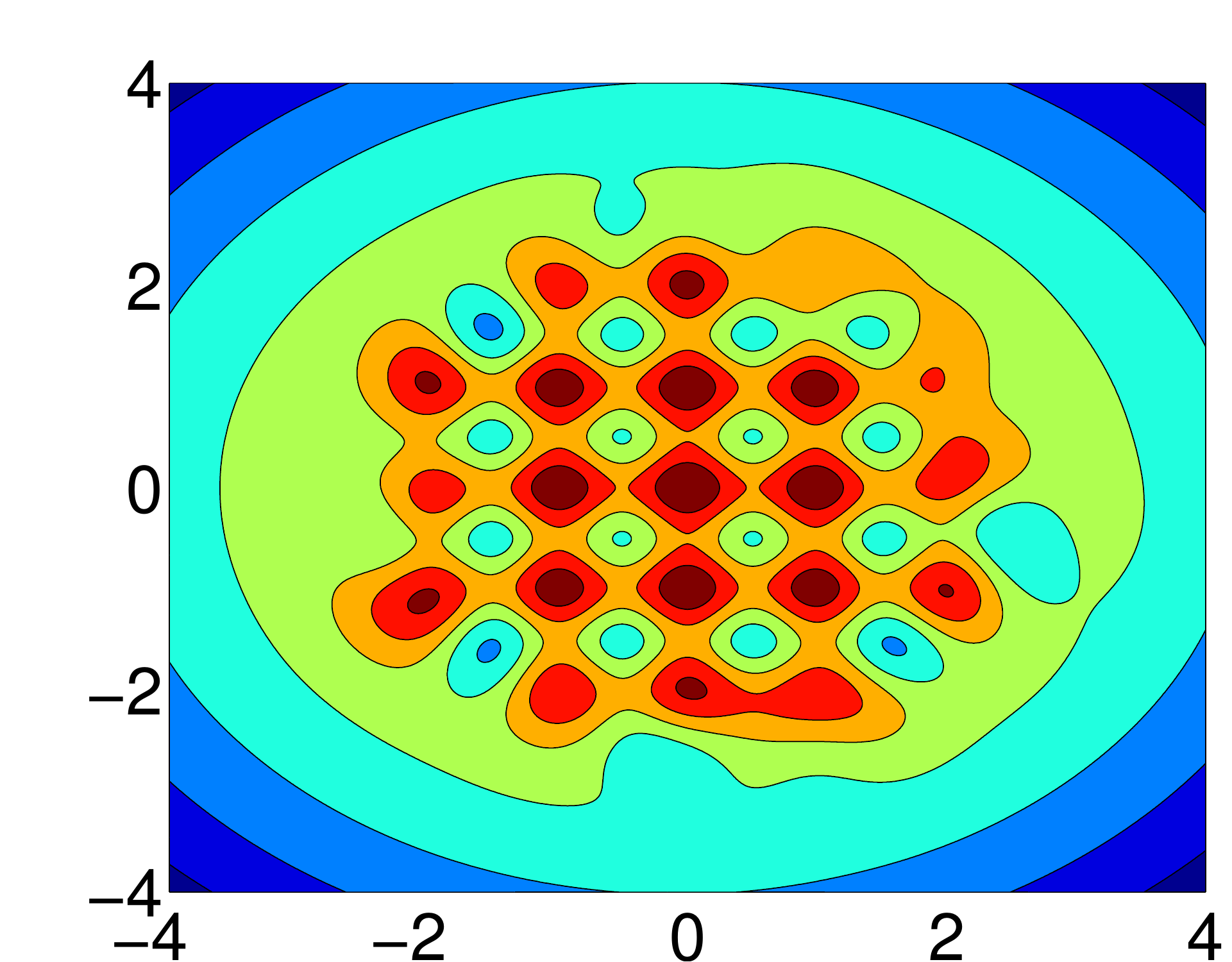}
  }
  \caption{\label{fig:behaviour} Rastrigin function -- Visual composition of PC-Kriging}
\end{figure}

The different types of meta-models are
illustrated in Fig.~\ref{fig:behaviour} which shows a 2-dimensional
contour plot of the output of the Rastrigin function ($N=128$~samples)
(Fig.~\ref{fig:behaviour:a}) and its approximations by PCE, ordinary
Kriging and PC-Kriging
(Fig.~\ref{fig:behaviour:b}-\ref{fig:behaviour:d}). The Rastrigin
function has a highly oscillatory behavior on the entire input space as
seen in Fig.~\ref{fig:behaviour:a}. This behavior is difficult to
meta-model with a small number of samples because many local
minima/maxima are missed out.

The analytical formulation of the Rastrigin function is a combination of
a quadratic component and a high-frequency trigonometric component. The
PCE model in Fig.~\ref{fig:behaviour:c} captures the global
characteristic of the function, \emph{i.e.} the quadratic component,
whereas the ordinary Kriging model in Fig.~\ref{fig:behaviour:b}
approximates the local characteristics, \emph{i.e.} the high-frequency
trigonometric component. Finally, the combination of PCE and Kriging
leads to a better meta-model as shown in Fig.~\ref{fig:behaviour:d}.

Note that the meta-models in Fig.~\ref{fig:behaviour} have a high
accuracy around the origin of the coordinate system due to the
definition of the input vector PDF as standard normal distributions
($X_i\sim\cN(0,1)$).

\subsubsection{Small experimental design}
\begin{figure} [!ht]
  \centering
  \subfigure[Ordinary Kriging]{
    \includegraphics[width=0.23\linewidth, angle=0]{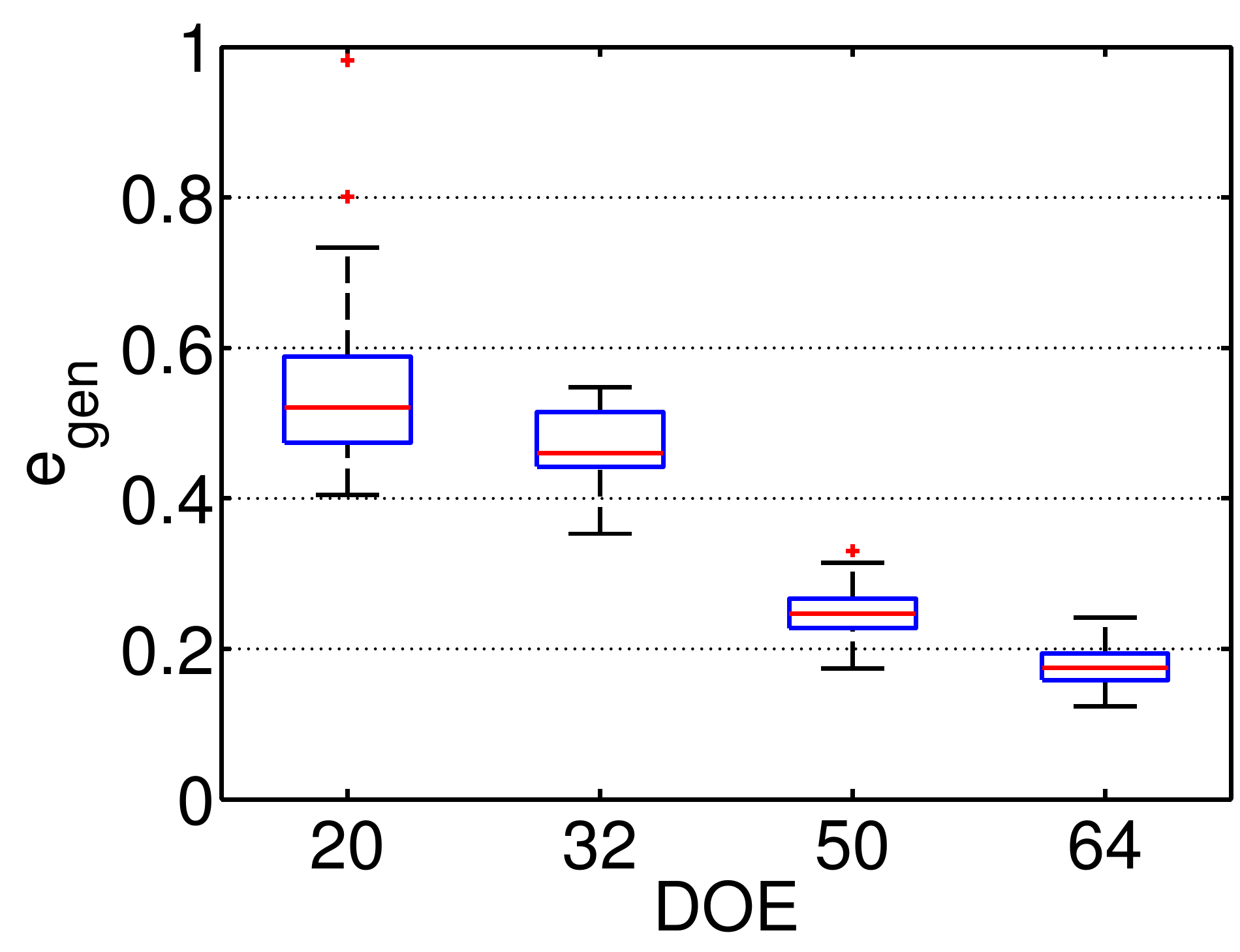}
  }
  \subfigure[PCE]{
    \includegraphics[width=0.23\linewidth, angle=0]{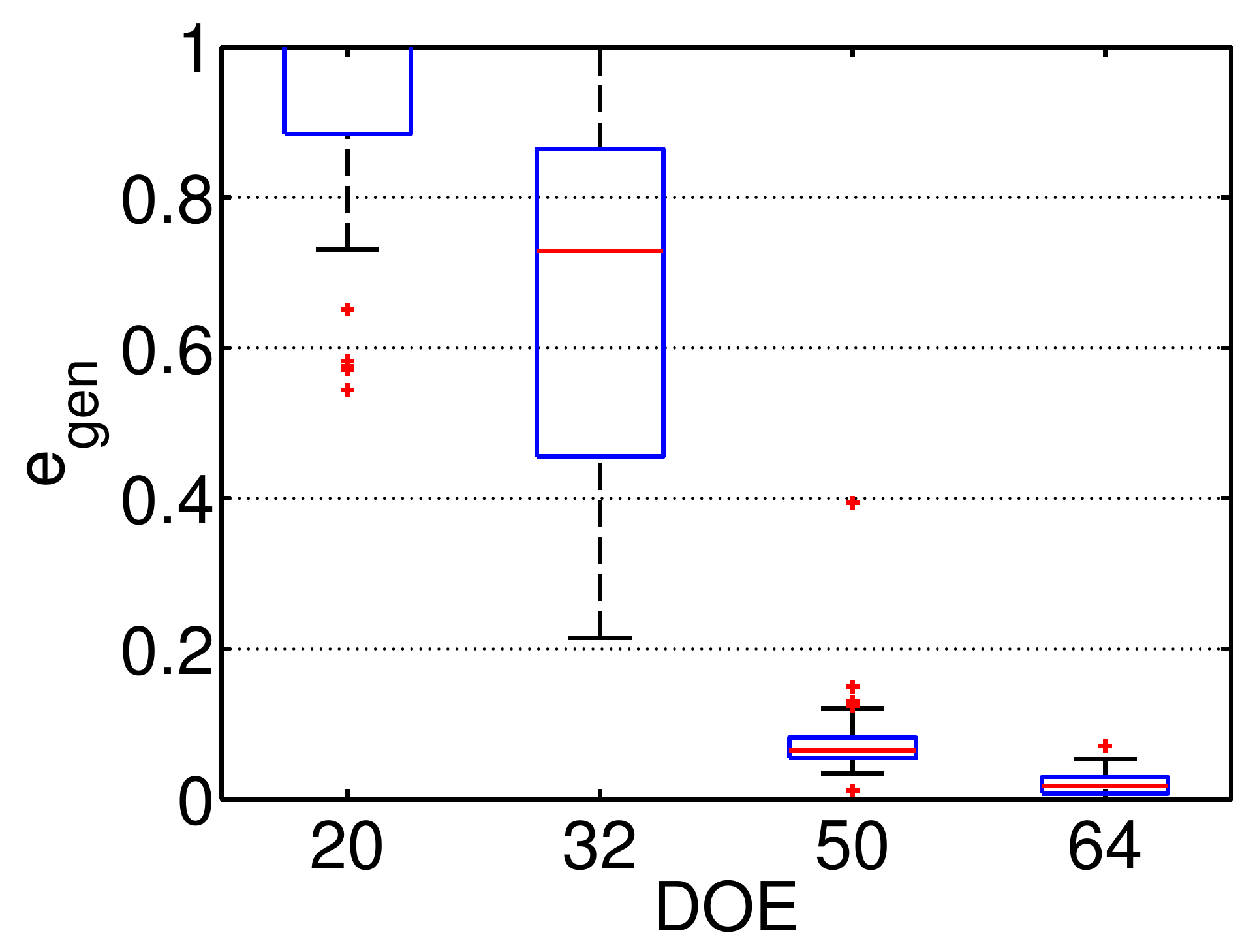}
  } 
  \subfigure[SPC-Kriging]{
    \includegraphics[width=0.23\linewidth, angle=0]{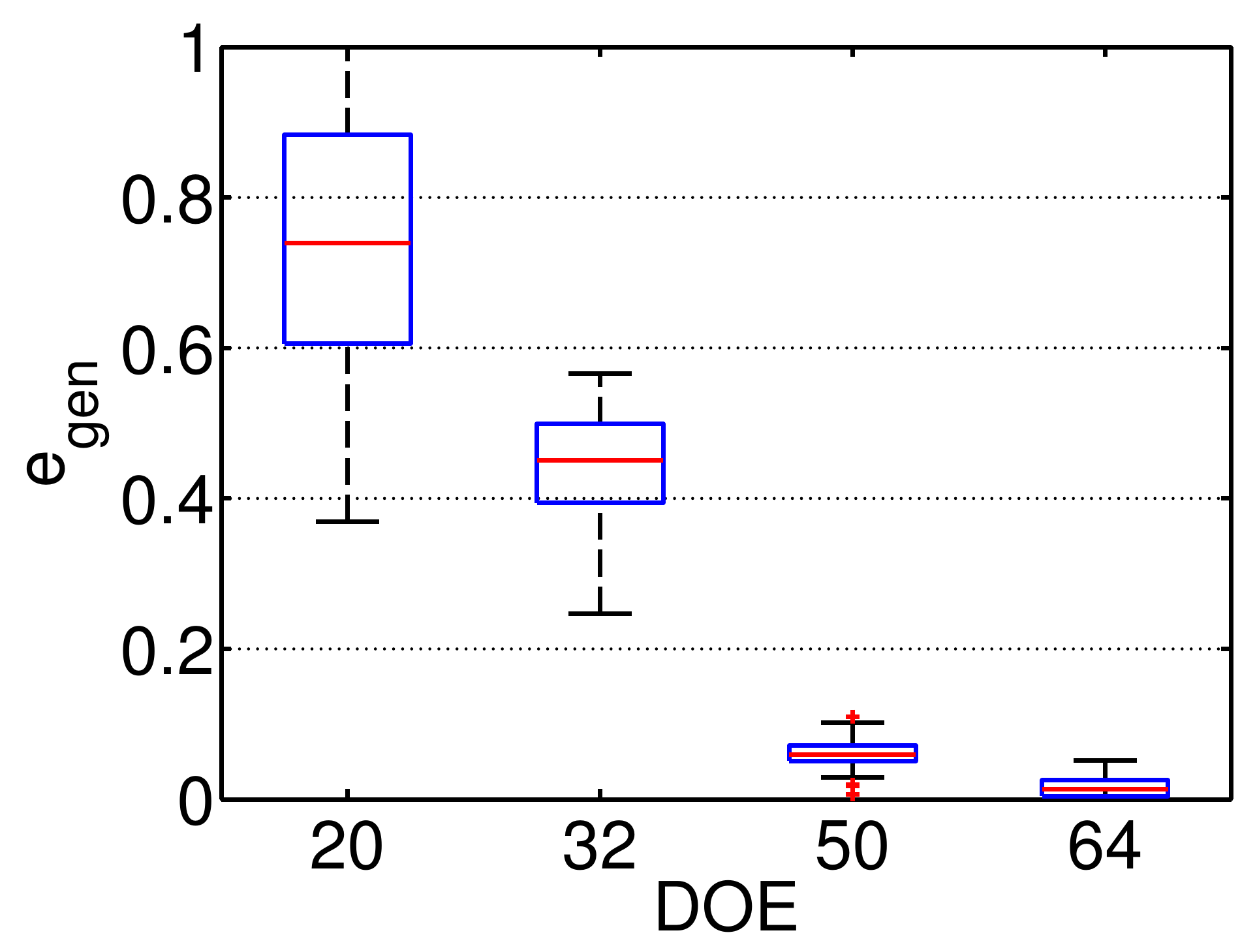}
  }
  \subfigure[OPC-Kriging]{
    \includegraphics[width=0.24\linewidth, angle=0]{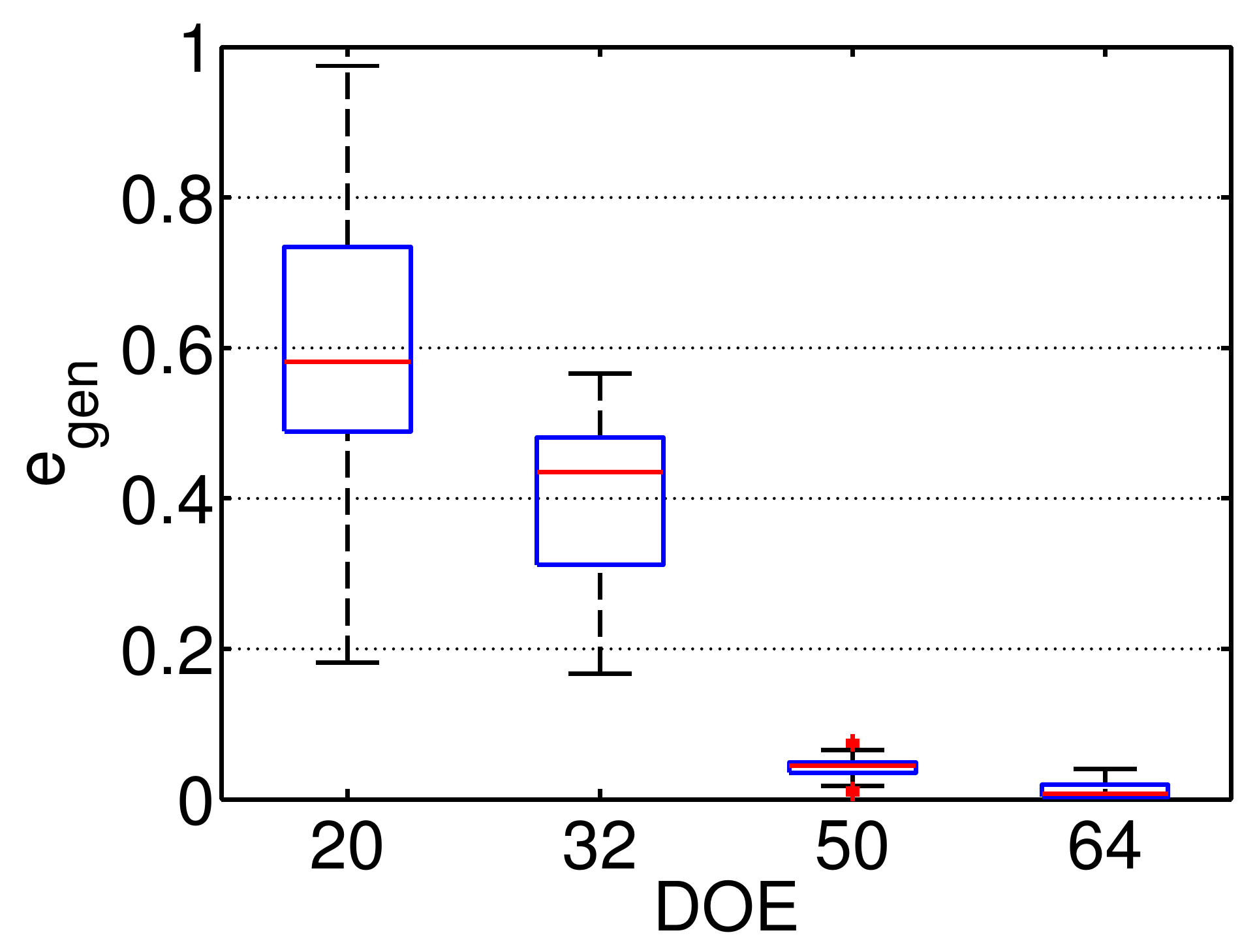}
  }
  \caption{\label{fig:ishigami}  Ishigami function -- Relative generalization error (Eq.~(\ref{eq:errgen})) for the various meta-modeling approaches}
\end{figure}

\begin{figure} [!ht]
  \centering
  \subfigure[Ordinary Kriging]{
    \includegraphics[width=0.23\linewidth, angle=0]{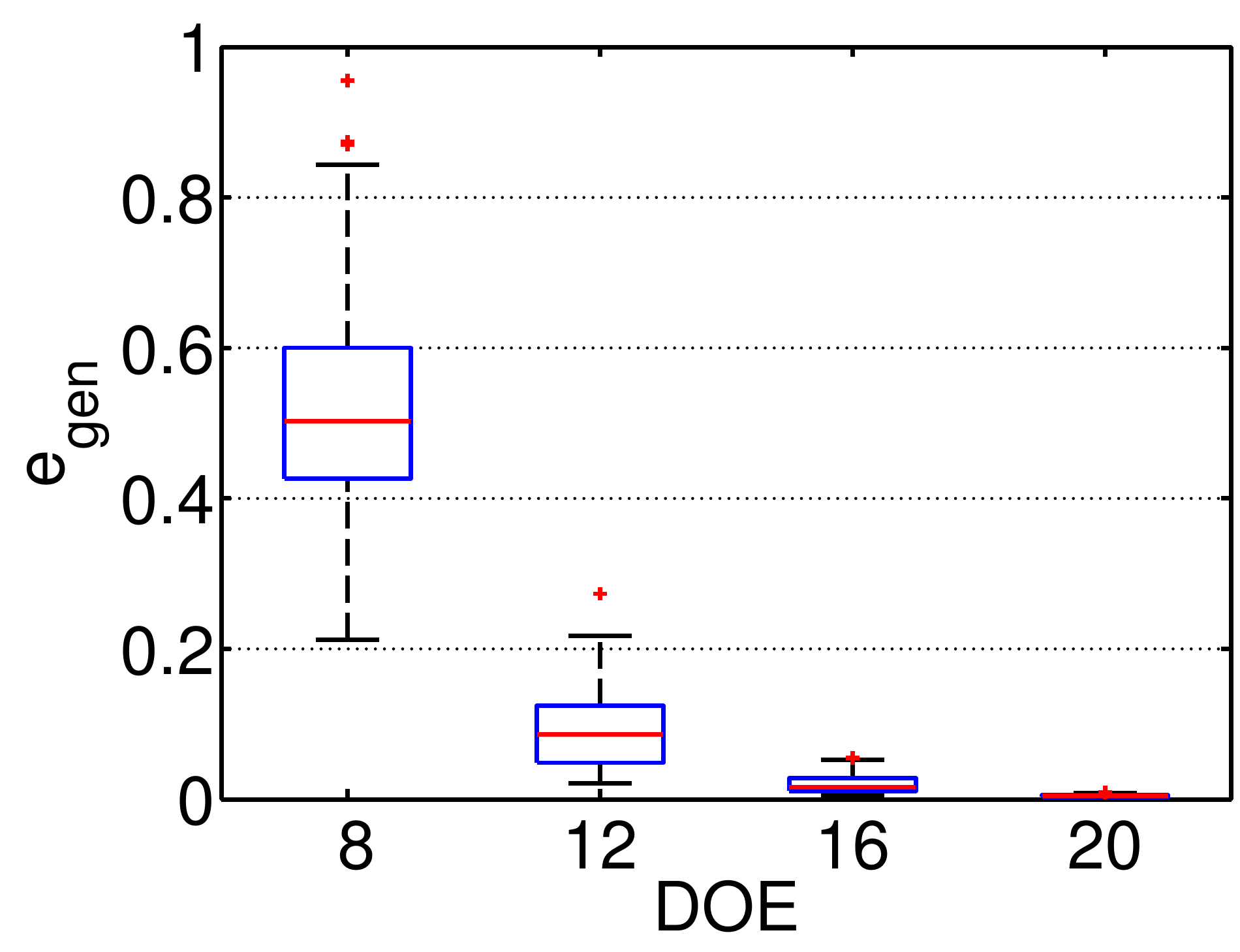}
  }
  \subfigure[PCE]{
    \includegraphics[width=0.23\linewidth, angle=0]{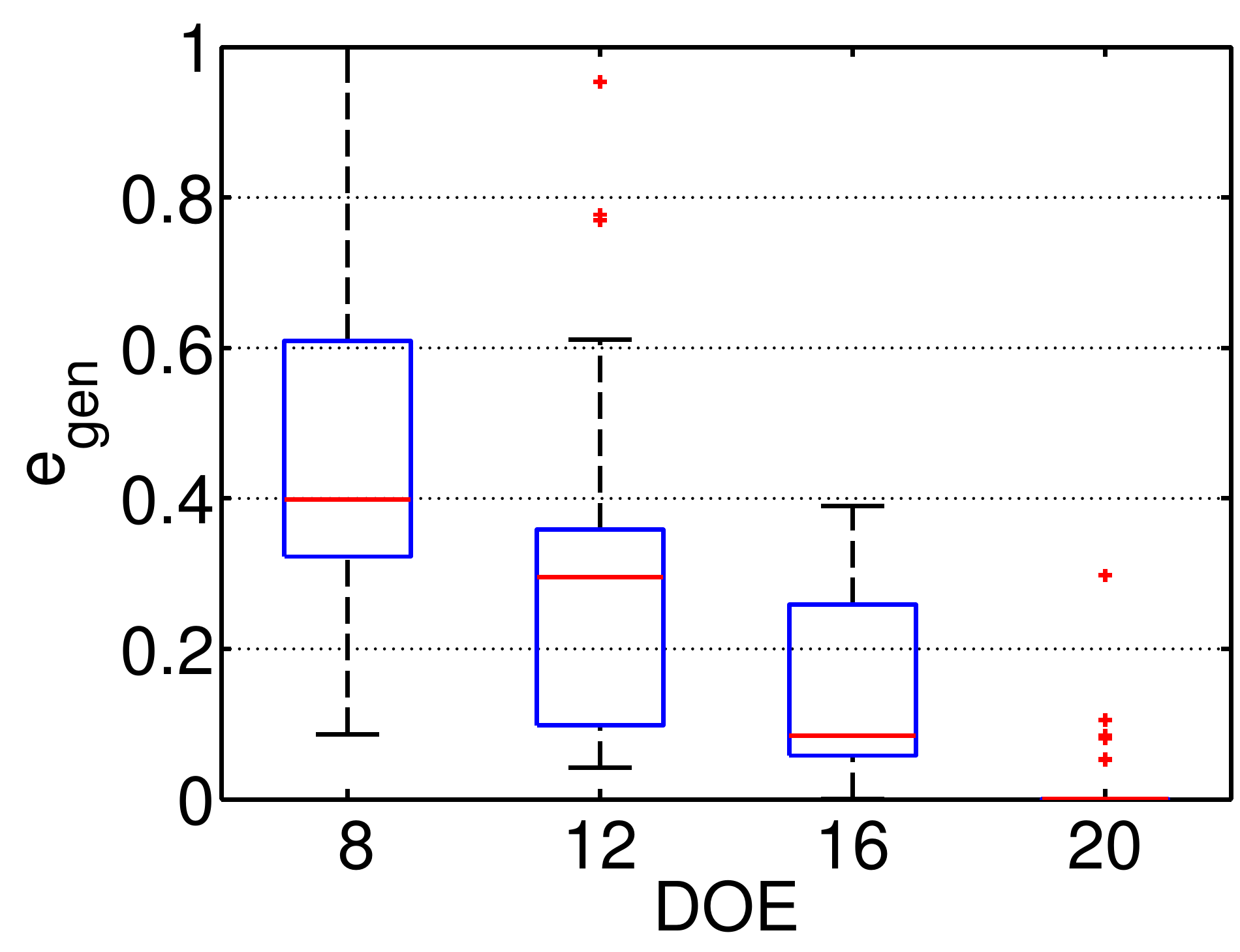}
  } 
  \subfigure[SPC-Kriging]{
    \includegraphics[width=0.23\linewidth, angle=0]{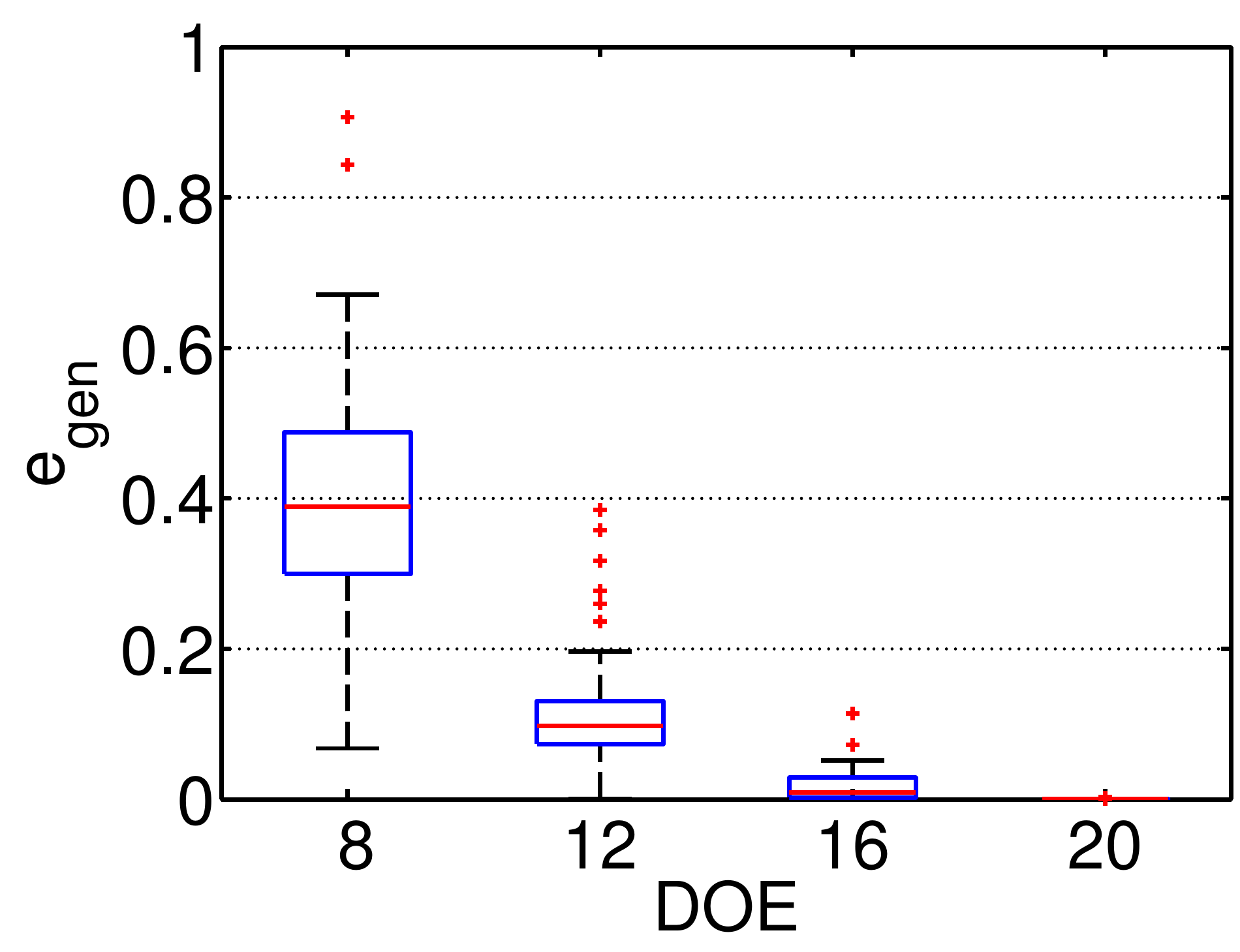}
  }
  \subfigure[OPC-Kriging]{
    \includegraphics[width=0.23\linewidth, angle=0]{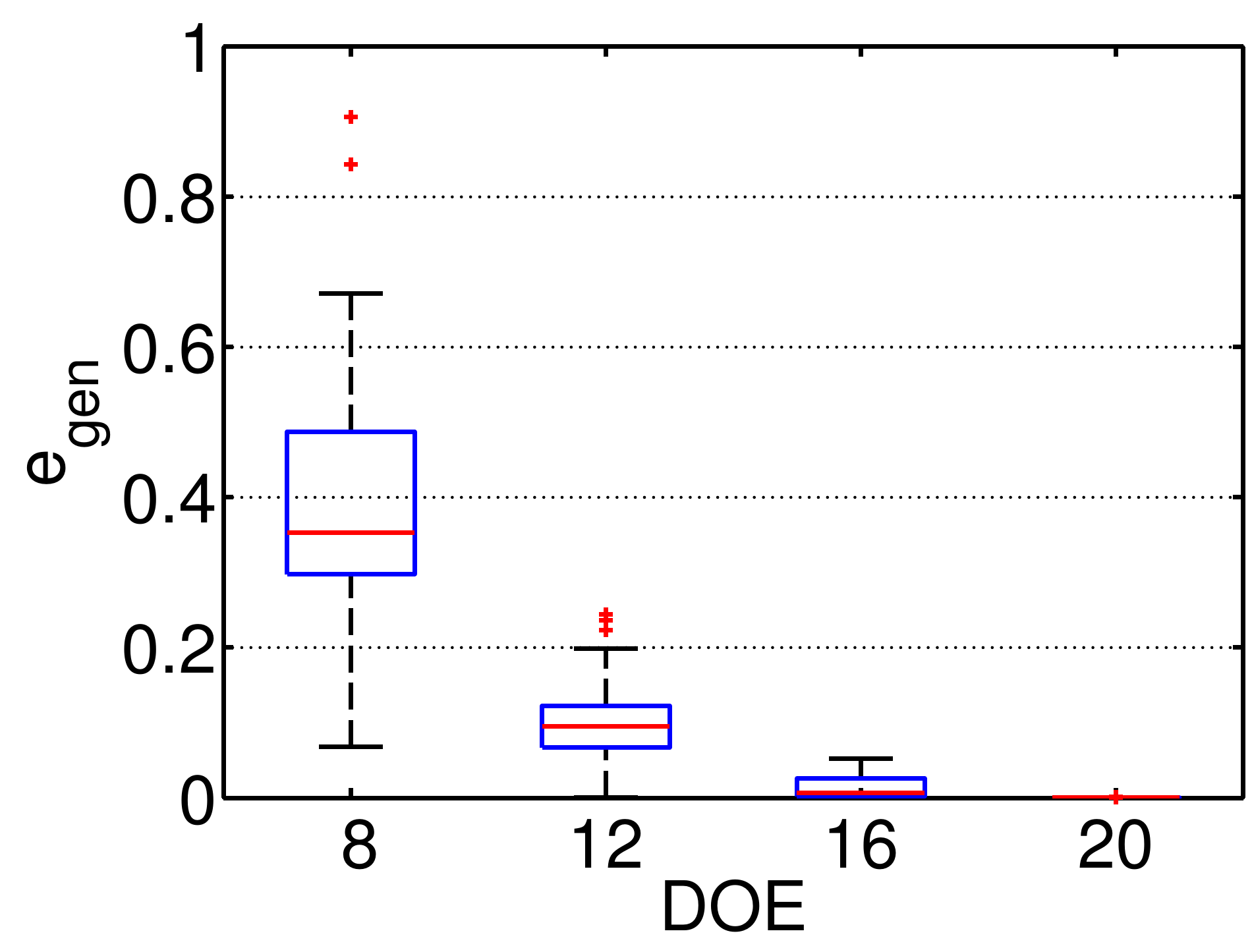}
  }
  \caption{\label{fig:rosenbrock}  Rosenbrock function  -- Relative generalization error (Eq.~(\ref{eq:errgen})) for the various meta-modeling approaches}
\end{figure}

\begin{figure} [!ht]
  \centering
  \subfigure[Ordinary Kriging]{
    \includegraphics[width=0.23\linewidth, angle=0]{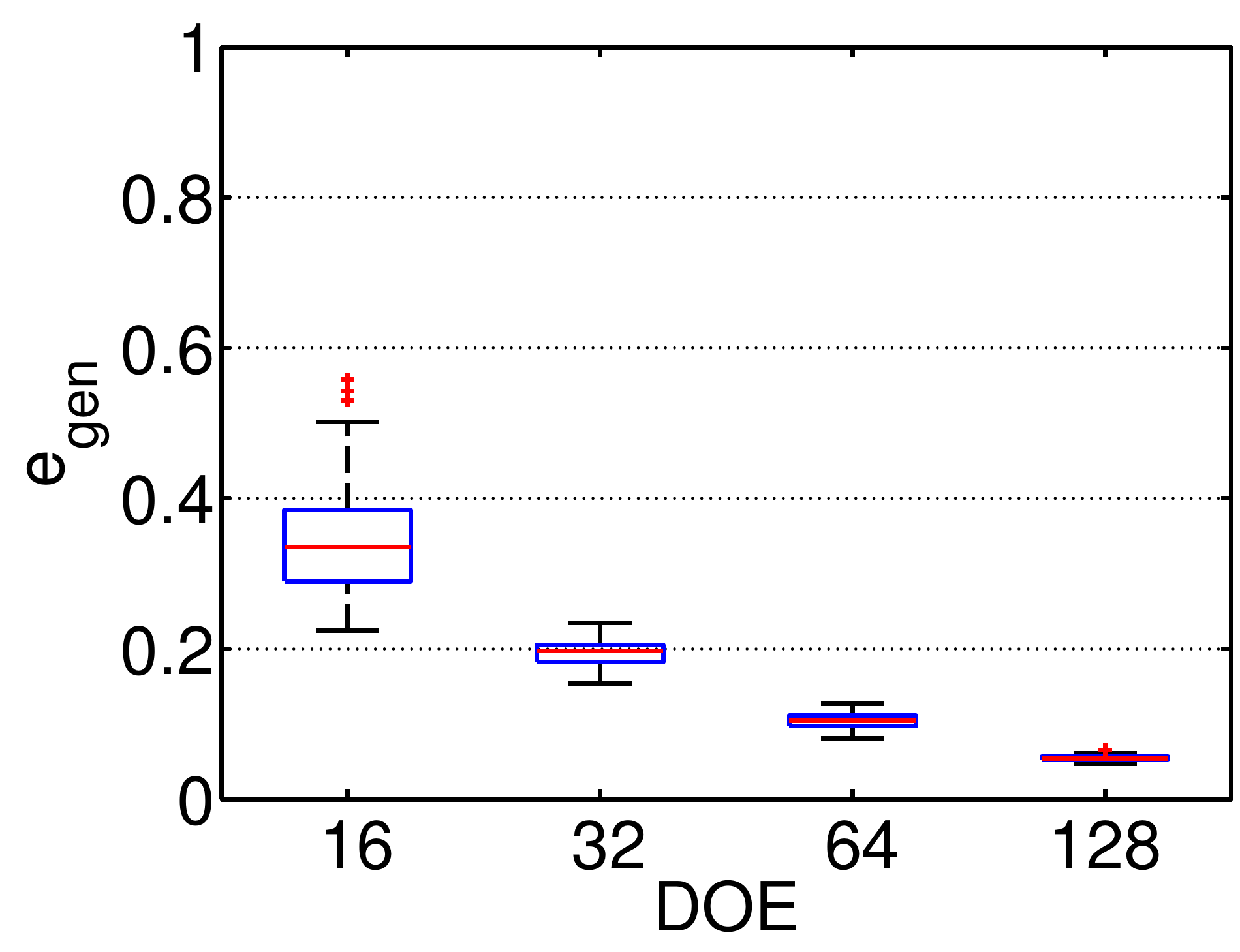}
  }
  \subfigure[PCE]{
    \includegraphics[width=0.23\linewidth, angle=0]{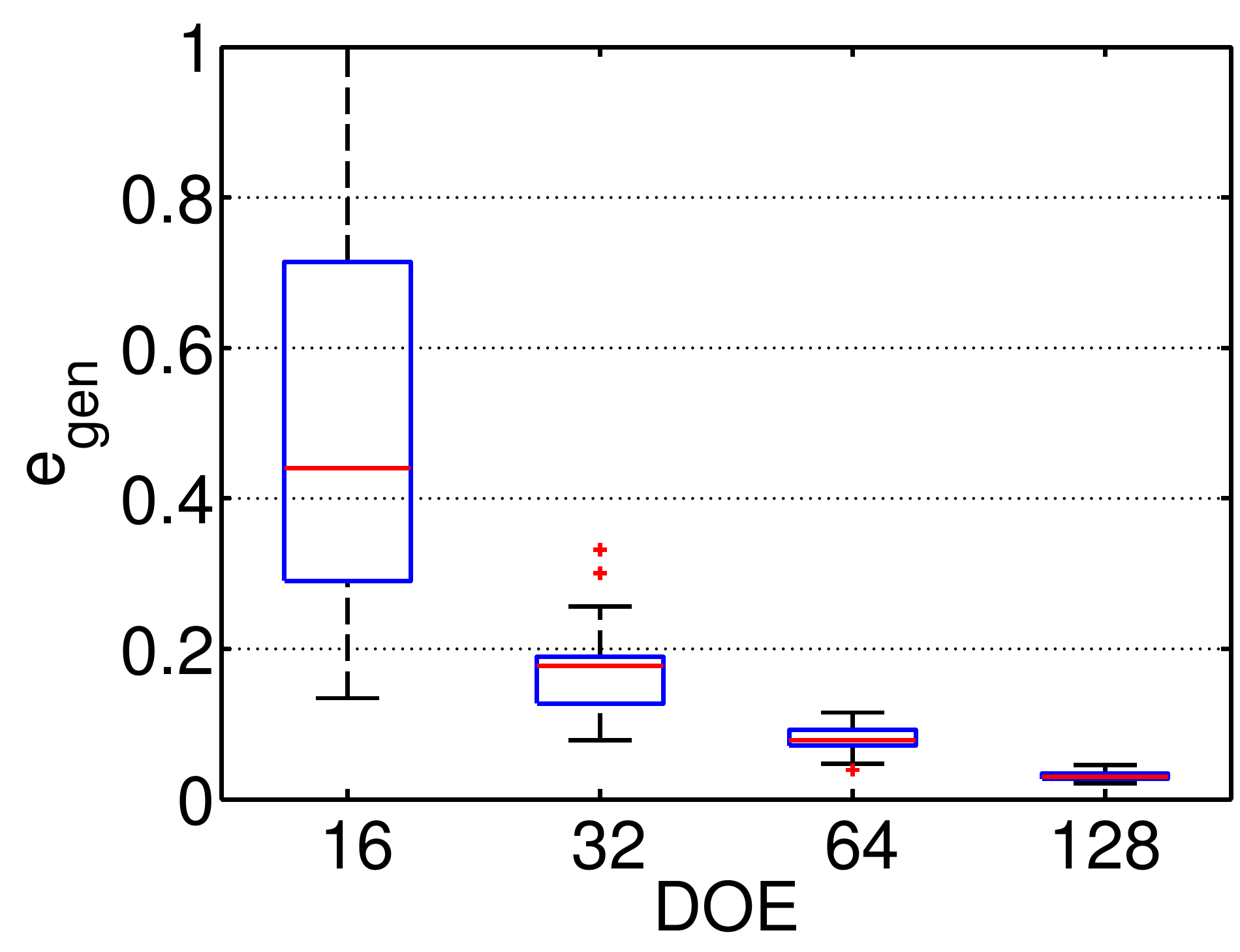}
  } 
  \subfigure[SPC-Kriging]{
    \includegraphics[width=0.23\linewidth, angle=0]{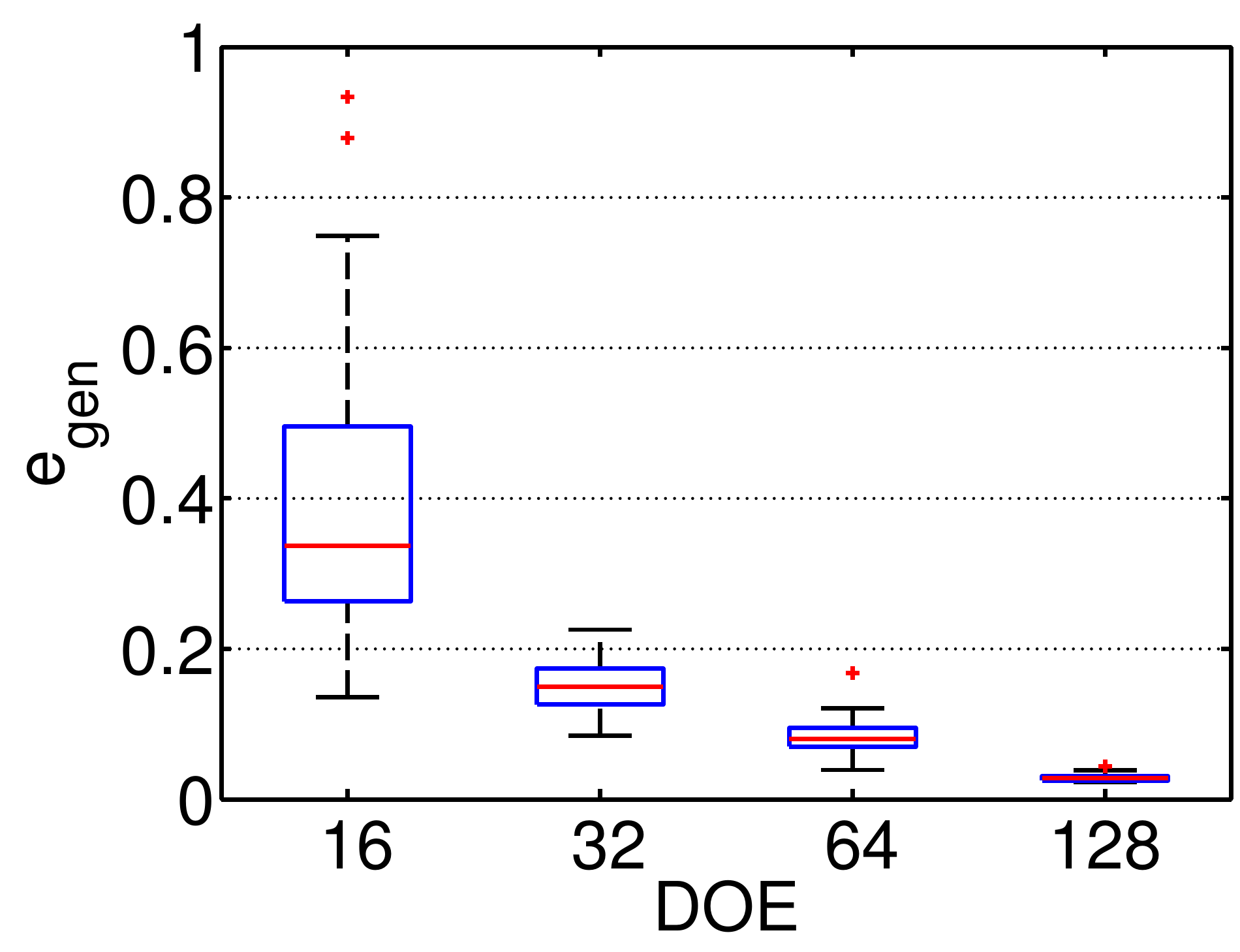}
  }
  \subfigure[OPC-Kriging]{
    \includegraphics[width=0.23\linewidth, angle=0]{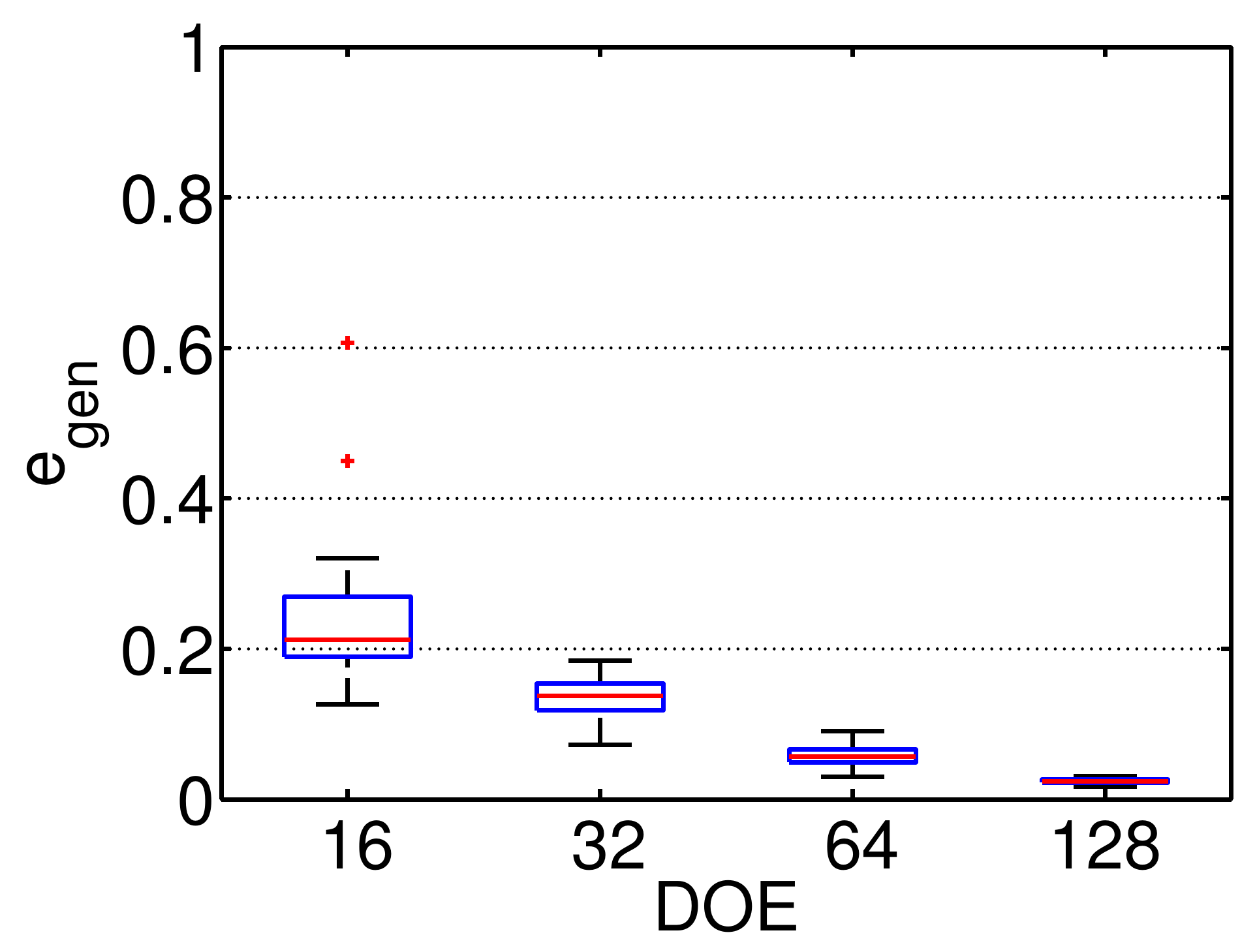}
  }
  \caption{\label{fig:sobol}  Sobol' function  -- Relative generalization error (Eq.~(\ref{eq:errgen})) for the various meta-modeling approaches}
\end{figure}

\begin{figure} [!ht]
  \centering
  \subfigure[Ordinary Kriging]{
    \includegraphics[width=0.23\linewidth, angle=0]{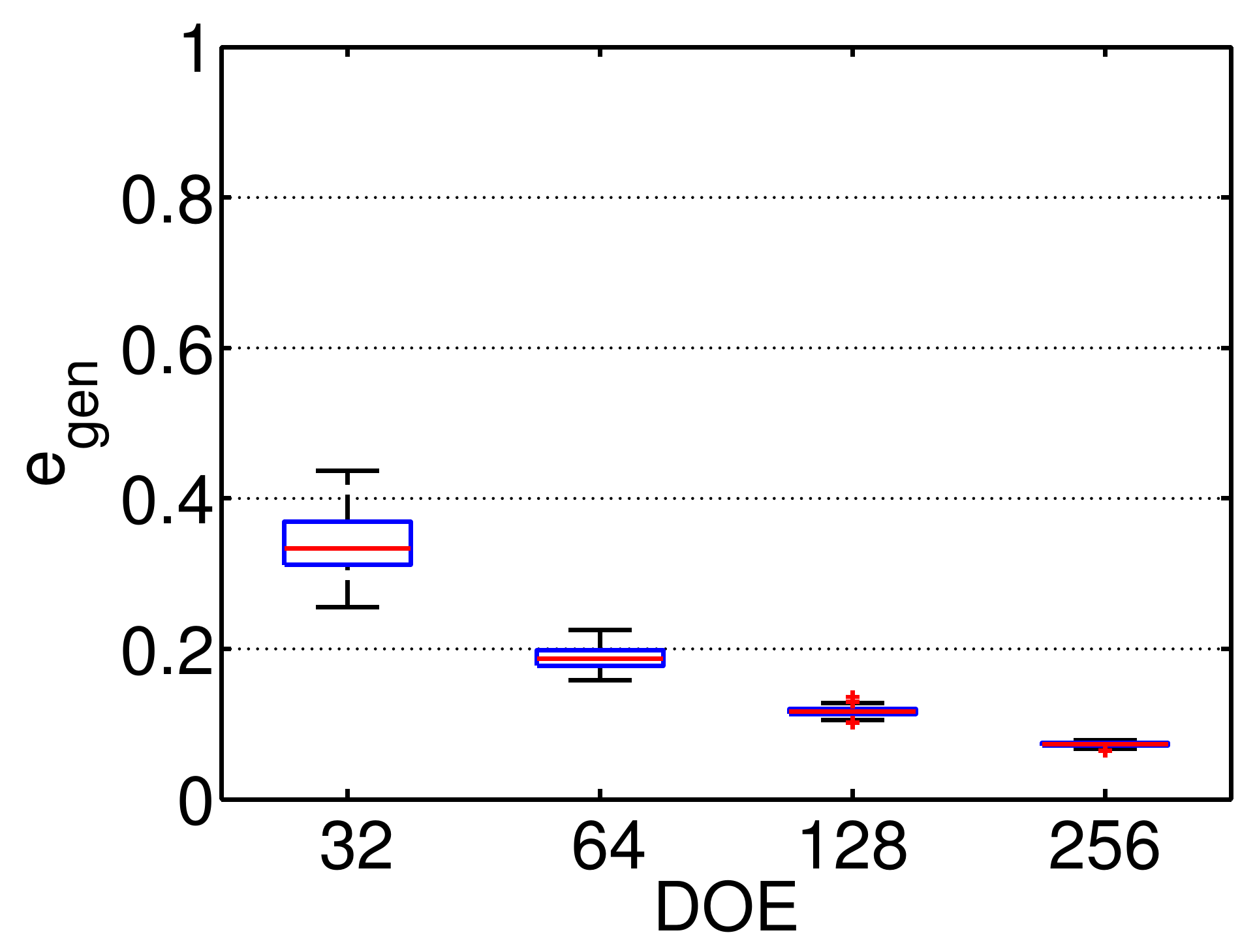}
  }
  \subfigure[PCE]{
    \includegraphics[width=0.23\linewidth, angle=0]{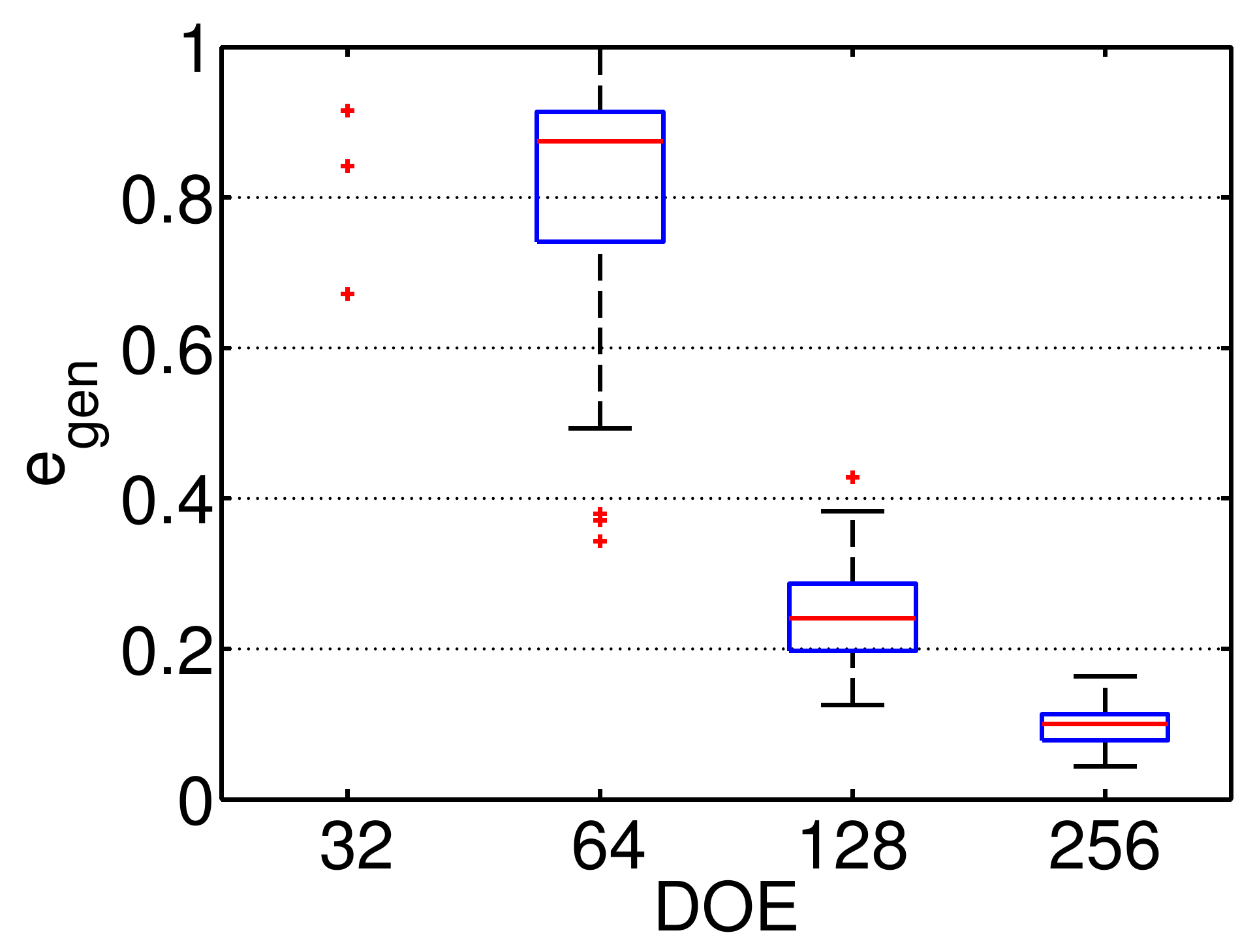}
  } 
  \subfigure[SPC-Kriging]{
    \includegraphics[width=0.23\linewidth, angle=0]{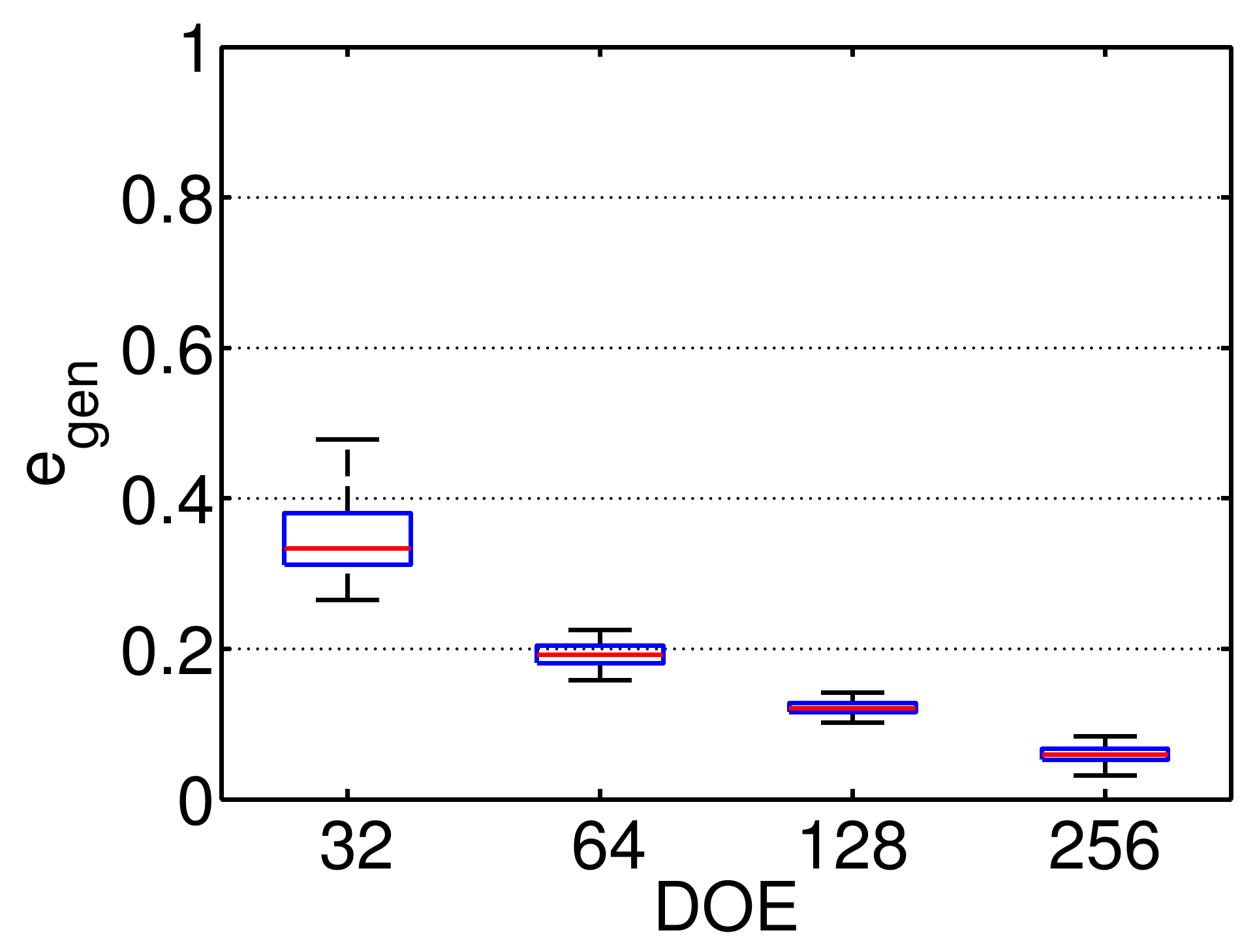}
  }
  \subfigure[OPC-Kriging]{
    \includegraphics[width=0.23\linewidth, angle=0]{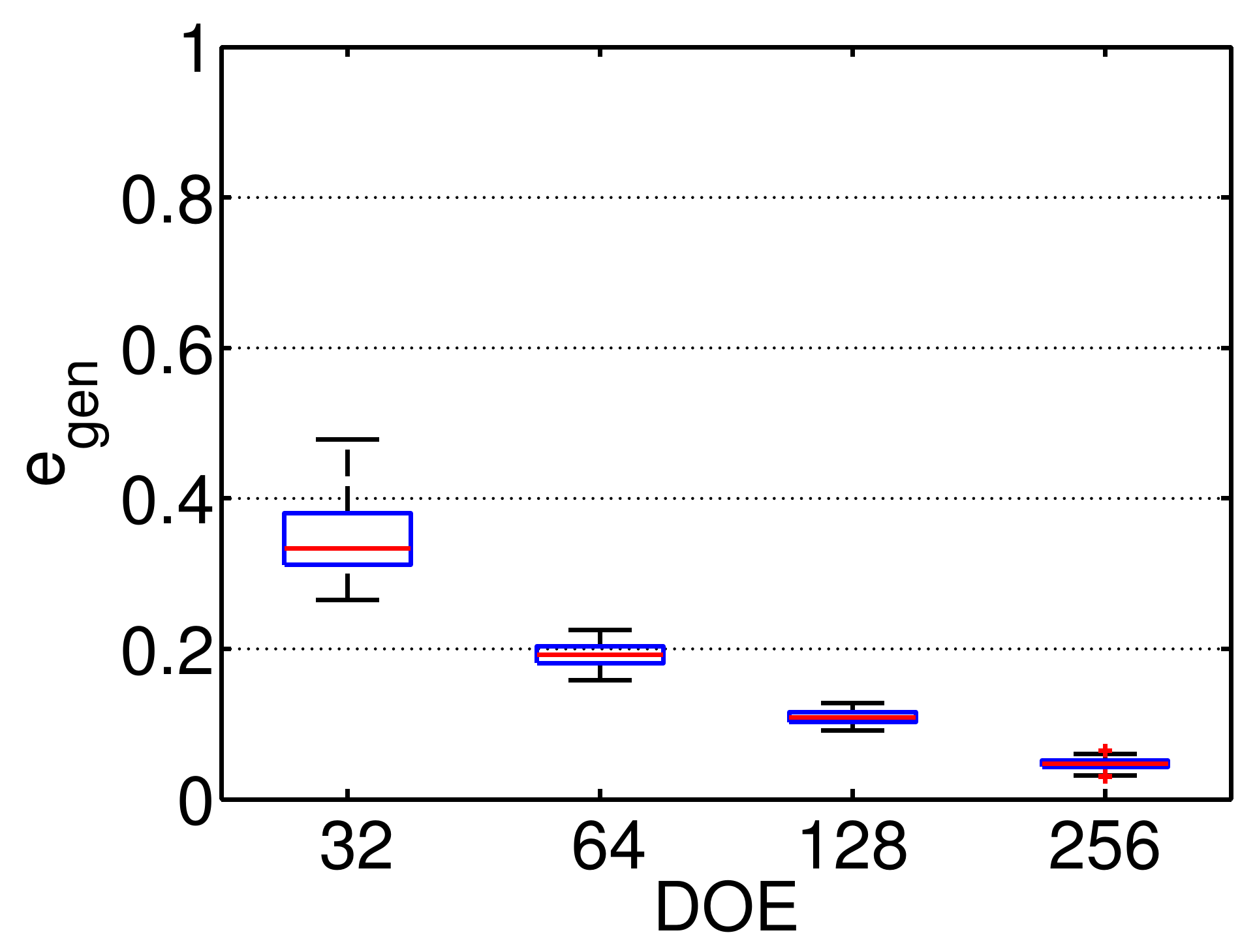}
  }
  \caption{\label{fig:morris}  Morris function -- Relative generalization error (Eq.~(\ref{eq:errgen})) for the various meta-modeling approaches}
\end{figure}

\begin{figure} [!ht]
  \centering
  \subfigure[Ordinary Kriging]{
    \includegraphics[width=0.23\linewidth, angle=0]{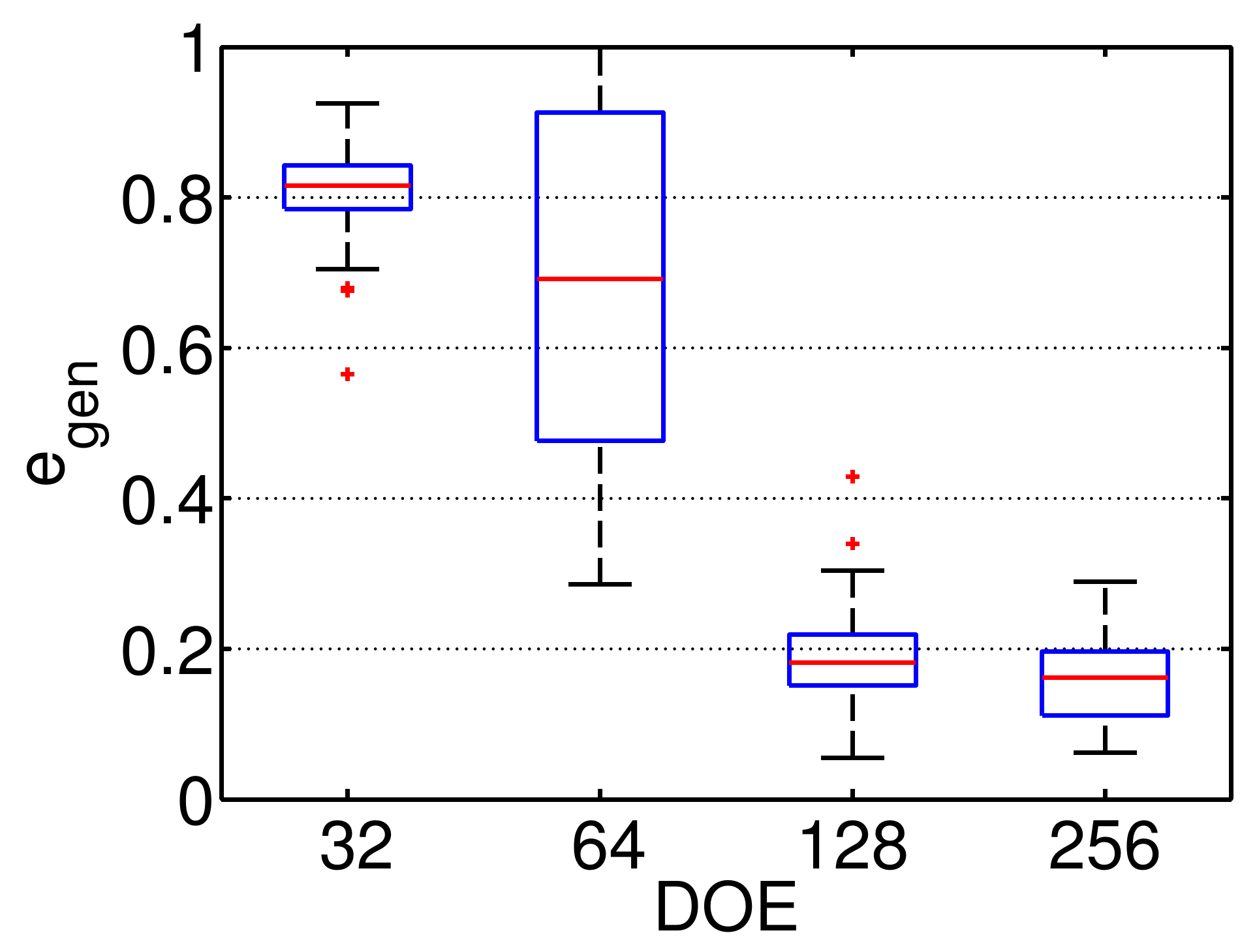}
  }
  \subfigure[PCE]{
    \includegraphics[width=0.23\linewidth, angle=0]{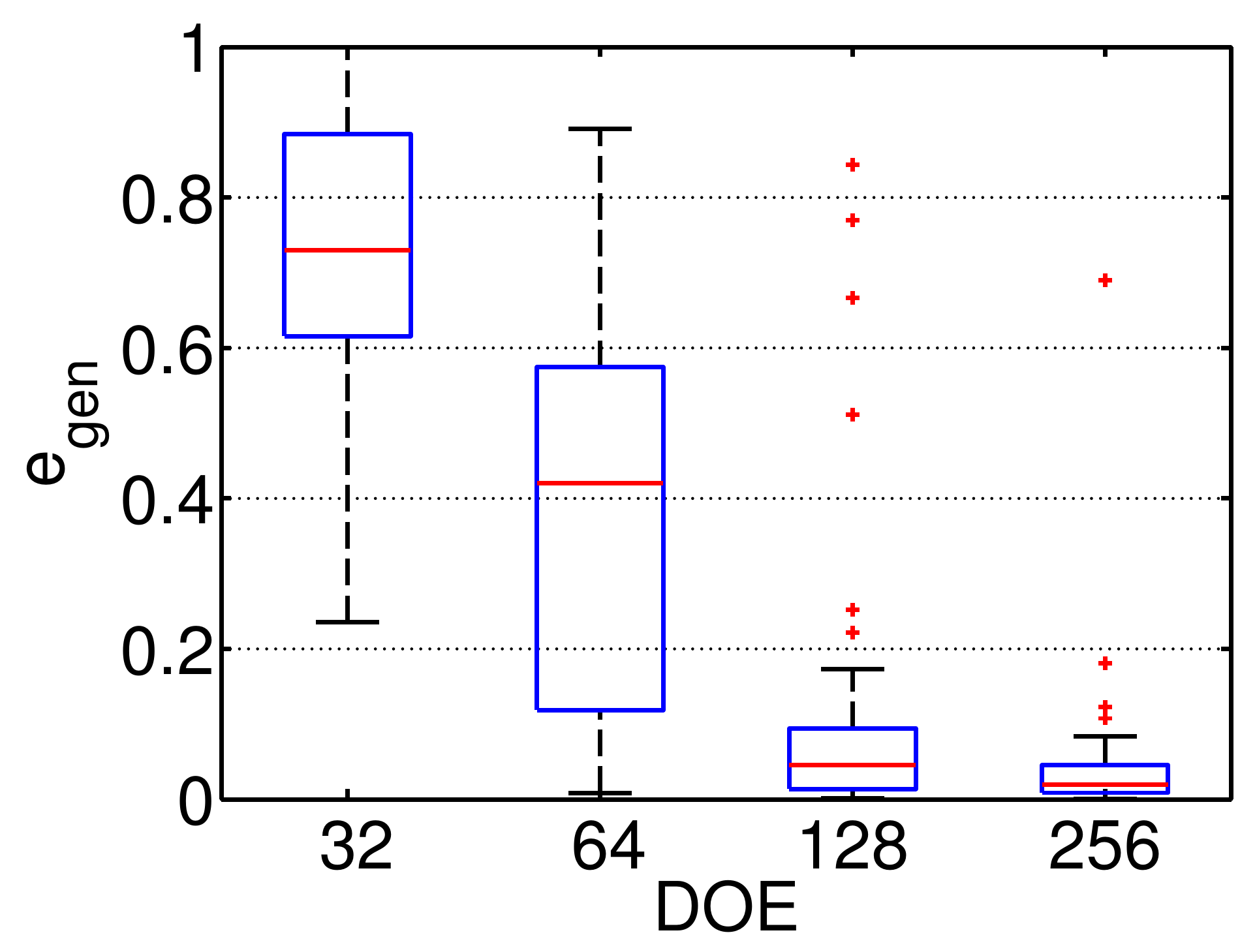}
  } 
  \subfigure[SPC-Kriging]{
    \includegraphics[width=0.23\linewidth, angle=0]{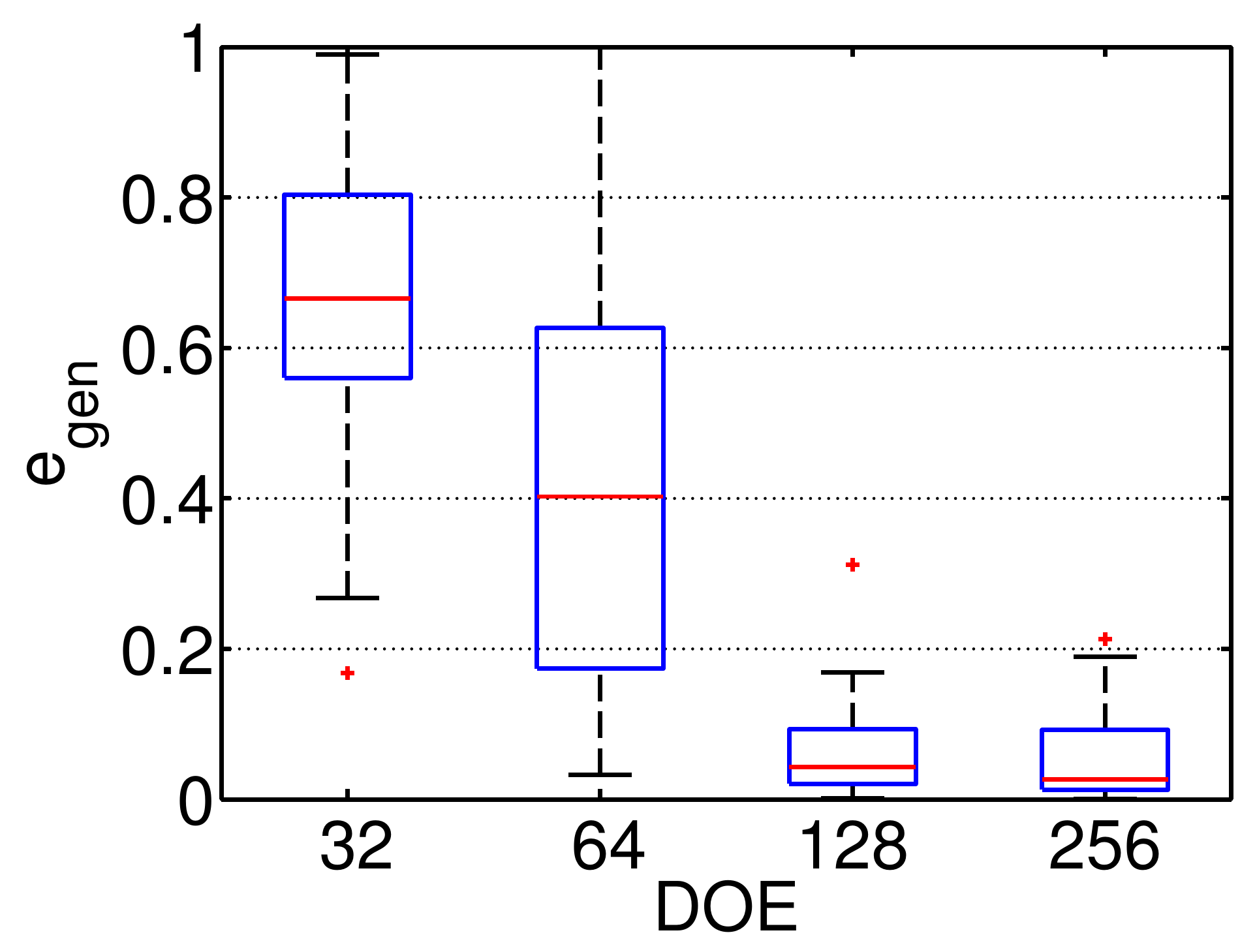}
  }
  \subfigure[OPC-Kriging]{
    \includegraphics[width=0.23\linewidth, angle=0]{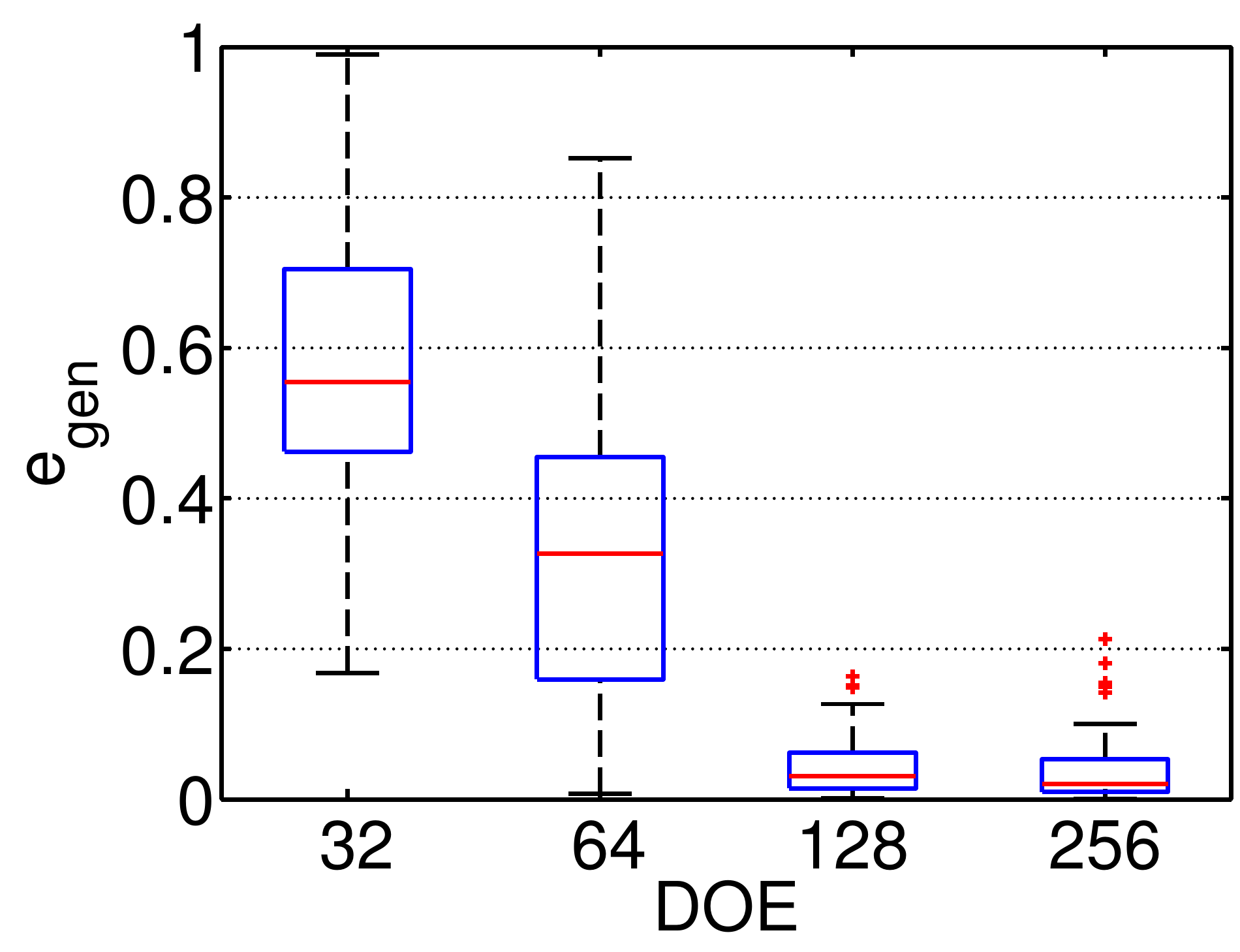}
  }
  \caption{\label{fig:rastrigin}  Rastrigin function -- Relative generalization error (Eq.~(\ref{eq:errgen})) for the various meta-modeling approaches}
\end{figure}

\begin{figure} [!ht]
  \centering
  \subfigure[Ordinary Kriging]{
    \includegraphics[width=0.23\linewidth, angle=0]{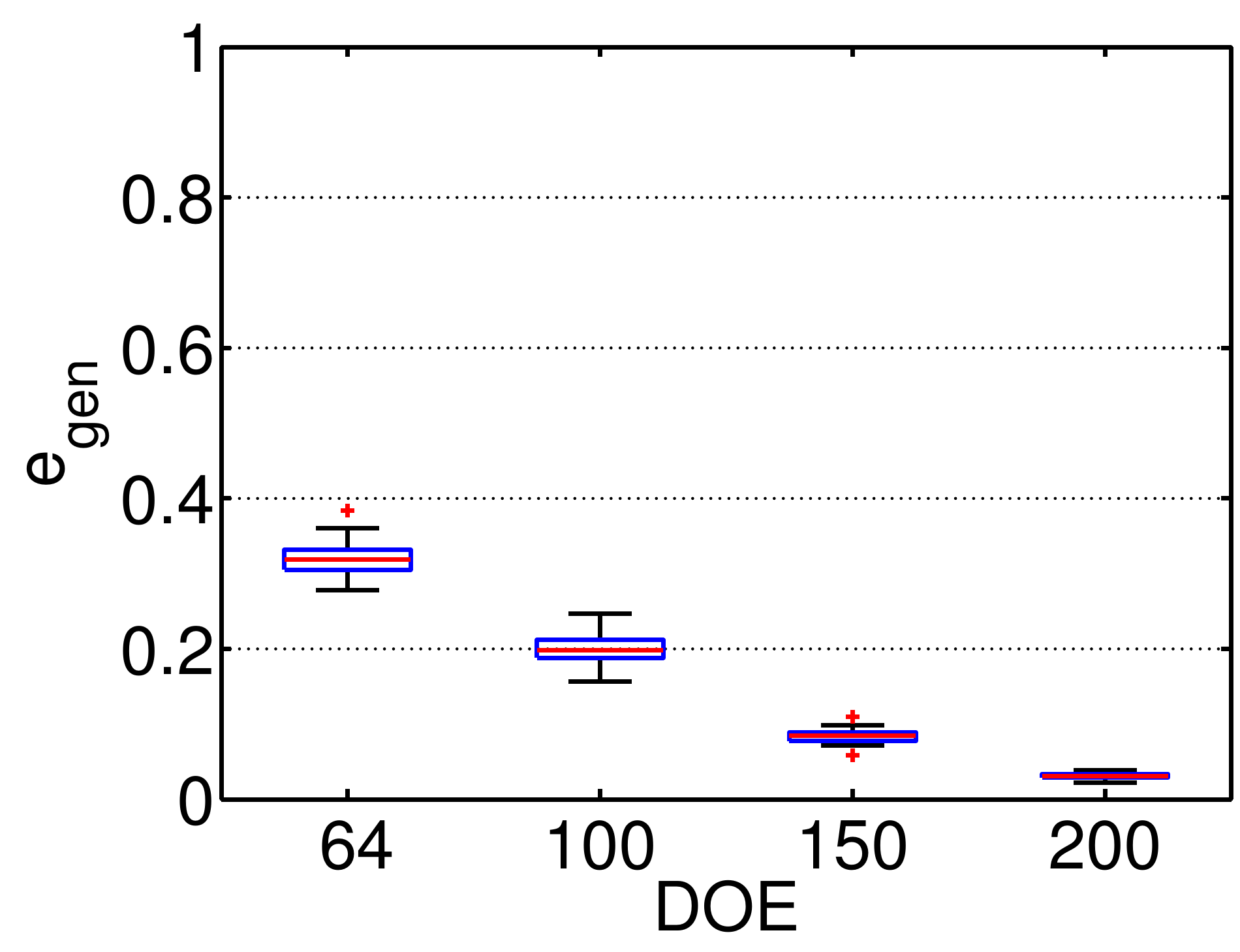}
  }
  \subfigure[PCE]{
    \includegraphics[width=0.23\linewidth, angle=0]{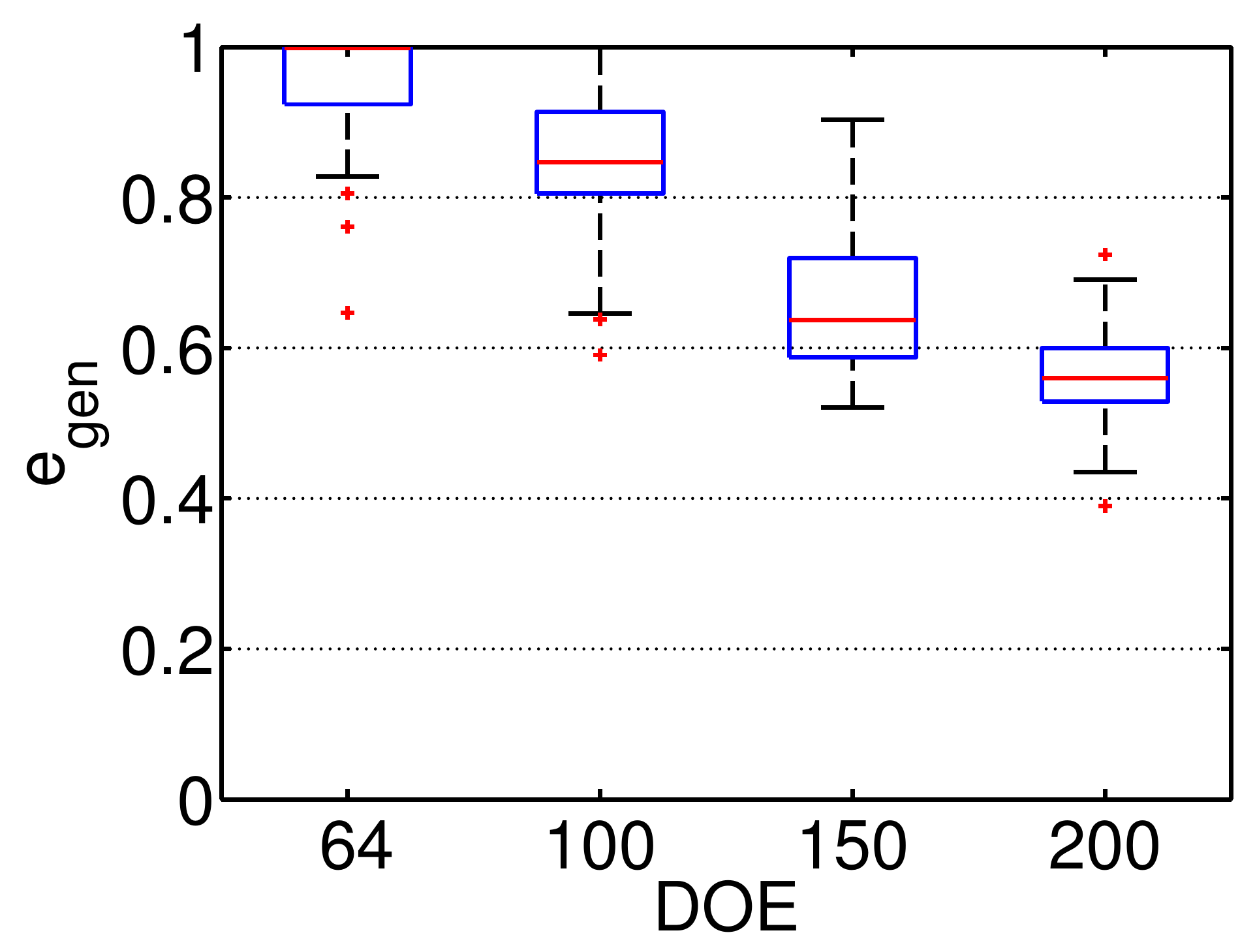}
  } 
  \subfigure[SPC-Kriging]{
    \includegraphics[width=0.23\linewidth, angle=0]{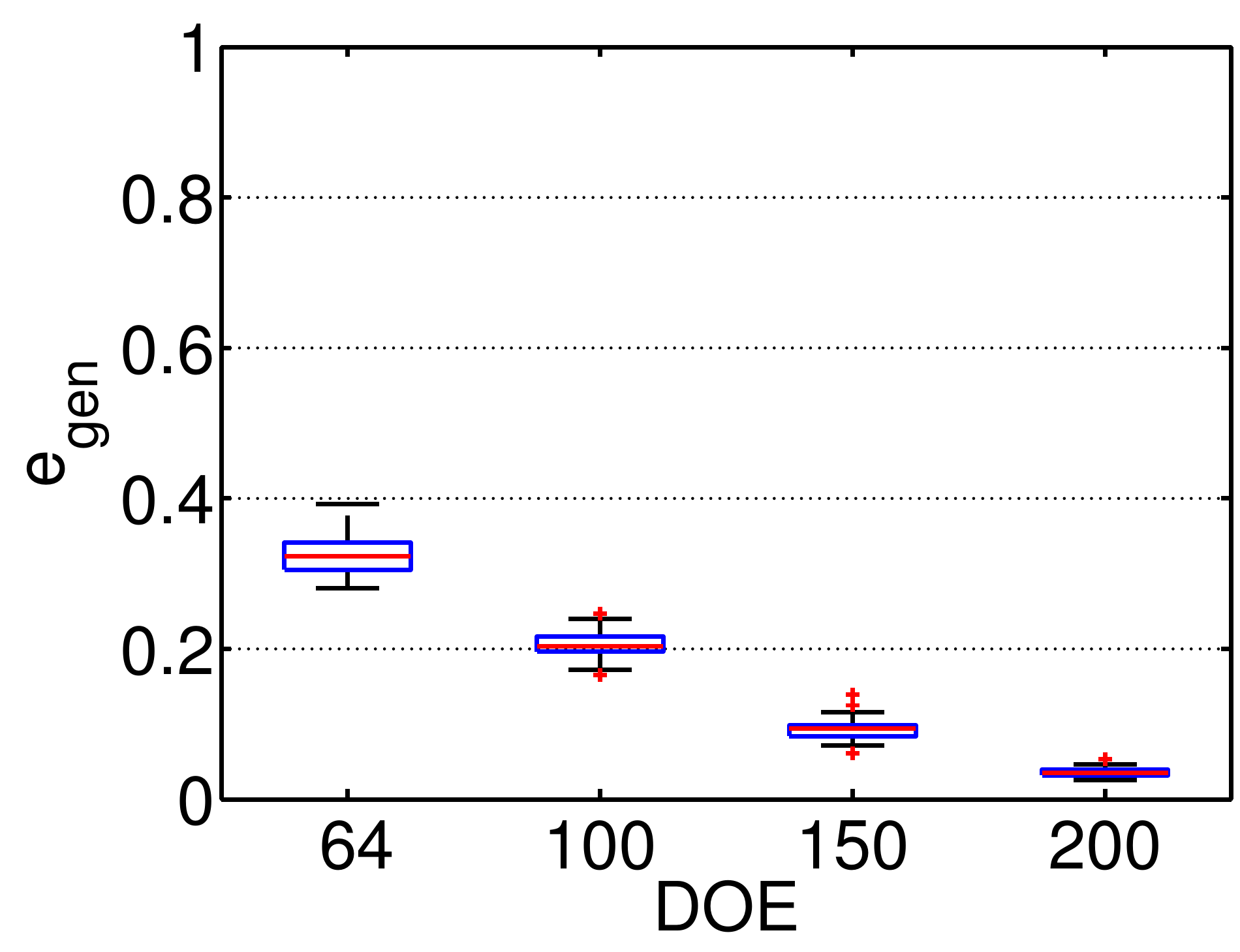}
  }
  \subfigure[OPC-Kriging]{
    \includegraphics[width=0.23\linewidth, angle=0]{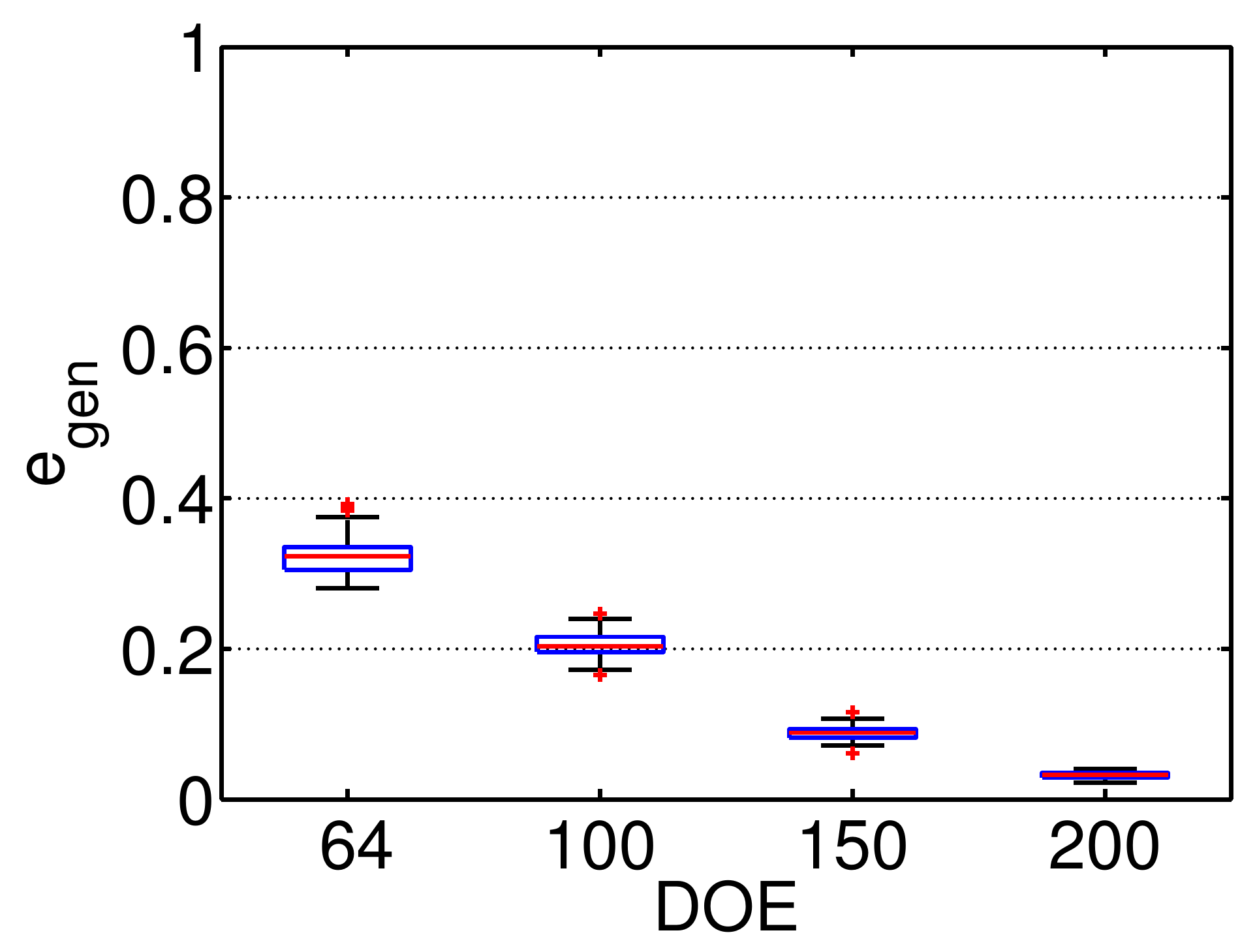}
  }
  \caption{\label{fig:ohagan}  O'Hagan function -- Relative generalization error (Eq.~(\ref{eq:errgen})) for the various meta-modeling approaches}
\end{figure}

The four meta-modeling techniques are compared for the six analytical
functions using experimental designs of increasing size. The number of
samples is chosen so that it yields a large range of relative
generalization errors on the second axis. The results are illustrated in
Fig.~\ref{fig:ishigami}-\ref{fig:ohagan}. In each figure (A) shows the
ordinary Kriging model and (B) shows the PCE model. The new approaches
SPC-Kriging and OPC-Kriging are shown in (C) and (D) respectively.

Figure~\ref{fig:ishigami} shows the relative generalization error for
the various meta-modeling techniques for the Ishigami function. For a
small sample size of $N=20$ samples, ordinary Kriging performs best with
respect to the median value of the box plots. { SPC-Kriging and OPC-Kriging perform worse due to the effect of overfitting (see also Section \ref{sec:errorestimation}). The number of parameters to be estimated in the case of PC-Kriging is larger than in the case of ordinary Kriging due to the number of polynomials in the trend of the Kriging model. Thus, OPC-Kriging (and also SPC-Kriging) are more prone to overfitting than ordinary Kriging for small experimental designs.  } 
When the number of samples
is increased, { however}, the two PC-Kriging approaches perform better than the
traditional approaches because their median value and their variation of
the error are lower. For the large sample sizes ($N \geq 50$),
PC-Kriging performs similarly to PCE, though slightly better.
OPC-Kriging is slightly more accurate than SPC-Kriging over the whole
range of sample sizes.

Figure~\ref{fig:rosenbrock} presents the results of the Rosenbrock
function, which is a purely polynomial function and can be modeled
accordingly with a small number of polynomials based on a small number
of points in the experimental design (\emph{e.g.} in the case of PCE). This is the reason
why the number of points lies within $N=8,\ldots,20$. Highly accurate
surrogate models are obtained with only 20 samples. For small sample
sizes OPC-Kriging performs best among the four techniques in terms of
the relative generalization error.

The Sobol' function is more complex than the two previous functions
because of the dimensionality ($M=8$) and the non-smooth behavior at
$x_i=0.5$. Thus more samples are needed to obtain a similar range of
relative generalization errors compared to the previous functions, as
seen in Fig.~\ref{fig:sobol}. Behaviors for the larger sample sizes
($N=64,\,128$) are very similar among the meta-modeling approaches,
although PC-Kriging performs slightly better than the traditional PCE and
ordinary Kriging approaches. For very small sample sizes ($N=16,\, 32$),
OPC-Kriging performs significantly better than the others.

Figure~\ref{fig:morris} shows the results for the Morris function. A
large experimental design is required to properly surrogate the
computational model because of the high dimensionality of the input
vector $\vX$ and the amount of interactive terms of different input
variables $X_i$ in the analytical formulation (see Eq.~(\ref{eq:f4})).
The relative generalization error of the two PC-Kriging approaches
resembles more the one of ordinary Kriging than the one of PCE in this
case. PCE is not capable of modeling this analytical function with a
small number of samples.

The results associated with the Rastrigin function are shown in
Fig.~\ref{fig:rastrigin}. Despite the low dimensionality of the input
($M=2$), many samples are needed to obtain small error estimates. This
is due to the fact that the function output is highly oscillatory over
the entire input space as previously illustrated in
Fig.~\ref{fig:behaviour}. In comparison to
Section~\ref{sec:results:viz}, which describes the qualitative
performance of PC-Kriging on the Rastrigin function, the quantitative
benefit of combining PCE and Kriging becomes visible in
Fig.~\ref{fig:rastrigin}: PC-Kriging performs better than the
traditional approaches. Ordinary Kriging performs the worst followed by
PCE. OPC-Kriging has statistically the lowest relative generalization
errors over the whole range of experimental design sizes.

Figure~\ref{fig:ohagan} displays the results associated with the O'Hagan
function. Similarly to the Morris function, the performance of
PC-Kriging in the case of the O'Hagan function resembles that of
ordinary Kriging whereas PCE performs worse than the other three
approaches. Over the entire displayed range of experimental designs in
Fig.~\ref{fig:ohagan}, the performance of OPC-Kriging is slightly better
than the performance of SPC-Kriging and ordinary Kriging. Note that
meta-modeling the O'Hagan function requires less samples in the
experimental design to obtain the same accuracy as in the case of the
Rastrigin function despite the fact that the O'Hagan function has a
15-dimensional input space and that both functions are very smooth.

Summarizing the results of all six analytical functions in
Fig.~\ref{fig:ishigami}-\ref{fig:ohagan}, the proposed PC-Kriging
approaches perform better than or at least as good as the traditional
PCE and Kriging approaches. Note that for functions like O'Hagan
(Fig.~\ref{fig:ohagan}) and Morris (Fig.~\ref{fig:morris}) the
performance of PC-Kriging is more similar to Kriging than PCE, whereas
for the other functions the performance of PC-Kriging resembles more
that of PCE. As one could expect, there is no general rule so as to
decide whether PCE or Kriging provide the most accurate meta-models for
a given experimental design. The advantage of PC-Kriging is to perform
as least as well as the best of the two.

The combination of PCE and Kriging and its increased accuracy comes with
a higher computational cost. The traditional ordinary Kriging and PCE
approaches have the lowest computational cost, SPC-Kriging has an
intermediate and OPC-Kriging has the highest cost. The high cost of
OPC-Kriging originates from the iterative character of the algorithm and
the accompanying repetitive calibration of Kriging models.  OPC-Kriging
computes LAR and calibrates $P$ Kriging models with increasing complex
trend, whereas ordinary Kriging consists of a single calibration of a
Kriging model with constant trend.  For a single calibration of a
surrogate of the Ishigami function (experimental design of size
$N=128$~samples) the ratio of computational times when comparing PCE to
ordinary Kriging, SPC-Kriging and OPC-Kriging is approximately $1:5$,
$1:20$ and $1:200$, respectively.

Note that it is intended to apply these techniques to realistic problems
where the evaluation of the exact computational model response lasts
much longer than the computation of a meta-model. The apparent
computational overload of OPC-Kriging will not be anymore an issue in
many practical applications.

\subsubsection{Large experimental designs}
\begin{figure} [!ht]
  \centering
  \subfigure[Ordinary Kriging]{
    \includegraphics[width=0.23\linewidth, angle=0]{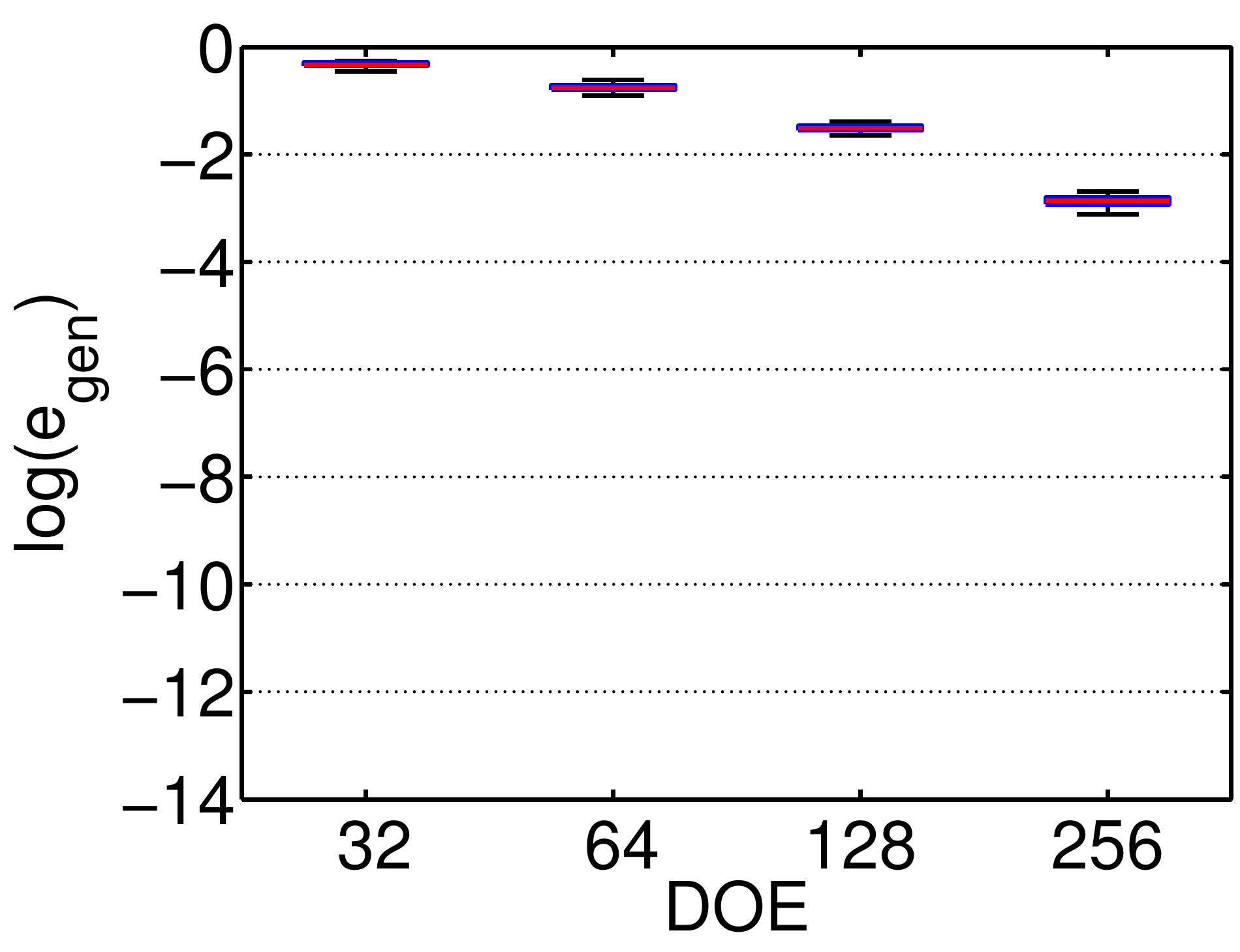}
  }
  \subfigure[PCE]{
    \includegraphics[width=0.23\linewidth, angle=0]{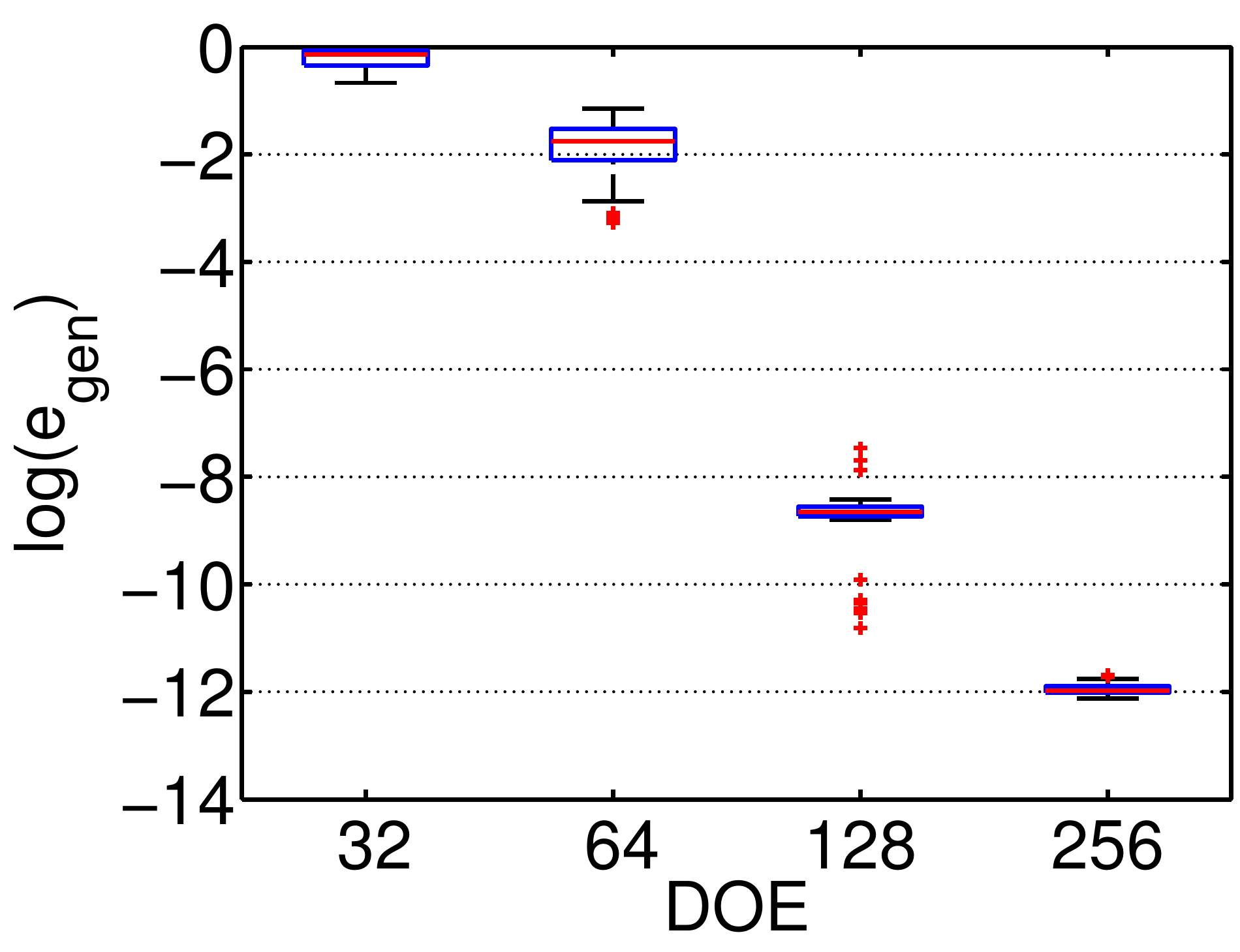}
  } 
  \subfigure[SPC-Kriging]{
    \includegraphics[width=0.23\linewidth, angle=0]{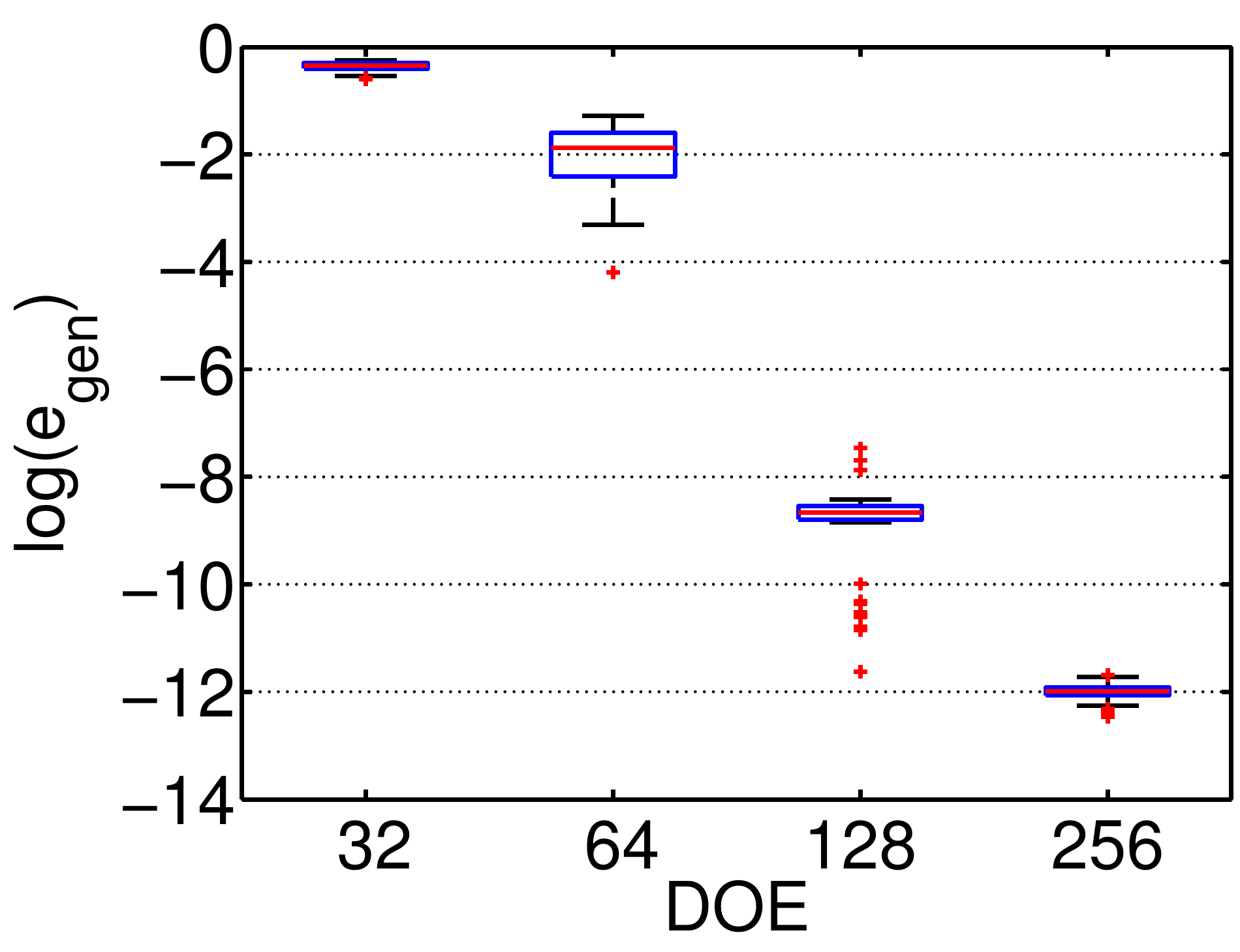}
  }
  \subfigure[OPC-Kriging]{
    \includegraphics[width=0.23\linewidth, angle=0]{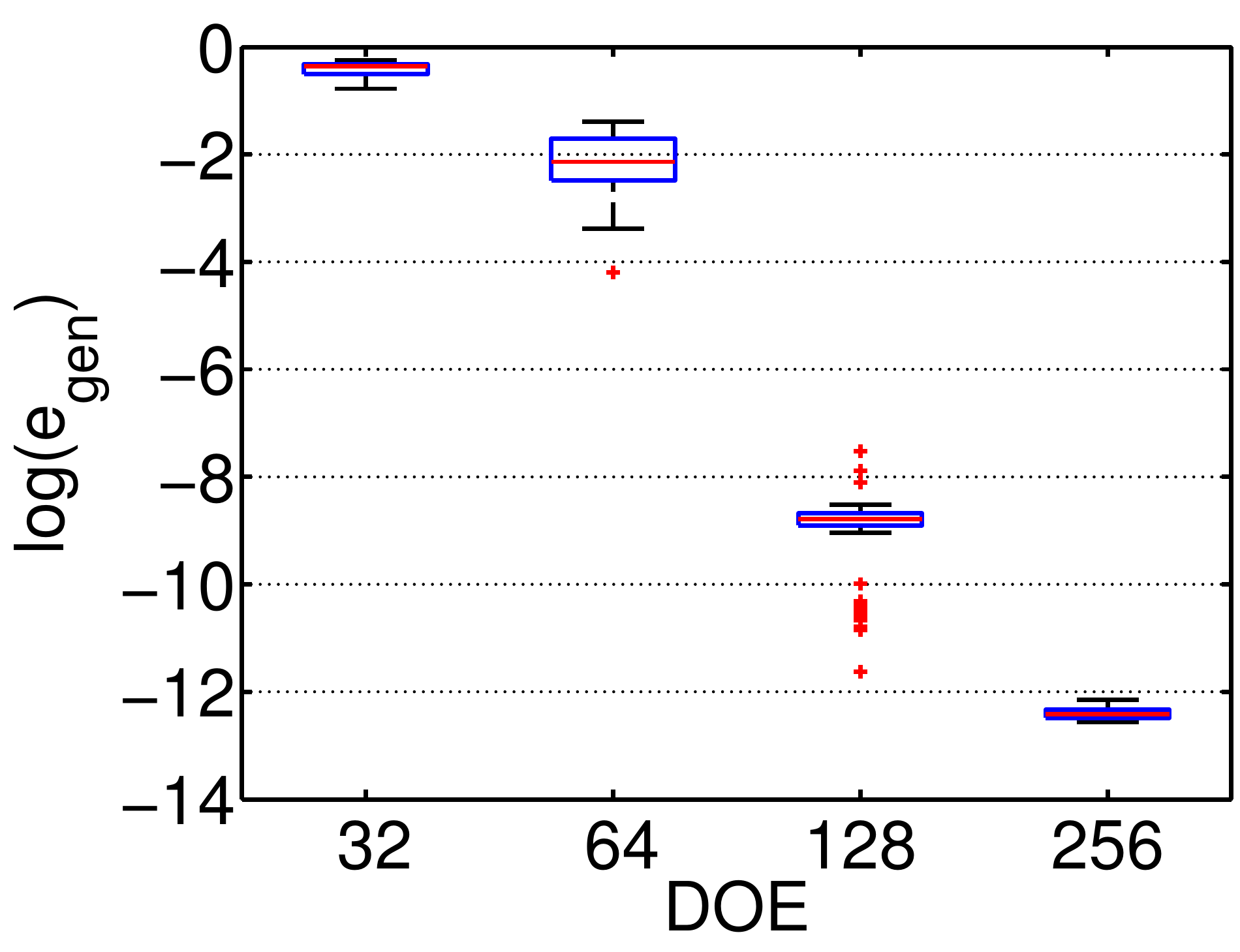}
  }
  \caption{\label{fig:large} The Ishigami function -- Relative
    generalization error (Eq.~(\ref{eq:errgen})) associated with large
    experimental designs for the various meta-modeling approaches}
\end{figure}

When the resources for experiments are limited the focus lies on doing
as few computational model runs as possible as discussed in the previous
section. In order to describe the entire behavior of the meta-modeling
approaches, the Ishigami function is also studied now for larger
experimental designs in order to assess the convergence of the various
schemes. Results are shown in Fig.~\ref{fig:large}. This figure
illustrates the evolution of the relative generalization error from
small to large sample sizes on the \emph{logarithmic} (base 10) scale.
The error measure decreases fast when enlarging the sample set because
the Ishigami function is composed of sine and cosine functions which can
be approximated well with series of polynomials, \emph{e.g.} as in a
Taylor expansion. Thus PCE is the dominating effect for PC-Kriging.

For large sample sizes, ordinary Kriging is outperformed by the other
three approaches. Kriging in general works well with small sample sizes.
If too many samples are used, the interpolation algorithm becomes
unstable due to singularities and bad conditioning in the
auto-correlation matrix. If a large sample size is available, regional
Kriging models on a subset of samples, \emph{e.g.} the neighboring
samples, are more suitable \cite{Dubrule1983}.

The performance of PC-Kriging and also PCE for a large number of samples
(here 128 and 256~samples in Fig.~\ref{fig:large}) is in the order of
magnitude of the machine computation precision. Errors around
$\epsilon_{gen}\approx 10^{-12}$ originate from numerical round-off
errors which are not reducible by adding more samples. It is
questionable though, whether in reality such a high accuracy in the
meta-model prediction is needed.

\subsubsection{Evolution of the error measures}
\begin{figure}
  \centering
  \includegraphics[width=0.6\linewidth, angle=0]{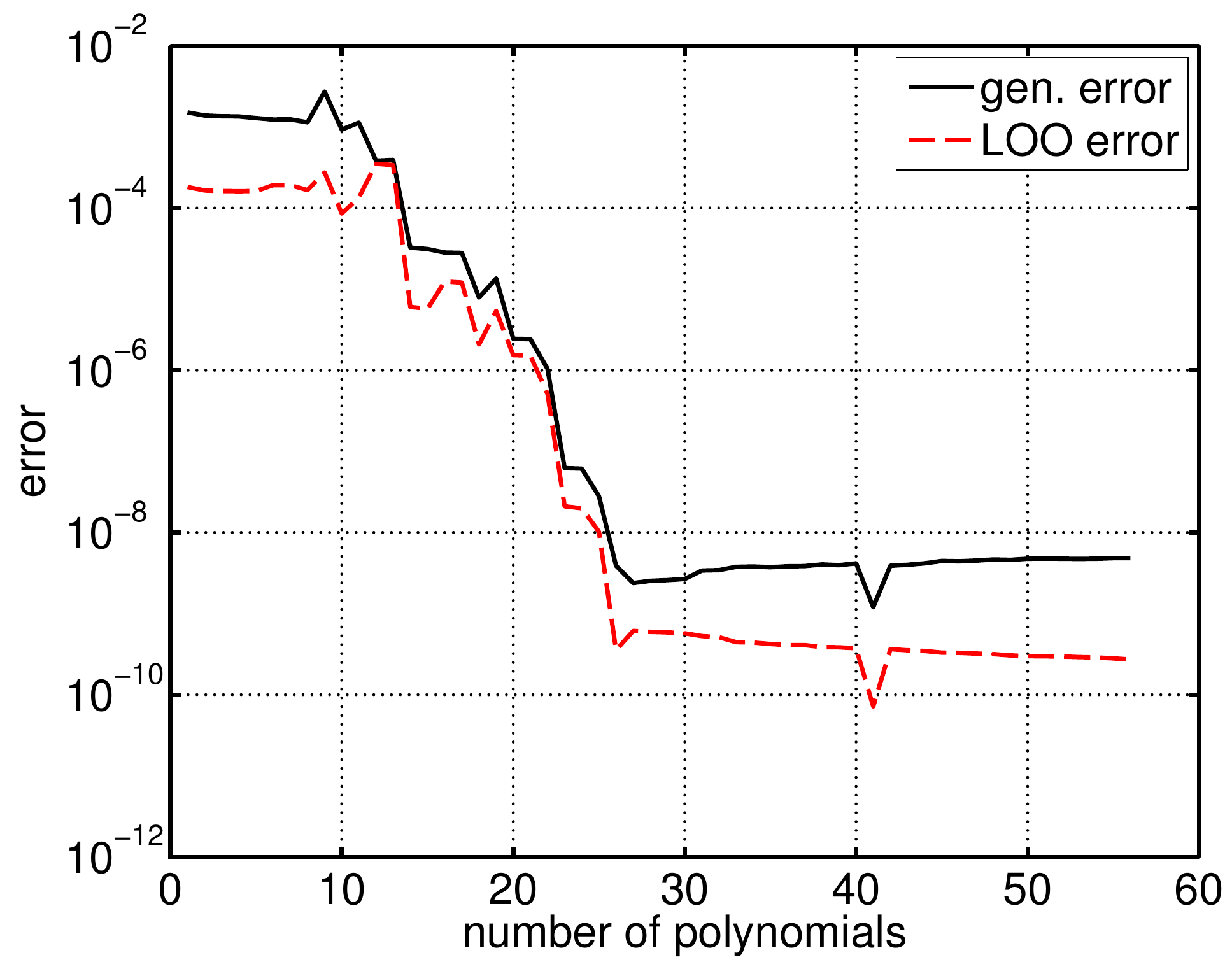}
  \caption{\label{fig:loocvpck}  Evolution of the leave-one-out error (LOO) and relative generalization error inside the OPC-Kriging algorithm as a function of the number of polynomials in the regression part}
\end{figure}

The OPC-Kriging algorithm includes the tracking of the LOO error to
optimally choose the sparse set of orthonormal polynomials. The
evolution of the LOO error for the Ishigami function and a sample size
of $N=128$ samples is presented in Fig.~\ref{fig:loocvpck}. The
experimental-design-based LOO error (dashed, red line) is compared to
the relative generalization error which is shown as the solid black
line.

The first point to notice is that the LOO error slightly under-predicts
the true value of the relative generalization error for all sizes of
polynomial sets. This is due to the fact that the LOO error is based
solely on the information contained in the experimental design samples
whereas the relative generalization error is based on a large validation
set ($n=10^5$). Although there is an inherent under-prediction, the
overall behavior of the two error measures is similar. Thus the choice
of the optimal set of polynomials for the OPC-Kriging can be based on
the LOO error, which is obtained as a by-product of the procedure used
to fit the parameters of the PC-Kriging model. In the example case of
Fig.~\ref{fig:loocvpck}, choosing only half of the polynomials,
\emph{i.e.} $P=27$ leads to a meta-model which is almost as accurate as
using all $56$~polynomials. The optimal set of polynomials can be chosen
at the point where the decrease in LOO error becomes insignificant. This
reduces the number of polynomials needed and thus also reduces the
complexity of the OPC-Kriging meta-model.

\section{Conclusion}\label{sec:c}
In the context of increasing computational power, computational models
in engineering sciences have become more and more complex. Many analyses
such as reliability assessment or design optimization, require repeated
runs of such computational models which may be infeasible due to
resource limitations. To overcome these limitations computational models
are nowadays approximated by easy-to-evaluate functions called
meta-models.

This paper summarized the principles of two popular non-intrusive
meta-modeling techniques, namely Polynomial Chaos Expansions (PCE) and
Kriging (also called Gaussian process modeling). Then the combination of
the two approaches to a new meta-modeling approach called
Polynomial-Chaos-Kriging (PC-Kriging) is proposed. Two formulations of
PC-Kriging are introduced in this paper, namely Optimal-PC-Kriging
(OPC-Kriging) and Sequential-PC-Kriging (SPC-Kriging). SPC-Kriging
employs first a least-angle-regression (LAR) algorithm to determine the
optimal sparse set of orthonormal polynomials in the input space. Then,
the set of polynomials is used as the trend of a universal Kriging
meta-model. OPC-Kriging employs the same least-angle-regression
algorithm as SPC-Kriging, yet iteratively adds polynomials to the trend
part of the universal Kriging model one-by-one, and fit the
hyperparameters of the auto-correlation function in each iteration. In
this case polynomials are added to the trend in the order they are
selected by the LAR algorithm. Based on the LOO error the best
meta-model is found to be the OPC-Kriging meta-model.

The performance of the four approaches (ordinary Kriging, PCE,
SPC-Kriging, OPC-Kriging) is compared in terms of the relative
generalization error on benchmark analytical functions.  The results
show that PC-Kriging is better than, or at least as good as the distinct
approaches for small experimental designs. Specifically, OPC-Kriging is
preferable to SPC-Kriging as it reduces the number of polynomials in the
regression part and thus reduces the complexity of the meta-model, at a
computational calibration cost which is however higher than that of
SPC-Kriging.

The analysis of the performance of PC-Kriging is limited to some
benchmark analytical functions in this paper. The ongoing research
applies PC-Kriging to realistic engineering problems such as reliability
analysis or design optimization. The idea of adaptive experimental
designs (also called design enrichment) is introduced in order to
increase the accuracy of the surrogate in some specific regions of the
input space (\eg close to the zero-level of the limit state function in
reliability analysis) instead of everywhere. The initialization of the
iterative algorithm is a small initial experimental design to which
points are added in regions of interest. These added points will then
increase the quality of the meta-model specifically in those regions.
Preliminary ideas developed in \cite{SchoebiIFIP2014,SchoebiRouen2014,
  SchoebiCSM2014} are currently investigated in details.

\section*{Acknowledgements}
This research was carried out with the support of Orange Labs,
Issy-les-Moulineaux, France under the research agreement \#C12063. This
support is gratefully acknowledged.

\bibliographystyle{chicago}
\bibliography{IJ4UQ}

\end{document}